\documentclass[a4paper,11pt]{article}
\pdfoutput=1 

\usepackage{jheppub} 

\usepackage{graphicx} 
\usepackage{epstopdf}
\usepackage{amsmath,amssymb,mathrsfs}
\usepackage{ bbold }
\usepackage[dvipsnames]{xcolor}

\usepackage{tikz}\usetikzlibrary{decorations.pathmorphing}
\usetikzlibrary{shapes.geometric}
\usetikzlibrary{calc}
\usetikzlibrary{shapes.misc}
\usetikzlibrary{decorations.markings}
\tikzstyle{gluon}=[decorate, decoration={coil,aspect=0.8, amplitude=1.5pt,  segment length=3pt}]
\tikzstyle{lgluon}=[decorate, decoration={coil,aspect=-0.8, amplitude=1.5pt,  segment length=3pt}]
\tikzstyle{dabc}=[fill=black!30!white,star, star points=5, star point ratio=2.25, draw,inner sep=1.3pt]

\newcommand{\req}[1]{{Eq.~(\ref{#1})}}

\newcommand{\tr}{\mbox{tr}}
\newcommand{\dd}{\ensuremath{\text{d}^2}}
\newcommand{\Nc}{{\ensuremath{N_c}}}
\newcommand{\thalf}{{\tfrac{1}{2}}}
\newcommand{\hcalf}{{\ensuremath{\hat{\mathcal{F}}}}}
\newcommand{\hcalo}{{\ensuremath{\hat{\mathcal{O}}}}}
\newcommand{\cala}{{\ensuremath{\mathcal{A}}}}
\newcommand{\calb}{{\ensuremath{\mathcal{B}}}}
\newcommand{\calf}{{\ensuremath{\mathcal{F}}}}
\newcommand{\calo}{{\ensuremath{\mathcal{O}}}}
\newcommand{\calp}{{\ensuremath{\mathcal{P}}}}
\newcommand{\calv}{{\ensuremath{\mathcal{V}}}}

\newcommand{\calw}{{\ensuremath{\mathcal{W}}}}
\newcommand{\calg}{{\ensuremath{\mathcal{G}}}}
\newcommand{\cald}{{\ensuremath{\mathcal{D}}}}
\newcommand{\xvec}{{\ensuremath{\underline{x}}}}
\newcommand{\yvec}{{\ensuremath{\underline{y}}}}
\newcommand{\zvec}{{\ensuremath{\underline{z}}}}
\newcommand{\uvec}{{\ensuremath{\underline{u}}}}

\newcommand{\wvec}{{\ensuremath{\underline{w}}}}

\newcommand{\kvec}{{\ensuremath{\underline{k}}}}

\title{\boldmath Small-\texorpdfstring{$x$}{x} TMD  distributions initial condition: \texorpdfstring{$N_c$}{Nc}-dependence and Gaussian approximations}

\author[a,1]{Florian Cougoulic,\note{Corresponding author.}}
\author[a]{Piotr Korcyl}
\author[a]{and Tomasz Stebel}

\affiliation[a]{Institute of Theoretical Physics, Jagiellonian University, ul. \L ojasiewicza 11, 30-348 Krak\'ow, Poland}

\emailAdd{florian.cougoulic@uj.edu.pl}
\emailAdd{piotr.korcyl@uj.edu.pl}
\emailAdd{tomasz.stebel@uj.edu.pl}

\abstract{
We systematically derive expressions for ten small-$x$ Transverse Momentum Dependent (TMD) distributions in the Gaussian approximation, three in the quark-gluon sector and seven in the gluon-gluon sector; for the general $SU(N_c)$ gauge group. The derived formulae depend on the logarithm of the dipole amplitude. Using the McLerran-Venugopalan model for the initial condition, we simulate the dipole amplitude as well as estimate all ten TMD distributions for $N_c \in \{2,3,4,5\}$. We compare these explicit numerical results with the derived expressions and find very good agreement for all studied values of $N_c$. Consequently, we study the scaling with $N_c$ of the TMD distributions. Thanks to that, we are able to derive the large-$N_c$ limit of the Gaussian approximation results and show that they agree with expressions obtained in the mean-field approximation. By comparing with our numerical results, we are able to demonstrate the size, origin, and significance of subleading-$N_c$ corrections. This work sets the stage for a systematic study of subleading corrections induced by the JIMWLK evolution in the rapidity evolution of the TMD distributions. Interestingly, we find an exact sum rule that relates all seven gluon-gluon TMD operators at $N_c = 3$ and arbitrary $x$.}

\begin{document} 
\maketitle
\flushbottom

\section{Introduction}
\label{sec:intro}

The study of transverse momentum dependent (TMD) distributions is one of the pillars for the theoretical understanding of high-energy collisions, see \cite{Boussarie:2023izj} for a review. 
These distributions naturally emerge at high energies, where the large separation between the scales of the problem allows the observable to be written in a factorized form \cite{Collins:2011zzd}.
Over the years, they have been widely used to describe observables in QCD-induced processes, such as: 
distribution of heavy bosons at small transverse momentum 
\cite{Collins:1984kg,Qiu:2000ga,Ji:2004xq},
spin asymmetries 
\cite{Ralston:1979ys,Sivers:1989cc,Sivers:1990fh,Collins:1992kk,Boer:2003cm},
the scattering of hadrons at high-energy 
\cite{Lipatov:1985uk,Catani:1990eg}, and many others.

In recent years, the TMD distributions have also attracted a lot of attention in the context of small-$x$ physics. \footnote{For example, a lot of effort is currently being put into joint resummation of Collins–Soper–Sterman soft gluon emission corrections \cite{Collins:1981uk,Collins:1981uw,Collins:1984kg} and small $x$ gluon emissions \cite{Mueller:2012uf,Mueller:2013wwa,Zhou:2018lfq,Xiao:2017yya,Hatta:2020bgy,Hatta:2021jcd,Hentschinski:2021lsh,Taels:2022tza,Caucal:2022ulg,Caucal:2023nci,Caucal:2023fsf,Caucal:2024bae,Caucal:2024vbv,Caucal:2024nsb,Caucal:2025mth,Duan:2024nlr,Duan:2024qev,Mukherjee:2023snp}. {Another direction is analysis of the sea-quark TMD distributions \cite{Gelis:2006hy, Marquet:2009ca,Kang:2012vm,Tong:2022zwp,Hatta:2022lzj,Hauksson:2024bvv,Caucal:2025xxh}.}}
In the limit of low Bjorken $x$, TMD distributions are expressed in terms of operators evaluated against a target state described by a \textit{color glass condensate} (CGC). See the reviews \cite{Gelis:2010nm,Albacete:2014fwa,Blaizot:2016qgz,Morreale:2021pnn, Raj:2025hse} and the references therein.
The dependence of the dipole and TMD distributions on rapidity can be obtained using the JIMWLK equation \cite{Jalilian-Marian:1997dw,Jalilian-Marian:1997gr,Weigert:2000gi,Iancu:2001ad,Iancu:2000hn,Ferreiro:2001qy}, alternatively, the Balitsky hierarchy \cite{Balitsky:1995ub,Balitsky:1998kc,Balitsky:1998ya}, or its large-$N_c$ truncation: the BK equation \cite{Balitsky:1995ub,Balitsky:1998ya,Kovchegov:1999yj,Kovchegov:1999ua}.
These sum the logarithmically enhanced parameter, $\alpha_s \ln\frac{1}{x}$, associated with each gluon emission, see for example \cite{Kovchegov:2012mbw}.
For distributions that can be expressed from the dipole and do not contain any strong $1/N_c$ correction, it has been studied in \cite{Rummukainen:2003ns,Kovchegov:2008mk} that the agreement between the solutions of the two evolution equations is better than the naive counting $\calo(1/N_c^2)$ by an order of magnitude.
This is quantified by the evaluation of $\Delta$ defined as:
\begin{align*}
    \Delta_{\xvec,\yvec;\zvec} = \langle \hat{S}_{\xvec\zvec} \hat{S}_{\zvec\yvec} \rangle - \langle \hat{S}_{\xvec\zvec} \rangle \langle \hat{S}_{\zvec\yvec} \rangle \sim \calo(1/N_c^2) \overset{N_c=3}{\longrightarrow} \calo(10\%) \overset{\text{observed}}{\longrightarrow} \calo(1\%)
\end{align*}
which appears when one subtracts the BK equation from the JIMWLK equation for the evolution of a dipole $\langle \hat{S}_{\xvec\yvec}\rangle$.
It is thus natural to question how this conclusion extends to distributions whose initial conditions may contain large $1/N_c$ corrections. 
Interesting candidates, which we focus on in this manuscript, are leading twist TMD distributions.
Similar to the dipole, these TMD distributions depend only on two transverse coordinates. The novelty is that the operators depend on two local insertions of the field strength tensor $F^{\mu\nu}$, in addition to Wilson Lines (WL). Those can be related, in the appropriate gauge, to derivatives acting on WLs.
Another interpretation for some TMD distribution, such as the Weizs\"acker-Williams gluon distribution, is the relation to coincidence limits of a quadrupole operator \cite{Dominguez:2011wm}. 
Thus, the study of the $N_c$-dependence of the solutions of the evolution equations for TMD distributions relates to the analysis performed \cite{Dumitru:2011vk} for the quadrupole. It has been shown in the latter that the solution of the JIMWLK equation for the quadrupole agrees with a careful analysis of the Gaussian approximation at $N_c = 3$. While the agreement has been confirmed at short distances with the $N_c \rightarrow \infty$ limit, significant deviations are anticipated at large distances.
This motivates the analysis of TMD distributions' initial condition and, in particular, the large separation region in which strong $1/N_c$ corrections can appear.

In this work, we achieve several goals. First, we derive expressions for the Gaussian approximation of the TMD distributions, which turn out to agree with the explicit numerical evaluation for the McLerran--Venugopalan (MV) initial condition \cite{McLerran:1993ni,McLerran:1993ka,McLerran:1994vd} in a wide range of separations. 
Second, we introduce appropriate $N_c$ dependent rescalings that allow us to identify building blocks which are $N_c$-independent (or almost $N_c$-independent) and can be explicitly related to irreducible representations {(irrep)} and their charges. We express all studied TMD operators {as linear combination of } these blocks
{, which we label $\hat{\Omega}$. The relation to irrep allows us to systematize the calculation of the Gaussian approximation by making use of an invariant of the theory, namely the quadratic Casimir operator (the projection onto an irrep of a set of Wilson lines at the same coordinate, is understood as a single Wilson line at the same coordinate in the corresponding irrep).}
Third, based on these results, we work out the large-$N_c$ limit, which we compare with the approximations obtained using the mean-field approach. This allows us to identify the subleading corrections together with their magnitude and range of separations where they can be dominant. By identifying the sources of $1/N_c$ corrections in the MV initial condition and in the very definition of the TMD distributions, we lay the groundwork for investigating $1/N_c$ corrections induced by the evolution. 

Our manuscript is composed as follows. In section \ref{sec:Op_small-x}, we introduce the notation for small-$x$ operators such as the dipole and TMD operators, as well as their birdtrack representation \cite{Cvitanovic:1976am,Cvitanovic:2008zz,Dokshitzer:1995fv,Keppeler:2017kwt,Peigne:2024srm}, which allows us to introduce the operators $\hat{\Omega}$ for which the interpretation in terms of the quadratic Casimir is explicit.
In section \ref{Sec:Gaussian_approx}, we compute the expectation value of operators $\hat{\Omega}$ within the Gaussian approximation for any values of $N_c$. The methodology used is similar to that used in previous quadrupole calculations, employing a pictorial representation, as seen in, {e.g.} \cite{Gelis:2001da,Dominguez:2008aa,Dominguez:2011wm}, or a more recent generalization at next-to-eikonal order \cite{Agostini:2025vvx}.
In section \ref{sec:num_result_omega}, we present the numerical results obtained for $\langle \hat{\Omega} \rangle$ from simulations according to the MV model for $N_c \in \{2,3,4,5\}$, and compare them to the results of the Gaussian approximation. Particular emphasis is placed on the scaling obtained in those distributions.
In section \ref{sec: tmds from omega}, we relate the expectation values of operators $\hat{\Omega}$s to those of the TMD operators and explicitly calculate the large-$N_c$ limit of the Gaussian approximation for the TMD distributions.
In section \ref{sec:num_result_tmds}, we present the numerical results for TMD distributions obtained from the MV model and compare them to the Gaussian approximation and the results from the mean-field approximation in the literature, as discussed in \cite{Caucal:2025zkl}. 
{The reader may also consult similar results in \cite{Dominguez:2010xd,Dominguez:2011wm,vanHameren:2016ftb} addressing some of those TMD distributions in the Gaussian approximation.}
As we evaluate the TMD distributions over the range $N_c \in \{2,3,4,5\}$, we explicitly demonstrate the convergence toward large-$N_c$ results. However, we also illustrate the significance of the $1/N_c$ contributions at $N_c = 3$.
In section \ref{sec: Conclusion and prospects}, we conclude with a summary of the findings and highlight some prospects for future research.

\section{Operators at small-\texorpdfstring{$x$}{x}: TMD operators and irreps}
\label{sec:Op_small-x}

In this section, we list the small-$x$ TMD operators of interest for this study. We also introduce the operator $\hat{\Omega}^\omega$ and how it relates to TMD operators. This operator has a simple interpretation in terms of color charges, which will be useful for highlighting the $N_c$ scaling in the following sections.

\subsection{Wilson lines and dipole}

We introduce a first light-cone direction $n^\mu$, where $n \cdot n =0$.
The opposite light-cone $\overline{n}^\mu$ is denoted by $\overline{n}$ such that $n \cdot \overline{n} = 1$ and $\overline{n}\cdot \overline{n} = 0$.
The projectile light-cone gauge reads $A \cdot n= 0$, and we denote the large (\textit{eikonal}) component of the gauge field as $\alpha = \overline{n}\cdot A$. In this gauge, we have
\begin{equation}
    A^\mu dx_\mu \sim (A\cdot \overline{n})\ d(n\cdot x) \equiv \alpha\ dt
\end{equation}
where the corrections are beyond the eikonal approximation and the scope of the manuscript.
The symbol $t$ denotes the light-cone time along the propagation of the projectile $x_{proj}^\mu = (n \cdot x)\, \overline{n}^\mu$, and will be used instead of explicit $x^\pm$ in order to be agnostic over the choice of coordinate frame.

\noindent
Wilson lines (WLs) in the fundamental (resp.~adjoint) irreducible representation (irrep) are denoted by $V$ (resp.~$U$)\footnote{The sign convention is related to the Covariant derivative, which we write as $D^\mu = \partial^\mu - ig A^\mu$ in the fundamental representation.}:
\begin{equation}
    V_\xvec[t_1,t_0] \equiv P \exp \left\{ ig \int_{t_0}^{t_1} dt\ \alpha^a(t,\xvec) t^a_F \right\},
\end{equation}
where the path-ordering symbol $P$ orders from right to left insertions of the field $\alpha^a t^a$ along the light-cone trajectory from $t_0$ to $t_1$.
The position $\xvec$ is in the plane transverse to both $n$ and $\overline{n}$.
Infinite WLs are written with the following short-handed notation
\begin{equation}
    V_\xvec \equiv V_\xvec[\infty,-\infty].
\end{equation}
The following notation for the hermitian conjugate of a WL proves to be convenient to underline the contours involved in this study:
\begin{equation}
    V^\dagger_\xvec[a,b] \equiv \left( V_\xvec[b,a] \right)^\dagger.
\end{equation}
All WLs written in the rest of the manuscript will follow those conventions.
We define the dipole operator as:
\begin{equation}
    \hat{S}(\xvec,\yvec) = \frac{1}{\Nc} \tr \left[V_\xvec V^\dagger_\yvec \right].
\end{equation}
By convention, we will reserve the hat symbol to denote operators that will be averaged. The same quantity without the hat symbol will denote the averaged quantity when it applies.

Let us mention that we will only focus on operators that are not explicitly gauge-invariant. The scope of this manuscript is about operators encountered at small-$x$, and their numerical evaluation, for which the gauge is usually chosen such that the transverse gauge link at $\pm \infty$ can be neglected, i.e., the projectile light-cone gauge, or the Feynman gauge. Gauge invariance can easily be recovered from the forms employed in this manuscript and are already available in the literature\footnote{Some care must be taken with conventions for Wilson lines whenever traces of four or more WL are considered.} \cite{Marquet:2016cgx,Bury:2018kvg}.

\subsection{TMD operators}
We focus on leading-twist unpolarized TMD distributions evaluated at small-$x$. Corresponding TMD operators contain two insertions of the field strength tensor within gauge link contours.
Due to our gauge choice, they can be related to the action of the derivative on a WL.
As an illustration, let us consider the past pointing staple. It involves the following operator
\begin{equation}
    \overline{n}_\mu \overline{n}_\nu \delta_{jk}^\perp \int dt_x dt_y\  F^{j\mu}(t_x,\xvec) U_\xvec[t_x,-\infty] U^\dagger_\yvec[-\infty,t_y] F^{k\nu}(t_y, \yvec)
\end{equation}
for which the transverse link at $-\infty$ is assumed to be unity by gauge choice.
To extract the derivatives of WLs, we note
\begin{equation}
   \overline{n}_\mu F^{j\mu} = \partial^j \alpha.
\end{equation}
Using this simplification, for the part of the contour at the transverse position $\xvec$, we write
\begin{subequations}
\begin{align}
    \overline{n}_\mu\int dt\ F^{j\mu}(t,\xvec) U_\xvec[t,-\infty]
    &= \int dt\ (\partial^j \alpha) U_\xvec[t,-\infty] \\
    &= \int dt\  V^\dagger_\xvec[-\infty,\infty]V_\xvec[\infty,t] (\partial^j \alpha) V_\xvec[t,-\infty] \\
    &= V^\dagger_\xvec \left( \tfrac{1}{ig} \partial^j V_\xvec \right)
\end{align}
\end{subequations}
This last structure is of particular interest, and some care must be taken when it is evaluated numerically as discretization effects can break some of its symmetries, such as anti-hermicity or belonging to the adjoint irrep of $SU(N_c)$.

\paragraph{Substitution for derivatives.}
Following the remark in \cite{Marquet:2016cgx} about the numerical implementation of the derivative on the lattice, we define the finite difference symbol $D_i$ to be given by\footnote{The central difference is evaluated to second order accuracy using the coefficients $(1;0;-1)$. We also check the implementation to fourth order accuracy using the coefficients $(-1/6;4/3;0;-4/3;1/6)$.}
\begin{equation}
D_i V_\xvec \equiv \frac{1}{2} \left( V_{\xvec + e_i} - V_{\xvec - e_i} \right).
\end{equation}
where $e_i$ denotes the lattice step in the direction $i$.
Each derivatives of a WL appearing in the combination $\left( V^\dagger_\xvec \partial^i_\xvec V_\xvec \right)$ are replaced according to:
\begin{equation}
\left( V^\dagger_\xvec \partial^i_\xvec V_\xvec \right) 
\longrightarrow
\mathcal{A}_\xvec^i = \frac{1}{2a} \left[ \left( V_\xvec^\dagger D_i V_\xvec \right) - \left( V_\xvec^\dagger D_i V_\xvec \right)^\dagger \right].
\end{equation}
where $a$ is the lattice spacing. This replacement is made to ensure anti-hermicity of the operator. We observe that the endpoints of the operator $\cala^i$ are both at $-\infty$. 
\begin{equation}
    \left( V^\dagger_\xvec \partial^i_\xvec V_\xvec \right)_{j\ell}  = \left(V_\xvec^\dagger[-\infty,+\infty]\right)_{jk} \left(\partial^i V_\xvec[+\infty,-\infty] \right)_{k\ell}
\end{equation}
This operator is thus conveniently used to express backward staples.
For convenience, we introduce the operator  $\mathcal{B}^i$ to be the reflection of $\cala^i$:
\begin{equation}
    \calb^i = (\partial^i V) V^\dagger = \left(\partial^i V[+\infty,-\infty]\right) V^\dagger[-\infty,+\infty],
\end{equation}
which has both endpoints at $+\infty$. This operator will be convenient to express forward staples.
To enforce the operators $\cala^i$ and $\calb^i$ to be in the adjoint irrep, we also introduce the corresponding projection:
\begin{subequations}
\begin{align}
\cala^i &\longrightarrow 2\, \tr \left[ \cala^i t^a\right] t^a = \cala^i - \frac{1}{N_c} \tr \left( \cala^i \right) \mathbb{1}\\
\calb^i &\longrightarrow 2\, \tr \left[ \calb^i t^a\right] t^a = \calb^i - \frac{1}{N_c} \tr \left( \calb^i \right) \mathbb{1}
\end{align}
\end{subequations}
where we used Fierz identity on the {r.h.s.} of the equal sign. This will ensure that traces of $\cala$ and $\calb$ vanish for any discretization length $a = L/N$ used for the numerical evaluations.

\paragraph{Quark - Gluon:} according to \cite{Marquet:2016cgx,Bury:2018kvg}, we write TMD operators as:
\begin{subequations}\label{def_TMDqg_WL}
\begin{align}
    \hcalf_{qg}^{(1)}(\xvec,\yvec) 
    &= \tr \left[ (\partial_i V^\dagger_\xvec) (\partial_i V_\yvec) \right] 
    = - \tr \left[\cala^i_\xvec V^\dagger_\xvec V_\yvec \cala^i_\yvec \right] \notag \\
    &= - \tr \left[V_\xvec^\dagger \calb^i_\xvec \calb^i_\yvec V_\yvec\right]
    = \tr \left[ (\calb^i_\xvec V_\xvec)^\dagger \calb^i_\yvec V_\yvec \right],\\
    -N_c \hcalf_{qg}^{(2)}(\xvec,\yvec) 
    &=  \tr \left[ (\partial V_\xvec) V^\dagger_\xvec (\partial V_\yvec) V^\dagger_\yvec \right] \tr \left[ V_\yvec V^\dagger_\xvec \right] \notag \\
    &= \tr\left[\calb^i_\xvec \calb^i_\yvec\right] \tr \left[ V_\yvec V^\dagger_\xvec \right]\\
    -\hcalf_{qg}^{(3)}(\xvec,\yvec) 
    &= \tr \left[ (\partial V_\xvec) V^\dagger_\xvec \left[ V_\yvec V^\dagger_\xvec \right](\partial V_\yvec) V^\dagger_\yvec \right] \notag \\
    &= \tr \left[ \calb^i_\xvec V_\yvec V^\dagger_\xvec \calb^i_\yvec \right]
\end{align}
\end{subequations}
Let us remark that in the rhs of those three equations, we have the same propagating systems $q\overline{q} \rightarrow qgg\overline{q}$. The only distinction being the way traces are being performed at $+\infty$. This observation will prove to be powerful in order to isolate involved charges.

\paragraph{Gluon - Gluon:} according to \cite{Marquet:2016cgx,Bury:2018kvg}, we write TMD operators as:
\begin{subequations}\label{def_TMDgg_WL}
\begin{align}
    N_c \hcalf_{gg}^{(1)}(\xvec,\yvec)
    &=  \tr \left[ (\partial_i V^\dagger_\xvec) (\partial_i V_\yvec) \right] \tr \left[ V_\xvec V^\dagger_\yvec\right] 
    = - \tr \left[\cala^i_\xvec V^\dagger_\xvec V_\yvec \cala^i_\yvec \right] \tr \left[ V_\xvec V^\dagger_\yvec\right] \notag \\
    &= - \tr \left[V_\xvec^\dagger \calb^i_\xvec \calb^i_\yvec V_\yvec\right] \tr \left[ V_\xvec V^\dagger_\yvec\right] 
    = \hcalf_{qg}^{(1)} \ \tr \left[ V_\xvec V^\dagger_\yvec\right], \\
    -N_c \hcalf_{gg}^{(2)}(\xvec,\yvec)
    &= \tr \left[ (\partial_i V_\yvec) V_\xvec^\dagger\right]  \tr \left[ (\partial_i V_\xvec) V^\dagger_\yvec\right] \notag \\
    &= \tr \left[ V_\yvec \cala^i_\yvec V_\xvec^\dagger\right]  \tr \left[ V_\xvec \cala^i_\xvec V^\dagger_\yvec\right] 
    = \tr \left[ \calb^i_\yvec V_\yvec V_\xvec^\dagger\right]  \tr \left[ \calb^i_\xvec V_\xvec V^\dagger_\yvec\right],  \\
    - \hcalf_{gg}^{(3)}(\xvec,\yvec)
    &= \tr \left[ (\partial_i V_\xvec)V^\dagger_\xvec (\partial_i V_\yvec)V^\dagger_\yvec\right]
    = \tr \left[ \calb^i_\xvec \calb^i_\yvec\right], \\
    - \hcalf_{gg}^{(4)}(\xvec,\yvec)
    &= \tr \left[ V^\dagger_\xvec(\partial_i V_\xvec) V^\dagger_\yvec (\partial_i V_\yvec)\right]
    = \tr \left[ \cala^i_\xvec \cala^i_\yvec\right], \\
    -\hcalf_{gg}^{(5)}(\xvec,\yvec)
    &= \tr \left[ V_\yvec V^\dagger_\xvec (\partial_i V_\yvec) V^\dagger_\yvec V_\xvec V_\yvec^\dagger (\partial_i V_\xvec) V_\xvec^\dagger \right] 
    = \tr \left[ V_\yvec V^\dagger_\xvec \calb^i_\yvec V_\xvec V_\yvec^\dagger \calb^i_\xvec \right] \\
    -N_c^2 \hcalf_{gg}^{(6)}(\xvec,\yvec)
    &= \tr \left[ (\partial_i V_\xvec)V^\dagger_\xvec (\partial_i V_\yvec)V^\dagger_\yvec\right]  \tr \left[V_\xvec V^\dagger_\yvec\right]  \tr \left[ V_\yvec V_\xvec^\dagger \right] \notag \\
    &= \tr \left[ \calb^i_\xvec \calb^i_\yvec\right]  \tr \left[V_\xvec V^\dagger_\yvec\right]  \tr \left[ V_\yvec V_\xvec^\dagger \right] ,\\
    -N_c \hcalf_{gg}^{(7)}(\xvec,\yvec)
    &= \tr \left[ V_\xvec V^\dagger_\yvec (\partial_i V_\xvec) V^\dagger_\xvec (\partial_i V_\yvec)V^\dagger_\yvec\right] \tr\left[V_\yvec V_\xvec^\dagger \right] \notag \\
    &= \tr \left[ V_\xvec V^\dagger_\yvec  \calb_\xvec^i \calb^i_\yvec\right] \tr\left[V_\yvec V_\xvec^\dagger \right].
\end{align}
\end{subequations}
Massaging the operator $\hcalf^{(3)}_{gg}$ into (a similar exercise can be done for $\hcalf^{(4)}_{gg}$)
\begin{equation}
    - \hcalf_{gg}^{(3)}(\xvec,\yvec)
    = \tr \left[ \calb^i_\xvec \calb^i_\yvec\right]
    = \tr \left[ \calb^i_\xvec \calb^i_\yvec\right] \frac{1}{N_c} \tr \left[ V^\dagger_\xvec V_\xvec V^\dagger_\yvec V_\yvec\right],
\end{equation}
we notice that all gluon-gluon TMD operators are projections of the tensor product:
\begin{equation}
    \left(\calb^i_\xvec \otimes V_\xvec  \otimes V^\dagger_\xvec \right) \otimes \left( \calb^i_\yvec \otimes V_\yvec  \otimes V^\dagger_\yvec \right).
\end{equation}

\subsection{Birdtrack representation: link between TMD operators and irreps}
\label{sec_birdtrack}
Let us introduce a birtrack \cite{Cvitanovic:1976am,Cvitanovic:2008zz,Dokshitzer:1995fv,Keppeler:2017kwt,Peigne:2024srm} representation of the previously defined operators.
We start with the definition of the operator $\calb^i$
\begin{equation}
\calb^i = 
\vcenter{\hbox{
\begin{tikzpicture}[scale=0.3]
    \draw[thick] (-5,1) -- ++ (10,0);\draw[thick,->] (-5,1) -- ++ (3,0);
    \draw[thick] (-5,1.5) -- ++ (10,0);\draw[thick,-<] (-5,1.5) -- ++ (3,0);
    \draw[fill=Blue!50!white] (1,1) circle (0.2);
    \draw[thick] (-5,1) to[out=180,in=180] (-5,1.5);
\end{tikzpicture}}}
=
\vcenter{\hbox{
\begin{tikzpicture}[scale=0.3]
    \draw[gluon] (-5,0) -- ++ (7,0);
    \draw[fill=Blue!50!white] (-5,0) circle (0.2);
    \draw[thick,->] (3,-.5) -- ++(1,0);
    \draw[thick,-<] (3,+.5) -- ++(1,0);
    \draw[thick] (5,-.5) -- (3,-.5) to[out=180,in=-90] (2,0) to[out=90,in=180] (3,.5) -- (5,.5);
\end{tikzpicture}}}.
\end{equation}
WLs are represented as horizontal plain lines with an arrow indicating the representation being fundamental or dual, or coiling for the adjoint irrep.
The bullet denotes the action of the derivative on the corresponding Wilson line, or equivalently, the insertion of $\partial^j \alpha$.
The state at $+\infty$ light-cone time is on the right side, while the state at the $-\infty$ light-cone time is on the left. \textit{vertical} lines represent the way the WLs are being traced over into the operator of interest, i.e., they are simply Kronecker deltas of the associated irrep.

Birdtracks are used to highlight the relation between the $3+7$ previously listed TMD operators and the corresponding projection into irreps.
This is a key step in order to identify potential $N_c$ - scaling relations between TMD distributions.

\paragraph{Quark - Gluon:}
In the following, we assume the top (resp. bottom) group of WLs to be at the coordinate $\xvec$ (resp. $\yvec$); and to the left (resp. right) are the traces performed at $-\infty$ (resp. $+\infty$) to define the TMD operators.
\begin{align}
-\, \hcalf^{(1)}_{qg} &= 
\vcenter{\hbox{
\begin{tikzpicture}[scale=0.3]
    \draw[thick] (-5,1.5) -- ++ (10,0);\draw[thick,->] (-5,1.5) -- ++ (3,0);
    \draw[thick] (-5,2) -- ++ (10,0);\draw[thick,-<] (-5,2) -- ++ (3,0);
    \draw[fill=Blue!50!white] (1,1.5) circle (0.2);
    \draw[thick] (-5,1) -- ++ (10,0);\draw[thick,-<] (-5,1) -- ++ (4,0);
    \draw[thick] (-5,-1) -- ++ (10,0);\draw[thick,->] (-5,-1) -- ++ (6,0);
    \draw[thick] (-5,-1.5) -- ++ (10,0);\draw[thick,-<] (-5,-1.5) -- ++ (7,0);
    \draw[thick] (-5,-2) -- ++ (10,0);\draw[thick,->] (-5,-2) -- ++ (7,0);
    \draw[fill=Blue!50!white] (-1,-2) circle (0.2);
    \draw[thick] (5,1.5) to[out=0,in=0] (5,1);
    \draw[thick] (5,-1) to[out=0,in=0] (5,-1.5);
    \draw[thick] (5,2) to[out=0,in=0] (5,-2);
    \draw[thick] (-5,2) to[out=180,in=180] (-5,1.5);
    \draw[thick] (-5,-2) to[out=180,in=180] (-5,-1.5);
    \draw[thick] (-5,1) to[out=180,in=180] (-5,-1);
\end{tikzpicture}
}}
= - 
\vcenter{\hbox{
\begin{tikzpicture}[scale=0.3]
    \draw[thick] (-5,1) -- ++ (10,0);\draw[thick,-<] (-5,1) -- ++ (4,0);
    \draw[fill=Blue!50!white] (1,1) circle (0.2);
    \draw[thick] (-5,-2) -- ++ (10,0);\draw[thick,->] (-5,-2) -- ++ (7,0);
    \draw[fill=Blue!50!white] (-1,-2) circle (0.2);
    \draw[thick] (5,1) to[out=0,in=0] (5,-2);
    \draw[thick] (-5,1) to[out=180,in=180] (-5,-2);
\end{tikzpicture}
}} \\[2ex]
-N_c\, \hcalf^{(2)}_{qg} &= 
\vcenter{\hbox{
\begin{tikzpicture}[scale=0.3]
    \draw[thick] (-5,1.5) -- ++ (10,0);\draw[thick,->] (-5,1.5) -- ++ (3,0);
    \draw[thick] (-5,2) -- ++ (10,0);\draw[thick,-<] (-5,2) -- ++ (3,0);
    \draw[fill=Blue!50!white] (1,1.5) circle (0.2);
    \draw[thick] (-5,1) -- ++ (10,0);\draw[thick,-<] (-5,1) -- ++ (4,0);
    \draw[thick] (-5,-1) -- ++ (10,0);\draw[thick,->] (-5,-1) -- ++ (6,0);
    \draw[thick] (-5,-1.5) -- ++ (10,0);\draw[thick,-<] (-5,-1.5) -- ++ (7,0);
    \draw[thick] (-5,-2) -- ++ (10,0);\draw[thick,->] (-5,-2) -- ++ (7,0);
    \draw[fill=Blue!50!white] (-1,-2) circle (0.2);
    \draw[thick] (5,1) to[out=0,in=0] (5,-1);
    \draw[thick] (5,2) to[out=0,in=0] (5,-2);
    \draw[thick] (5,1.5) to[out=0,in=0] (5,-1.5);
    \draw[thick] (-5,2) to[out=180,in=180] (-5,1.5);
    \draw[thick] (-5,-2) to[out=180,in=180] (-5,-1.5);
    \draw[thick] (-5,1) to[out=180,in=180] (-5,-1);
\end{tikzpicture}
}} \\[2ex]
- \hcalf^{(3)}_{qg} &= 
\vcenter{\hbox{
\begin{tikzpicture}[scale=0.3]
    \draw[thick] (-5,1.5) -- ++ (10,0);\draw[thick,->] (-5,1.5) -- ++ (3,0);
    \draw[thick] (-5,2) -- ++ (10,0);\draw[thick,-<] (-5,2) -- ++ (3,0);
    \draw[fill=Blue!50!white] (1,1.5) circle (0.2);
    \draw[thick] (-5,1) -- ++ (10,0);\draw[thick,-<] (-5,1) -- ++ (4,0);
    \draw[thick] (-5,-1) -- ++ (10,0);\draw[thick,->] (-5,-1) -- ++ (6,0);
    \draw[thick] (-5,-1.5) -- ++ (10,0);\draw[thick,-<] (-5,-1.5) -- ++ (7,0);
    \draw[thick] (-5,-2) -- ++ (10,0);\draw[thick,->] (-5,-2) -- ++ (7,0);
    \draw[fill=Blue!50!white] (-1,-2) circle (0.2);
    \draw[thick] (5,-1) to[out=0,in=0,looseness=2] (5,2);
    \draw[thick] (5,-1.5) to[out=0,in=0,looseness=1] (5,1.5);
    \draw[thick] (5,1) to[out=0,in=0,looseness=2] (5,-2);
    \draw[thick] (-5,2) to[out=180,in=180] (-5,1.5);
    \draw[thick] (-5,-2) to[out=180,in=180] (-5,-1.5);
    \draw[thick] (-5,1) to[out=180,in=180] (-5,-1);
\end{tikzpicture}
}}
\end{align}
Since $\calb^i$ is traceless, the projection onto the trivial triplet from $1 \otimes \overline{3}$ does not contribute, and we are left with only the states from $8 \otimes \overline{3}$:
\begin{equation}
   (\overline{3}\otimes 3 ) \otimes \overline{3} \quad \supset \quad  8 \otimes \overline{3} = \overline{3} \oplus 6 \oplus \overline{15}
\end{equation}
In order to single out the irrep, it is most convenient to consider the operator $\hat{\Omega}_{qg}$
\begin{align}
\hat{\Omega}_{qg}^\omega = 
{2} \times 
\vcenter{\hbox{
\begin{tikzpicture}[scale=0.3]
    \draw[thick] (-5,1.5) -- ++ (10,0);\draw[thick,->] (-5,1.5) -- ++ (3,0);
    \draw[thick] (-5,2) -- ++ (10,0);\draw[thick,-<] (-5,2) -- ++ (3,0);
    \draw[fill=Blue!50!white] (1,1.5) circle (0.2);
    \draw[thick] (-5,1) -- ++ (10,0);\draw[thick,-<] (-5,1) -- ++ (4,0);
    \draw[thick] (-5,-1) -- ++ (10,0);\draw[thick,->] (-5,-1) -- ++ (6,0);
    \draw[thick] (-5,-1.5) -- ++ (10,0);\draw[thick,-<] (-5,-1.5) -- ++ (7,0);
    \draw[thick] (-5,-2) -- ++ (10,0);\draw[thick,->] (-5,-2) -- ++ (7,0);
    \draw[fill=Blue!50!white] (-1,-2) circle (0.2);
    \draw[thick] (-5,2) to[out=180,in=180] (-5,1.5);
    \draw[thick] (-5,-2) to[out=180,in=180] (-5,-1.5);
    \draw[thick] (-5,1) to[out=180,in=180] (-5,-1);
    \draw[thick] (5.5,2) to[out=0,in=90,looseness=1.5] (6,1.75) to [out=-90,in=0, looseness=1.5] (5.5,1.5);
    \draw[thick] (5.5,-2) to[out=0,in=-90,looseness=1.5] (6,-1.75) to [out=90,in=0, looseness=1.5] (5.5,-1.5);
    \draw[gluon] (6,1.75) to[out=0,in=0,looseness=2] (6,-1.75);
    \draw[thick] (5.5,1) -- (6,1) to[out=0,in=0,looseness=2] (6,-1) -- (5.5,-1);
    \draw[fill=white] (7.5,0) ellipse (1.2 and .5) node {\scriptsize $\omega$};
\end{tikzpicture}}}
\end{align}
where $\omega$ is an irrep of the decomposition $8 \otimes \overline{3}$, and the right ellipse denotes the corresponding projection.\footnote{Let us comment on the normalization of the operator $\hat{\Omega}_{qg}$. To the lowest order in the coupling, the only non-vanishing contribution implies the correlation $\langle (\partial\alpha^a)_\xvec (\partial \alpha^b)_\yvec \rangle \propto \delta^{ab}$ from the action of the derivative. Replacing the remaining WLs by the identity makes explicit a factor $\tr\, \calp_{\omega} = K_\omega$, where $\calp_\omega$ is the projection operator onto the irrep $\omega$. This factor will naturally emerge when we perform the Gaussian approximation.}
The projections as linear maps $8\otimes \overline{3} \rightarrow 8\otimes \overline{3}$ reads:\footnote{We use the birdtracks representation for convenience, as it does not flood expressions with irrelevant indices. In conventional notation, one could write: $\left[\calp_{\overline{3}}\right]^{ba}_{ji} = \frac{1}{C_F} (t^bt^a)_{ji}$ and $\left[ \calp_{\overline{15}/6}\right]^{ba}_{ji} = \thalf \delta^{ba} \delta_{ji} \pm (t^at^b)_{ji} - \frac{1}{N_c \pm 1} (t^bt^a)_{ji}$. }
\begin{align}
   \calp_{\overline{3} \subset 8\otimes\overline{3}} &=  \frac{1}{C_F}
\vcenter{\hbox{
\begin{tikzpicture}[scale=0.3]
    \draw[gluon] (-3,1) to[out=0,in=100] (-1,-1);
    \draw[gluon] (1,-1) to[in=-180,out=80] (3,1);
    \draw[thick] (-3,-1) -- (3,-1);
    \draw[thick,-<] (-1,-1) -- (0,-1);
\end{tikzpicture}}}, \qquad C_F = \frac{N_c^2-1}{2N_c}, \\[2ex]
    \calp_{\overline{15}/6 \subset 8\otimes\overline{3}} &= 
    \frac{1}{2}
\vcenter{\hbox{
\begin{tikzpicture}[scale=0.3]
    \draw[gluon] (-3,1) -- (3,1);
    \draw[thick] (-3,-1) -- (3,-1);
    \draw[thick,-<] (-1,-1) -- (0,-1);
\end{tikzpicture}}} 
\pm 
\vcenter{\hbox{
\begin{tikzpicture}[scale=0.3]
    \draw[gluon] (-3,1) to[out=0,in=100] (1,-1);
    \draw[gluon] (-1,-1) to[in=-180,out=80] (3,1);
    \draw[thick] (-3,-1) -- (3,-1);
    \draw[thick,-<] (-1,-1) -- (0,-1);
\end{tikzpicture}}}
- \frac{1}{N_c \pm 1}
\vcenter{\hbox{
\begin{tikzpicture}[scale=0.3]
    \draw[gluon] (-3,1) to[out=0,in=100] (-1,-1);
    \draw[gluon] (1,-1) to[in=-180,out=80] (3,1);
    \draw[thick] (-3,-1) -- (3,-1);
    \draw[thick,-<] (-1,-1) -- (0,-1);
\end{tikzpicture}}}
\end{align}
As expected, one can explicitly check orthogonality and idempotence of those projection operators.
Alternatively, they can be expressed in the vector space $V^{\otimes 3} \otimes \overline{V}^{\otimes 3}$ by contracting the two gluon legs with a generator and multiplying by $2$.
As an illustration, for the irrep $\overline{3}$ belonging to the decomposition $8 \otimes \overline{3}$, the projector as a linear map in $8 \otimes \overline{3} \rightarrow 8 \otimes \overline{3}$ is related to the linear map in $ (\overline{3} \otimes 3) \otimes \overline{3} \rightarrow (\overline{3} \otimes 3) \otimes \overline{3}$ according to
\begin{equation}
   \calp_{\overline{3} \subset 8\otimes\overline{3}} = \frac{1}{C_F}
\vcenter{\hbox{
\begin{tikzpicture}[scale=0.3]
    \draw[gluon] (-3,1) to[out=0,in=100] (-1,-1);
    \draw[gluon] (1,-1) to[in=-180,out=80] (3,1);
    \draw[thick] (-3,-1) -- (3,-1);
    \draw[thick,-<] (-1,-1) -- (0,-1);
\end{tikzpicture}}}
\longrightarrow \frac{2}{C_F}
\vcenter{\hbox{
\begin{tikzpicture}[scale=0.25]
    \draw[gluon] (-3,1) to[out=0,in=100] (-1,-1);
    \draw[gluon] (1,-1) to[in=-180,out=80] (3,1);
    \draw[thick] (-4,-1) -- (4,-1);
    \draw[thick,-<] (-1,-1) -- (0,-1);
    \draw[thick] (-4,1.5) to[out=0,in=90] (-3,1) to[out=-90,in=0] (-4,0.5);
    \draw[thick] (4,1.5) to[out=180,in=90] (3,1) to[out=-90,in=180] (4,0.5);
\end{tikzpicture}}}
\longrightarrow \frac{1}{2 C_F}
\vcenter{\hbox{
\begin{tikzpicture}[scale=0.3]
    \draw[thick] (-3,1) to[out=0,in=90] (-1,-1) -- (1,-1) to[out=90,in=180] (3,1);
    \draw[thick,-<] (-1,-1) -- (0,-1);
    \draw[thick] (-3,.5) to[out=0,in=90] (-1.5,-1) -- (-3,-1);
    \draw[thick] (3,.5) to[out=180,in=90] (1.5,-1) -- (3,-1);
\end{tikzpicture}}}
\end{equation}
Since the top $\overline{q}q$ are implied to be in an adjoint irrep from the insertions of $\calb^i$, we can replace the gluon lines with $\overline{q}q$ lines, which yields the right-most result. 
We identify the rightmost tensor as being the one involved in the operator $\hcalf^{(1)}_{qg}$, thus we have
\begin{equation}
    \hat{\Omega}_{qg}^{\overline{3}} = - \frac{1}{2C_F} \hcalf_{qg}^{(1)}.
\end{equation}

In order to relate the two remaining projectors over irreps $\overline{15}$ and $6$ to TMD operators, we evaluate the identity tensor $I_{8\otimes\overline{3}}$ and the crossing tensor $X_{8\otimes\overline{3}}$ in the following expression:
\begin{equation}
\calp_{\overline{15}/6 \subset 8\otimes\overline{3}} = 
\frac{1}{2} I_{8\otimes\overline{3}}
\pm X_{8\otimes\overline{3}}
- \frac{1}{N_c \pm 1} C_F \calp_{\overline{3} \subset 8\otimes\overline{3}}.
\end{equation}
The identity tensor can be written in the following form
\begin{equation}
I_{8\otimes \overline{3}} =
\vcenter{\hbox{
\begin{tikzpicture}[scale=0.3]
    \draw[gluon] (-3,1) -- (3,1);
    \draw[thick] (-3,-1) -- (3,-1);
    \draw[thick,-<] (-1,-1) -- (0,-1);
\end{tikzpicture}}} 
\longrightarrow 2
\vcenter{\hbox{
\begin{tikzpicture}[scale=0.3]
    \draw[gluon] (-3,1) -- (3,1);
    \draw[thick] (-4,-1) -- (4,-1);
    \draw[thick,-<] (-1,-1) -- (0,-1);
    \draw[thick] (-4,1.5) to[out=0,in=90] (-3,1) to[out=-90,in=0] (-4,0.5);
    \draw[thick] (4,1.5) to[out=180,in=90] (3,1) to[out=-90,in=180] (4,0.5);
\end{tikzpicture}}}
\longrightarrow 
\vcenter{\hbox{
\begin{tikzpicture}[scale=0.3]
    \draw[thick] (-4,-1) -- (4,-1);
    \draw[thick,-<] (-1,-1) -- (0,-1);
    \draw[thick,->] (-1,.5) -- (0,.5);
    \draw[thick,-<] (-1,1.5) -- (0,1.5);
    \draw[thick] (-4,.5) -- (4,.5);
    \draw[thick] (-4,1.5) -- (4,1.5);
\end{tikzpicture}}}
\end{equation}
Again, we used for the rightmost contribution the fact that we have implicit adjoint systems for the top $\overline{q}q$-systems. We recognize the rightmost tensor as being the one involved in the operator $\hcalf^{(2)}_{qg}$. Working out the last tensor in the following fashion, we have
\begin{equation}
    X_{8\otimes \overline{3}} = 
\vcenter{\hbox{
\begin{tikzpicture}[scale=0.3]
    \draw[gluon] (-3,1) to[out=0,in=100] (1,-1);
    \draw[gluon] (-1,-1) to[in=-180,out=80] (3,1);
    \draw[thick] (-3,-1) -- (3,-1);
    \draw[thick,-<] (-1,-1) -- (0,-1);
\end{tikzpicture}}}
\longrightarrow 2 
\vcenter{\hbox{
\begin{tikzpicture}[scale=0.3]
    \draw[gluon] (-3,1) to[out=0,in=100] (1,-1);
    \draw[gluon] (-1,-1) to[in=-180,out=80] (3,1);
    \draw[thick] (-4,-1) -- (4,-1);
    \draw[thick,-<] (-1,-1) -- (0,-1);
    \draw[thick] (-4,1.5) to[out=0,in=90] (-3,1) to[out=-90,in=0] (-4,0.5);
    \draw[thick] (4,1.5) to[out=180,in=90] (3,1) to[out=-90,in=180] (4,0.5);
\end{tikzpicture}}}
\longrightarrow \frac{1}{2}
\vcenter{\hbox{
\begin{tikzpicture}[scale=0.3]
    \draw[thick] (-4,-1) -- (-2,-1) to[out=0,in=180] (2,1.5) -- (4,1.5);
    \draw[thick] (-4,.5) -- (4,.5);
    \draw[thick] (-4,1.5) -- (-2,1.5) to[out=0,in=180] (2,-1) -- (4,-1);
    \draw[thick,-<] (-4,-1) -- ++(1,0);
    \draw[thick,->] (-4,.5) -- ++(1,0);
    \draw[thick,-<] (-4,1.5) -- ++(1,0);
\end{tikzpicture}}}
\end{equation}
We readily see the link with the operator $\hcalf^{(3)}_{qg}$. Thus, the linear combination of interest is
\begin{subequations}\label{eq:Omega_to_Fqg}
\begin{align}
    -\hat{\Omega}_{qg}^{\overline{3}} &= \frac{1}{2C_F} \hcalf_{qg}^{(1)}, \\
    -\hat{\Omega}_{qg}^{\overline{15} / 6} &= \frac{N_c}{2} \hcalf^{(2)}_{qg} \pm \frac{1}{2}\hcalf^{(3)}_{qg} -\frac{\hcalf^{(1)}_{qg}}{2(N_c\pm1)}.
\end{align}
\end{subequations}

\textit{Remark.}
In the basis $\hat{\Omega}_{qg}$, each operators represent an irrep transition for the system at $\xvec$ (and $\yvec$ provided one take the conjugated irreps): $\overline{3} \rightarrow \overline{3}$, $\overline{3} \rightarrow \overline{15}$, and $\overline{3} \rightarrow 6$ between the projection at $-\infty$ and the projection at $+\infty$.
Furthermore, since the representation $6$ does not appear for $SU(2)$, the linear combination of TMD operators with $\pm \rightarrow -$ vanishes.

\paragraph{Gluon - Gluon:}
\begin{align}
- N_c\hcalf^{(1)}_{gg} &=
\vcenter{\hbox{
\begin{tikzpicture}[scale=0.3]
    \draw[thick] (-5,2) -- ++ (10,0);\draw[thick,->] (-5,2) -- ++ (3,0);
    \draw[thick] (-5,2.5) -- ++ (10,0);\draw[thick,-<] (-5,2.5) -- ++ (3,0);
    \draw[fill=Blue!50!white] (1,2) circle (0.2);
    \draw[thick] (-5,1.5) -- ++ (10,0);\draw[thick,-<] (-5,1.5) -- ++ (4,0);
    \draw[thick] (-5,1) -- ++(10,0);\draw[thick,->] (-5,1) -- ++(4,0);
    \draw[thick] (-5,-1) -- ++(10,0);\draw[thick,-<] (-5,-1) -- ++(6,0);
    \draw[thick] (-5,-1.5) -- ++ (10,0);\draw[thick,->] (-5,-1.5) -- ++ (6,0);
    \draw[thick] (-5,-2) -- ++ (10,0);\draw[thick,-<] (-5,-2) -- ++ (7,0);
    \draw[thick] (-5,-2.5) -- ++ (10,0);\draw[thick,->] (-5,-2.5) -- ++ (7,0);
    \draw[fill=Blue!50!white] (-1,-2.5) circle (0.2);
    \draw[thick] (5,1.5) to[out=0,in=0,looseness=2] (5,2);
    \draw[thick] (5,-1.5) to[out=0,in=0,looseness=2] (5,-2);
    \draw[thick] (5,2.5) to[out=0,in=0,looseness=1] (5,-2.5);
    \draw[thick] (5,1) to[out=0,in=0,looseness=1] (5,-1);
    \draw[thick] (-5,2) to[out=180,in=180] (-5,2.5);
    \draw[thick] (-5,-2) to[out=180,in=180] (-5,-2.5);
    \draw[thick] (-5,1) to[out=180,in=180] (-5,-1);
    \draw[thick] (-5,1.5) to[out=180,in=180] (-5,-1.5);
\end{tikzpicture}}}
= -
\vcenter{\hbox{
\begin{tikzpicture}[scale=0.3]
    \draw[thick] (-5,1) -- ++ (10,0);\draw[thick,-<] (-5,1) -- ++ (4,0);
    \draw[fill=Blue!50!white] (1,1) circle (0.2);
    \draw[thick] (-5,-2) -- ++ (10,0);\draw[thick,->] (-5,-2) -- ++ (7,0);
    \draw[fill=Blue!50!white] (-1,-2) circle (0.2);
    \draw[thick] (5,1) to[out=0,in=0] (5,-2);
    \draw[thick] (-5,1) to[out=180,in=180] (-5,-2);
    \draw[thick,->] (-5,.5) -- (-1,.5);
    \draw[thick,-<] (-5,-1.5) -- (2,-1.5);
    \draw[thick] (-5,.5) -- (5,.5) to[out=0,in=0] (5,-1.5) -- (-5,-1.5) to[out=180,in=180] cycle;
\end{tikzpicture}}} \\[2ex]
- N_c \hcalf^{(2)}_{gg} &=
\vcenter{\hbox{
\begin{tikzpicture}[scale=0.3]
    \draw[thick] (-5,2) -- ++ (10,0);\draw[thick,->] (-5,2) -- ++ (3,0);
    \draw[thick] (-5,2.5) -- ++ (10,0);\draw[thick,-<] (-5,2.5) -- ++ (3,0);
    \draw[fill=Blue!50!white] (1,2) circle (0.2);
    \draw[thick] (-5,1.5) -- ++ (10,0);\draw[thick,-<] (-5,1.5) -- ++ (4,0);
    \draw[thick] (-5,1) -- ++(10,0);\draw[thick,->] (-5,1) -- ++(4,0);
    \draw[thick] (-5,-1) -- ++(10,0);\draw[thick,-<] (-5,-1) -- ++(6,0);
    \draw[thick] (-5,-1.5) -- ++ (10,0);\draw[thick,->] (-5,-1.5) -- ++ (6,0);
    \draw[thick] (-5,-2) -- ++ (10,0);\draw[thick,-<] (-5,-2) -- ++ (7,0);
    \draw[thick] (-5,-2.5) -- ++ (10,0);\draw[thick,->] (-5,-2.5) -- ++ (7,0);
    \draw[fill=Blue!50!white] (-1,-2.5) circle (0.2);
    \draw[thick] (5,2.5) to[out=0,in=0,looseness=2] (5,1);
    \draw[thick] (5,-1) to[out=0,in=0,looseness=2] (5,2);
    \draw[thick] (5,-2) to[out=0,in=0,looseness=2] (5,-1.5);
    \draw[thick] (5,1.5) to[out=0,in=0,looseness=1] (5,-2.5);
    \draw[thick] (-5,2) to[out=180,in=180] (-5,2.5);
    \draw[thick] (-5,-2) to[out=180,in=180] (-5,-2.5);
    \draw[thick] (-5,1) to[out=180,in=180] (-5,-1);
    \draw[thick] (-5,-1.5) to[out=180,in=180] (-5,1.5);
\end{tikzpicture}}} \\[2ex]
- N_c \hcalf^{(3)}_{gg} &=
\vcenter{\hbox{
\begin{tikzpicture}[scale=0.3]
    \draw[thick] (-5,2) -- ++ (10,0);\draw[thick,->] (-5,2) -- ++ (3,0);
    \draw[thick] (-5,2.5) -- ++ (10,0);\draw[thick,-<] (-5,2.5) -- ++ (3,0);
    \draw[fill=Blue!50!white] (1,2) circle (0.2);
    \draw[thick] (-5,1.5) -- ++ (10,0);\draw[thick,-<] (-5,1.5) -- ++ (4,0);
    \draw[thick] (-5,1) -- ++(10,0);\draw[thick,->] (-5,1) -- ++(4,0);
    \draw[thick] (-5,-1) -- ++(10,0);\draw[thick,-<] (-5,-1) -- ++(6,0);
    \draw[thick] (-5,-1.5) -- ++ (10,0);\draw[thick,->] (-5,-1.5) -- ++ (6,0);
    \draw[thick] (-5,-2) -- ++ (10,0);\draw[thick,-<] (-5,-2) -- ++ (7,0);
    \draw[thick] (-5,-2.5) -- ++ (10,0);\draw[thick,->] (-5,-2.5) -- ++ (7,0);
    \draw[fill=Blue!50!white] (-1,-2.5) circle (0.2);
    \draw[thick] (5,-1) to[out=0,in=0,looseness=2] (5,-1.5);
    \draw[thick] (5,1) to[out=0,in=0,looseness=1] (5,1.5);
    \draw[thick] (5,-2.5) to[out=0,in=0,looseness=1] (5,2.5);
    \draw[thick] (5,-2) to[out=0,in=0,looseness=1] (5,2);
    \draw[thick] (-5,2) to[out=180,in=180] (-5,2.5);
    \draw[thick] (-5,-2) to[out=180,in=180] (-5,-2.5);
    \draw[thick] (-5,1) to[out=180,in=180] (-5,-1);
    \draw[thick] (-5,1.5) to[out=180,in=180] (-5,-1.5);
\end{tikzpicture}}}
= N_c\,
\vcenter{\hbox{
\begin{tikzpicture}[scale=0.3]
    \draw[thick] (-5,1) -- ++ (10,0);\draw[thick,->] (-5,1) -- ++ (3,0);
    \draw[thick] (-5,1.5) -- ++ (10,0);\draw[thick,-<] (-5,1.5) -- ++ (3,0);
    \draw[fill=Blue!50!white] (1,1) circle (0.2);
    \draw[thick] (-5,-1) -- ++ (10,0);\draw[thick,-<] (-5,-1) -- ++ (7,0);
    \draw[thick] (-5,-1.5) -- ++ (10,0);\draw[thick,->] (-5,-1.5) -- ++ (7,0);
    \draw[fill=Blue!50!white] (-1,-1.5) circle (0.2);
    \draw[thick] (5,-1.5) to[out=0,in=0,looseness=1] (5,1.5);
    \draw[thick] (5,-1) to[out=0,in=0,looseness=1] (5,1);
    \draw[thick] (-5,1) to[out=180,in=180] (-5,1.5);
    \draw[thick] (-5,-1) to[out=180,in=180] (-5,-1.5);
\end{tikzpicture}}} \\[2ex]
- \hcalf^{(4)}_{gg} &=
\vcenter{\hbox{
\begin{tikzpicture}[scale=0.3]
    \draw[thick] (-5,2) -- ++ (10,0);\draw[thick,->] (-5,2) -- ++ (3,0);
    \draw[thick] (-5,2.5) -- ++ (10,0);\draw[thick,-<] (-5,2.5) -- ++ (3,0);
    \draw[fill=Blue!50!white] (1,2) circle (0.2);
    \draw[thick] (-5,1.5) -- ++ (10,0);\draw[thick,-<] (-5,1.5) -- ++ (4,0);
    \draw[thick] (-5,1) -- ++(10,0);\draw[thick,->] (-5,1) -- ++(4,0);
    \draw[thick] (-5,-1) -- ++(10,0);\draw[thick,-<] (-5,-1) -- ++(6,0);
    \draw[thick] (-5,-1.5) -- ++ (10,0);\draw[thick,->] (-5,-1.5) -- ++ (6,0);
    \draw[thick] (-5,-2) -- ++ (10,0);\draw[thick,-<] (-5,-2) -- ++ (7,0);
    \draw[thick] (-5,-2.5) -- ++ (10,0);\draw[thick,->] (-5,-2.5) -- ++ (7,0);
    \draw[fill=Blue!50!white] (-1,-2.5) circle (0.2);
    \draw[thick] (5,2) to[out=0,in=0,looseness=2] (5,1.5);
    \draw[thick] (5,2.5) to[out=0,in=0,looseness=2] (5,1);
    \draw[thick] (5,-2) to[out=0,in=0,looseness=2] (5,-1.5);
    \draw[thick] (5,-2.5) to[out=0,in=0,looseness=2] (5,-1);
    \draw[thick] (-5,2) to[out=180,in=180] (-5,2.5);
    \draw[thick] (-5,-2) to[out=180,in=180] (-5,-2.5);
    \draw[thick] (-5,1) to[out=180,in=180] (-5,-1);
    \draw[thick] (-5,1.5) to[out=180,in=180] (-5,-1.5);
\end{tikzpicture}}}
\ \  = \phantom{-}
\vcenter{\hbox{
\begin{tikzpicture}[scale=0.3]
    \draw[thick] (-5,1) -- ++ (10,0);\draw[thick,->] (-5,1) -- ++ (3,0);
    \draw[thick] (-5,1.5) -- ++ (10,0);\draw[thick,-<] (-5,1.5) -- ++ (3,0);
    \draw[fill=Blue!50!white] (1,1) circle (0.2);
    \draw[thick] (-5,-1) -- ++ (10,0);\draw[thick,-<] (-5,-1) -- ++ (7,0);
    \draw[thick] (-5,-1.5) -- ++ (10,0);\draw[thick,->] (-5,-1.5) -- ++ (7,0);
    \draw[fill=Blue!50!white] (-1,-1.5) circle (0.2);
    \draw[thick] (5,1) to[out=0,in=0,looseness=2] (5,1.5);
    \draw[thick] (5,-1.5) to[out=0,in=0,looseness=2] (5,-1);
    \draw[thick] (-5,1) to[out=180,in=180] (-5,-1);
    \draw[thick] (-5,-1.5) to[out=180,in=180] (-5,1.5);
\end{tikzpicture}}} \\[2ex]
- \hcalf^{(5)}_{gg} &=
\vcenter{\hbox{
\begin{tikzpicture}[scale=0.3]
    \draw[thick] (-5,2) -- ++ (10,0);\draw[thick,->] (-5,2) -- ++ (3,0);
    \draw[thick] (-5,2.5) -- ++ (10,0);\draw[thick,-<] (-5,2.5) -- ++ (3,0);
    \draw[fill=Blue!50!white] (1,2) circle (0.2);
    \draw[thick] (-5,1.5) -- ++ (10,0);\draw[thick,-<] (-5,1.5) -- ++ (4,0);
    \draw[thick] (-5,1) -- ++(10,0);\draw[thick,->] (-5,1) -- ++(4,0);
    \draw[thick] (-5,-1) -- ++(10,0);\draw[thick,-<] (-5,-1) -- ++(6,0);
    \draw[thick] (-5,-1.5) -- ++ (10,0);\draw[thick,->] (-5,-1.5) -- ++ (6,0);
    \draw[thick] (-5,-2) -- ++ (10,0);\draw[thick,-<] (-5,-2) -- ++ (7,0);
    \draw[thick] (-5,-2.5) -- ++ (10,0);\draw[thick,->] (-5,-2.5) -- ++ (7,0);
    \draw[fill=Blue!50!white] (-1,-2.5) circle (0.2);
    \draw[thick] (5,1) to[out=0,in=0,looseness=1] (5,-2);
    \draw[thick] (5,-2.5) to[out=0,in=0,looseness=1] (5,1.5);
    \draw[thick] (5,-1.5) to[out=0,in=0,looseness=2] (5,2.5);
    \draw[thick] (5,-1) to[out=0,in=0,looseness=2] (5,2);
    \draw[thick] (-5,2) to[out=180,in=180] (-5,2.5);
    \draw[thick] (-5,-2) to[out=180,in=180] (-5,-2.5);
    \draw[thick] (-5,1.5) to[out=180,in=180] (-5,-1.5);
    \draw[thick] (-5,1) to[out=180,in=180] (-5,-1);
\end{tikzpicture}}} \\[2ex]
- N_c^2\hcalf^{(6)}_{gg} &=
\vcenter{\hbox{
\begin{tikzpicture}[scale=0.3]
    \draw[thick] (-5,2) -- ++ (10,0);\draw[thick,->] (-5,2) -- ++ (3,0);
    \draw[thick] (-5,2.5) -- ++ (10,0);\draw[thick,-<] (-5,2.5) -- ++ (3,0);
    \draw[fill=Blue!50!white] (1,2) circle (0.2);
    \draw[thick] (-5,1.5) -- ++ (10,0);\draw[thick,-<] (-5,1.5) -- ++ (4,0);
    \draw[thick] (-5,1) -- ++(10,0);\draw[thick,->] (-5,1) -- ++(4,0);
    \draw[thick] (-5,-1) -- ++(10,0);\draw[thick,-<] (-5,-1) -- ++(6,0);
    \draw[thick] (-5,-1.5) -- ++ (10,0);\draw[thick,->] (-5,-1.5) -- ++ (6,0);
    \draw[thick] (-5,-2) -- ++ (10,0);\draw[thick,-<] (-5,-2) -- ++ (7,0);
    \draw[thick] (-5,-2.5) -- ++ (10,0);\draw[thick,->] (-5,-2.5) -- ++ (7,0);
    \draw[fill=Blue!50!white] (-1,-2.5) circle (0.2);
    \draw[thick] (5,2.5) to[out=0,in=0,looseness=1.5] (5,-2.5);
    \draw[thick] (5,2) to[out=0,in=0,looseness=1.5] (5,-2);
    \draw[thick] (5,-1) to[out=0,in=0,looseness=1] (5,1);
    \draw[thick] (5,-1.5) to[out=0,in=0,looseness=1] (5,1.5);
    \draw[thick] (-5,2) to[out=180,in=180] (-5,2.5);
    \draw[thick] (-5,-2) to[out=180,in=180] (-5,-2.5);
    \draw[thick] (-5,1) to[out=180,in=180] (-5,-1);
    \draw[thick] (-5,1.5) to[out=180,in=180] (-5,-1.5);
\end{tikzpicture}}} \\[2ex]
-N_c \hcalf^{(7)}_{gg} &=
\vcenter{\hbox{
\begin{tikzpicture}[scale=0.3]
    \draw[thick] (-5,2) -- ++ (10,0);\draw[thick,->] (-5,2) -- ++ (3,0);
    \draw[thick] (-5,2.5) -- ++ (10,0);\draw[thick,-<] (-5,2.5) -- ++ (3,0);
    \draw[fill=Blue!50!white] (1,2) circle (0.2);
    \draw[thick] (-5,1.5) -- ++ (10,0);\draw[thick,-<] (-5,1.5) -- ++ (4,0);
    \draw[thick] (-5,1) -- ++(10,0);\draw[thick,->] (-5,1) -- ++(4,0);
    \draw[thick] (-5,-1) -- ++(10,0);\draw[thick,-<] (-5,-1) -- ++(6,0);
    \draw[thick] (-5,-1.5) -- ++ (10,0);\draw[thick,->] (-5,-1.5) -- ++ (6,0);
    \draw[thick] (-5,-2) -- ++ (10,0);\draw[thick,-<] (-5,-2) -- ++ (7,0);
    \draw[thick] (-5,-2.5) -- ++ (10,0);\draw[thick,->] (-5,-2.5) -- ++ (7,0);
    \draw[fill=Blue!50!white] (-1,-2.5) circle (0.2);
    \draw[thick] (5,-1.5) to[out=0,in=0,looseness=1] (5,1.5);
    \draw[thick] (5,1) to[out=0,in=0,looseness=2] (5,-2);
    \draw[thick] (5,-1) to[out=0,in=0,looseness=2] (5,2);
    \draw[thick] (5,-2.5) to[out=0,in=0,looseness=2] (5,2.5);
    \draw[thick] (-5,2) to[out=180,in=180] (-5,2.5);
    \draw[thick] (-5,-2) to[out=180,in=180] (-5,-2.5);
    \draw[thick] (-5,1) to[out=180,in=180] (-5,-1);
    \draw[thick] (-5,1.5) to[out=180,in=180] (-5,-1.5);
\end{tikzpicture}}}
\end{align}
One easily concludes from the birdtracks that the tensor
\begin{equation}
    \left(\calb^i_\xvec \otimes V_\xvec  \otimes V^\dagger_\xvec \right) \otimes \left( \calb^i_\yvec \otimes V_\yvec  \otimes V^\dagger_\yvec \right)
\end{equation}
is being projected at $-\infty$ in the same way for all operators $\hcalf_{gg}$. Those operators only differ by their projection at $+\infty$ and normalization.

Let us first consider the trace being performed at $-\infty$ (left-hand side). One can use the Fierz identity and replace the identity by the singlet and adjoint projection.
The singlet contribution simplifies greatly in the previous birdtracks, and we recover the following relations:
\begin{align}
    \hcalf^{(4)}_{gg} \big|_{singlet}^{-\infty} &= \hcalf^{(5)}_{gg} \big|_{singlet}^{-\infty} = 0 \\
    \hcalf^{(1)}_{gg} \big|_{singlet}^{-\infty} &= \hcalf^{(2)}_{gg} \big|_{singlet}^{-\infty} = \hcalf^{(6)}_{gg} \big|_{singlet}^{-\infty} = \hcalf^{(7)}_{gg} \big|_{singlet}^{-\infty} = \frac{1}{N_c^2} \hcalf^{(3)}_{gg}
\end{align}
In particular, we identify that $\hcalf^{(3)}_{gg}$ corresponds to a contribution proportional to the \textit{last} adjoint irrep from the decomposition
\begin{equation}
    (3 \otimes \overline{3}) \otimes 8 = (1 \oplus 8 ) \otimes 8 = (8 \otimes 8) \oplus \textcolor{Red}{(8 \otimes 1)} = (8 \otimes 8) \oplus \textcolor{Red}{8}.
\end{equation}
Thus, it makes sense to isolate this irrep by subtracting it from TMD operators with non-vanishing overlap with the singlet projection at $-\infty$, the corresponding contribution proportional to $\hcalf^{(3)}_{gg}$.
This effectively splits the basis of TMD operators according to the projection at $-\infty$ being a singlet or an adjoint, indicated in the following along the green dashed line:
\begin{center}
\begin{tikzpicture}[scale=0.3]
    \draw[thick] (-5,2) -- ++ (10,0);\draw[thick,->] (-5,2) -- ++ (3,0);
    \draw[thick] (-5,2.5) -- ++ (10,0);\draw[thick,-<] (-5,2.5) -- ++ (3,0);
    \draw[fill=Blue!50!white] (1,2) circle (0.2);
    \draw[thick] (-5,1.5) -- ++ (10,0);\draw[thick,-<] (-5,1.5) -- ++ (4,0);
    \draw[thick] (-5,1) -- ++(10,0);\draw[thick,->] (-5,1) -- ++(4,0);
    \draw[thick] (-5,-1) -- ++(10,0);\draw[thick,-<] (-5,-1) -- ++(6,0);
    \draw[thick] (-5,-1.5) -- ++ (10,0);\draw[thick,->] (-5,-1.5) -- ++ (6,0);
    \draw[thick] (-5,-2) -- ++ (10,0);\draw[thick,-<] (-5,-2) -- ++ (7,0);
    \draw[thick] (-5,-2.5) -- ++ (10,0);\draw[thick,->] (-5,-2.5) -- ++ (7,0);
    \draw[fill=Blue!50!white] (-1,-2.5) circle (0.2);
    \draw[thick] (-5,2) to[out=180,in=180] (-5,2.5);
    \draw[thick] (-5,-2) to[out=180,in=180] (-5,-2.5);
    \draw[thick] (-5,1) to[out=180,in=180] (-5,-1);
    \draw[thick] (-5,1.5) to[out=180,in=180] (-5,-1.5);
    \draw[dashed,Green] (-5,0) -- (-7,0) node[left] {\scriptsize $8 \oplus 1$};
    \node at (7,0) {$\{\mathcal{T}\}$};
\end{tikzpicture}
\end{center}
In this sense, the interaction depicted by the operator $\hcalf^{(3)}_{gg}$ is of a dipole that rotates the irreps according to the following transition $ (1\otimes 1)_{singlet} \rightarrow (8 \otimes 8)_{singlet}$.

After the removal of this contribution, one is left with the evaluation of tensors $\mathcal{T}$ corresponding to singlets of the decomposition $8^{\otimes 4}$. Thus, the complete basis contains nine elements which we label:\footnote{Note that we abuse the terminology here. Degenerated irreps do not care about the symmetry under permutation; the projections do. In this case, there are two distinct adjoints in the decomposition $8 \otimes 8$ and four possible ways to combine them into singlets, which we label $8_{aa} \oplus 8_{ss} \oplus 8_{as} \oplus 8_{sa}$.}
\begin{equation}
    1 \oplus 27 \oplus 0 \oplus 10 \oplus \overline{10} \oplus 8_{aa} \oplus 8_{ss} \oplus 8_{as} \oplus 8_{sa}
\end{equation}
This is higher than the remaining number of independent TMD operators that we work with, however, one should consider: (i) for two point interactions, one is only sensitive to the quadratic color charge; (ii) the decuplet and anti-decuplet have the same charge and the same anti-symmetry property under exchange of partons, (iii) mixed left-right symmetry adjoint decouple from same left-right symmetry irrep for two point interactions.
Thus, we can write the block structure as
\begin{equation}
    \left[ 1 \oplus 27 \oplus 0 \oplus 8_{ss} \oplus (10 + \overline{10}) \oplus 8_{aa}\right] \oplus \left[ 8_{as} \oplus 8_{sa} \oplus (10 - \overline{10})\right]
\end{equation}
where $10 \pm \overline{10}$ denotes the hermitian or anti-hermitian contributions from the decuplets. 
The first block contains only six elements, which is enough for our purpose\footnote{While the basis is not complete, it still spans all of the TMD operators relevant to this study} (in addition to the \textit{last} adjoint previously mentioned). 
Let us introduce the operator $\hat{\Omega}_{gg}$ which project at $+\infty$ onto the irreps of the decomposition $8 \otimes 8$ corresponding to the first block:

\begin{equation}
\hat{\Omega}_{gg} = \textcolor{Red}{4} \times
\vcenter{\hbox{
\begin{tikzpicture}[scale=0.3]
    \draw[thick] (-5,2) -- ++ (10,0);\draw[thick,->] (-5,2) -- ++ (3,0);
    \draw[thick] (-5,2.5) -- ++ (10,0);\draw[thick,-<] (-5,2.5) -- ++ (3,0);
    \draw[fill=Blue!50!white] (1,2) circle (0.2);
    \draw[thick] (-5,1.5) -- ++ (10,0);\draw[thick,-<] (-5,1.5) -- ++ (4,0);
    \draw[thick] (-5,1) -- ++(10,0);\draw[thick,->] (-5,1) -- ++(4,0);
    \draw[thick] (-5,-1) -- ++(10,0);\draw[thick,-<] (-5,-1) -- ++(6,0);
    \draw[thick] (-5,-1.5) -- ++ (10,0);\draw[thick,->] (-5,-1.5) -- ++ (6,0);
    \draw[thick] (-5,-2) -- ++ (10,0);\draw[thick,-<] (-5,-2) -- ++ (7,0);
    \draw[thick] (-5,-2.5) -- ++ (10,0);\draw[thick,->] (-5,-2.5) -- ++ (7,0);
    \draw[fill=Blue!50!white] (-1,-2.5) circle (0.2);
    \draw[thick] (-5,2) to[out=180,in=180] (-5,2.5);
    \draw[thick] (-5,-2) to[out=180,in=180] (-5,-2.5);
    \draw[thick] (-5,1) to[out=180,in=180] (-5,-1);
    \draw[thick] (-5,1.5) to[out=180,in=180] (-5,-1.5);
    \draw[thick,Red] (5.5,2.5) to[out=0,in=90] ++(.5,-.25) to[out=-90,in=0] ++(-.5,-.25);
    \draw[thick,Red] (5.5,1.5) to[out=0,in=90] ++(.5,-.25) to[out=-90,in=0] ++(-.5,-.25);
    \draw[thick,Red] (5.5,-1) to[out=0,in=90] ++(.5,-.25) to[out=-90,in=0] ++(-.5,-.25);
    \draw[thick,Red] (5.5,-2) to[out=0,in=90] ++(.5,-.25) to[out=-90,in=0] ++(-.5,-.25);
    \draw[gluon,Red] (6,2.25) to[out=0,in=0,looseness=2] (6,-2.25);
    \draw[gluon,Red] (6,1.25) to[out=0,in=0,looseness=2] (6,-1.25);
    \draw[Red,fill=white] (8,0) ellipse (1.2 and .5) node {\scriptsize $\omega$};
\end{tikzpicture}}}
\end{equation}
A convenient notation to manipulate tensors of the basis $\mathcal{T}$ is the following:
\begin{equation}
\vcenter{\hbox{
\begin{tikzpicture}[scale=0.3]
    \draw[thick] (-5,2) -- ++ (10,0);\draw[thick,->] (-5,2) -- ++ (3,0);
    \draw[thick] (-5,2.5) -- ++ (10,0);\draw[thick,-<] (-5,2.5) -- ++ (3,0);
    \draw[fill=Blue!50!white] (1,2) circle (0.2);
    \draw[thick] (-5,1.5) -- ++ (10,0);\draw[thick,-<] (-5,1.5) -- ++ (4,0);
    \draw[thick] (-5,1) -- ++(10,0);\draw[thick,->] (-5,1) -- ++(4,0);
    \draw[thick] (-5,-1) -- ++(10,0);\draw[thick,-<] (-5,-1) -- ++(6,0);
    \draw[thick] (-5,-1.5) -- ++ (10,0);\draw[thick,->] (-5,-1.5) -- ++ (6,0);
    \draw[thick] (-5,-2) -- ++ (10,0);\draw[thick,-<] (-5,-2) -- ++ (7,0);
    \draw[thick] (-5,-2.5) -- ++ (10,0);\draw[thick,->] (-5,-2.5) -- ++ (7,0);
    \draw[fill=Blue!50!white] (-1,-2.5) circle (0.2);
    \draw[thick] (-5,2) to[out=180,in=180] (-5,2.5);
    \draw[thick] (-5,-2) to[out=180,in=180] (-5,-2.5);
    \draw[thick] (-5,1) to[out=180,in=180] (-5,-1);
    \draw[thick] (-5,1.5) to[out=180,in=180] (-5,-1.5);
    \draw[dashed,Green] (-5,0) -- (-7,0) node[left] {\scriptsize $8 \oplus 1$};
    \node at (7,0) {$\{\mathcal{T}\}$};
\end{tikzpicture}}}
=
\tr \left[ 
\vcenter{\hbox{
\begin{tikzpicture}[scale=0.3]
    \draw[thick] (-5,2) -- ++ (3,0);\draw[thick,-<] (-5,2) -- ++ (1.5,0);
    \draw[thick] (-5,1.5) -- ++ (3,0);\draw[thick,->] (-5,1.5) -- ++ (1.5,0);
    \draw[thick] (-5,1.5) to[out=180,in=180] (-5,2);
    \draw[fill=Blue!50!white] (-3,1.5) circle (0.2);
    \draw[thick,-<] (-6,.5) -- ++ (1.5,0);\draw[thick] (-6,.5) -- ++ (4,0);
    \draw[thick,->] (-6,0) -- ++ (1.5,0);\draw[thick] (-6,0) -- ++ (4,0);
    \draw[thick] (5,2) -- ++ (-3,0);\draw[thick,->] (5,2) -- ++ (-1.5,0);
    \draw[thick] (5,1.5) -- ++ (-3,0);\draw[thick,-<] (5,1.5) -- ++ (-1.5,0);
    \draw[thick] (5,1.5) to[out=0,in=0] (5,2);
    \draw[fill=Blue!50!white] (3,1.5) circle (0.2);
    \draw[thick,->] (6,.5) -- ++ (-1.5,0);\draw[thick] (6,.5) -- ++ (-4,0);
    \draw[thick,-<] (6,0) -- ++ (-1.5,0);\draw[thick] (6,0) -- ++ (-4,0);
    \node at (0,1) {$\{ \mathcal{T} \}$};
    \node at (-6,-.75) {\scriptsize $-\infty$};
    \node at (+6,-.75) {\scriptsize $-\infty$};
    \node at (-2,-.75) {\scriptsize $+\infty$};
    \node at (+2,-.75) {\scriptsize $+\infty$};
    \node at (-7,.75) {\scriptsize $\xvec$};
    \node at (7,.75) {\scriptsize $\yvec$};
\end{tikzpicture}}}
\,\right]\,.
\end{equation}
The seven gluon-gluon TMD operators are linear combinations of elements in $\{ \mathcal{T} \}$ projected according to the above diagram.
At $+\infty$ on the left and right of $\{ \mathcal{T}\}$ the top $q\otimes \overline{q} $ system are respectively effective point-like adjoint systems. This follows from the action of the derivatives denoted by the blue bullet.

\vspace{.5cm} \noindent
\textit{The singlet contribution.} It is straightforward to identify the vector corresponding to the singlet ($1 \subset 8 \otimes 8$):
\begin{equation}\label{eq:gg_singlet}
\hat{\Omega}_{gg}^{1} = 
S = \textcolor{Red}{\frac{4}{N_c^2-1}}
\tr \left[ 
\vcenter{\hbox{
\begin{tikzpicture}[scale=0.3]
    \draw[thick] (-5,2) -- ++ (3,0);\draw[thick,-<] (-5,2) -- ++ (1.5,0);
    \draw[thick] (-5,1.5) -- ++ (3,0);\draw[thick,->] (-5,1.5) -- ++ (1.5,0);
    \draw[thick] (-5,1.5) to[out=180,in=180] (-5,2);
    \draw[fill=Blue!50!white] (-3,1.5) circle (0.2);
    \draw[thick,-<] (-6,.5) -- ++ (1.5,0);\draw[thick] (-6,.5) -- ++ (4,0);
    \draw[thick,->] (-6,0) -- ++ (1.5,0);\draw[thick] (-6,0) -- ++ (4,0);
    \draw[thick] (5,2) -- ++ (-3,0);\draw[thick,->] (5,2) -- ++ (-1.5,0);
    \draw[thick] (5,1.5) -- ++ (-3,0);\draw[thick,-<] (5,1.5) -- ++ (-1.5,0);
    \draw[thick] (5,1.5) to[out=0,in=0] (5,2);
    \draw[fill=Blue!50!white] (3,1.5) circle (0.2);
    \draw[thick,->] (6,.5) -- ++ (-1.5,0);\draw[thick] (6,.5) -- ++ (-4,0);
    \draw[thick,-<] (6,0) -- ++ (-1.5,0);\draw[thick] (6,0) -- ++ (-4,0);
    \node at (-6,-.75) {\scriptsize $-\infty$};
    \node at (+6,-.75) {\scriptsize $-\infty$};
    \node at (-2,-.75) {\scriptsize $+\infty$};
    \node at (+2,-.75) {\scriptsize $+\infty$};
    \node at (-7,.75) {\scriptsize $\xvec$};
    \node at (7,.75) {\scriptsize $\yvec$};
    \draw[thick,Red] (-1.8,2) to[out=0,in=90] (-1.5,1.75) to[out=-90,in=0] (-1.8,1.5);
    \draw[thick,Red] (-1.8,.5) to[out=0,in=90] (-1.5,.25) to[out=-90,in=0] (-1.8,0);
    \draw[thick,Red] (1.8,2) to[out=180,in=90] (1.5,1.75) to[out=-90,in=180] (1.8,1.5);
    \draw[thick,Red] (1.8,.5) to[out=180,in=90] (1.5,.25) to[out=-90,in=180] (1.8,0);
    \draw[gluon,Red] (-1.5,1.75) to[out=0,in=0,looseness=2] (-1.5,.25);
    \draw[lgluon,Red] (1.5,1.75) to[out=180,in=180,looseness=2] (1.5,.25);
\end{tikzpicture}}}
\,\right]
= \frac{- \hcalf^{(4)}_{gg}}{N_c^2-1}
\end{equation}
Due to the derivative insertions enforcing the adjoint projection, we can replace gluons by simply quark anti-quark lines, and we recover the operator $\hcalf^{(4)}_{gg}$ in the rhs.
It is worth noting that by invariance under time-reversal, the expectation values $\langle \hcalf_{gg}^{(3)} \rangle$ and $\langle \hcalf_{gg}^{(4)} \rangle$ are equal. However, at the operator level, those two serve different purposes in extracting irreps in the gluon-gluon sector, and they will differ on a per-realization basis within numerical simulations.

\vspace{.5cm} \noindent
\textit{The identity contribution from $\{\mathcal{T}\}$.} Within the subspace $8 \otimes 8$, we can write the identity as:
\begin{equation}
I = 
\textcolor{Red}{4}\ 
    \tr \left[ 
\vcenter{\hbox{
\begin{tikzpicture}[scale=0.3]
    \draw[thick] (-5,2) -- ++ (3,0);\draw[thick,-<] (-5,2) -- ++ (1.5,0);
    \draw[thick] (-5,1.5) -- ++ (3,0);\draw[thick,->] (-5,1.5) -- ++ (1.5,0);
    \draw[thick] (-5,1.5) to[out=180,in=180] (-5,2);
    \draw[fill=Blue!50!white] (-3,1.5) circle (0.2);
    \draw[thick,-<] (-6,.5) -- ++ (1.5,0);\draw[thick] (-6,.5) -- ++ (4,0);
    \draw[thick,->] (-6,0) -- ++ (1.5,0);\draw[thick] (-6,0) -- ++ (4,0);
    \draw[thick] (5,2) -- ++ (-3,0);\draw[thick,->] (5,2) -- ++ (-1.5,0);
    \draw[thick] (5,1.5) -- ++ (-3,0);\draw[thick,-<] (5,1.5) -- ++ (-1.5,0);
    \draw[thick] (5,1.5) to[out=0,in=0] (5,2);
    \draw[fill=Blue!50!white] (3,1.5) circle (0.2);
    \draw[thick,->] (6,.5) -- ++ (-1.5,0);\draw[thick] (6,.5) -- ++ (-4,0);
    \draw[thick,-<] (6,0) -- ++ (-1.5,0);\draw[thick] (6,0) -- ++ (-4,0);
    \node at (-6,-.75) {\scriptsize $-\infty$};
    \node at (+6,-.75) {\scriptsize $-\infty$};
    \node at (-2,-.75) {\scriptsize $+\infty$};
    \node at (+2,-.75) {\scriptsize $+\infty$};
    \node at (-7,.75) {\scriptsize $\xvec$};
    \node at (7,.75) {\scriptsize $\yvec$};
    \draw[thick,Red] (-1.8,2) to[out=0,in=90] (-1.5,1.75) to[out=-90,in=0] (-1.8,1.5);
    \draw[thick,Red] (-1.8,.5) to[out=0,in=90] (-1.5,.25) to[out=-90,in=0] (-1.8,0);
    \draw[thick,Red] (1.8,2) to[out=180,in=90] (1.5,1.75) to[out=-90,in=180] (1.8,1.5);
    \draw[thick,Red] (1.8,.5) to[out=180,in=90] (1.5,.25) to[out=-90,in=180] (1.8,0);
    \draw[gluon,Red] (-1.5,1.75) -- (1.5,1.75);
    \draw[gluon,Red] (-1.5,.25) -- (1.5,.25);
\end{tikzpicture}}}
\,\right]
\end{equation}
The gluon at the top can be replaced by quark lines since the derivative enforces the adjoint projection. The second gluon is recast using Fierz identity and we have
\begin{align}
& = \tr \left[ 
\vcenter{\hbox{
\begin{tikzpicture}[scale=0.3]
    \draw[thick] (-5,2) -- ++ (3,0);\draw[thick,-<] (-5,2) -- ++ (1.5,0);
    \draw[thick] (-5,1.5) -- ++ (3,0);\draw[thick,->] (-5,1.5) -- ++ (1.5,0);
    \draw[thick] (-5,1.5) to[out=180,in=180] (-5,2);
    \draw[fill=Blue!50!white] (-3,1.5) circle (0.2);
    \draw[thick,-<] (-6,.5) -- ++ (1.5,0);\draw[thick] (-6,.5) -- ++ (4,0);
    \draw[thick,->] (-6,0) -- ++ (1.5,0);\draw[thick] (-6,0) -- ++ (4,0);
    \draw[thick] (5,2) -- ++ (-3,0);\draw[thick,->] (5,2) -- ++ (-1.5,0);
    \draw[thick] (5,1.5) -- ++ (-3,0);\draw[thick,-<] (5,1.5) -- ++ (-1.5,0);
    \draw[thick] (5,1.5) to[out=0,in=0] (5,2);
    \draw[fill=Blue!50!white] (3,1.5) circle (0.2);
    \draw[thick,->] (6,.5) -- ++ (-1.5,0);\draw[thick] (6,.5) -- ++ (-4,0);
    \draw[thick,-<] (6,0) -- ++ (-1.5,0);\draw[thick] (6,0) -- ++ (-4,0);
    \node at (-6,-.75) {\scriptsize $-\infty$};
    \node at (+6,-.75) {\scriptsize $-\infty$};
    \node at (-2,-.75) {\scriptsize $+\infty$};
    \node at (+2,-.75) {\scriptsize $+\infty$};
    \node at (-7,.75) {\scriptsize $\xvec$};
    \node at (7,.75) {\scriptsize $\yvec$};
    \draw[thick,Red] (-1.8,2) -- (1.8,2);
    \draw[thick,Red] (-1.8,1.5) -- (1.8,1.5);
    \draw[thick,Red] (-1.8,.5) -- (1.8,.5);
    \draw[thick,Red] (-1.8,0) -- (1.8,0);
\end{tikzpicture}}}
\,\right]
-\frac{1}{N_c}
\tr \left[ 
\vcenter{\hbox{
\begin{tikzpicture}[scale=0.3]
    \draw[thick] (-5,2) -- ++ (3,0);\draw[thick,-<] (-5,2) -- ++ (1.5,0);
    \draw[thick] (-5,1.5) -- ++ (3,0);\draw[thick,->] (-5,1.5) -- ++ (1.5,0);
    \draw[thick] (-5,1.5) to[out=180,in=180] (-5,2);
    \draw[fill=Blue!50!white] (-3,1.5) circle (0.2);
    \draw[thick,-<] (-6,.5) -- ++ (1.5,0);\draw[thick] (-6,.5) -- ++ (4,0);
    \draw[thick,->] (-6,0) -- ++ (1.5,0);\draw[thick] (-6,0) -- ++ (4,0);
    \draw[thick] (5,2) -- ++ (-3,0);\draw[thick,->] (5,2) -- ++ (-1.5,0);
    \draw[thick] (5,1.5) -- ++ (-3,0);\draw[thick,-<] (5,1.5) -- ++ (-1.5,0);
    \draw[thick] (5,1.5) to[out=0,in=0] (5,2);
    \draw[fill=Blue!50!white] (3,1.5) circle (0.2);
    \draw[thick,->] (6,.5) -- ++ (-1.5,0);\draw[thick] (6,.5) -- ++ (-4,0);
    \draw[thick,-<] (6,0) -- ++ (-1.5,0);\draw[thick] (6,0) -- ++ (-4,0);
    \node at (-6,-.75) {\scriptsize $-\infty$};
    \node at (+6,-.75) {\scriptsize $-\infty$};
    \node at (-2,-.75) {\scriptsize $+\infty$};
    \node at (+2,-.75) {\scriptsize $+\infty$};
    \node at (-7,.75) {\scriptsize $\xvec$};
    \node at (7,.75) {\scriptsize $\yvec$};
    \draw[thick,Red] (-1.8,2) -- (1.8,2);
    \draw[thick,Red] (-1.8,1.5) -- (1.8,1.5);
    \draw[thick,Red] (-1.8,.5) to[out=0,in=90] (-1.5,.25) to[out=-90,in=0] (-1.8,0);
    \draw[thick,Red] (1.8,.5) to[out=180,in=90] (1.5,.25) to[out=-90,in=180] (1.8,0);
\end{tikzpicture}}}
\,\right] \notag \\[2ex]
&= \left( -N_c^2 \hcalf^{(6)}_{gg} \right) - \frac{1}{N_c} \left( -N_c \hcalf^{(3)}_{gg}\right) = - \left[ N_c^2 \hcalf^{(6)}_{gg} - \hcalf^{(3)}_{gg}\right]
\end{align}

\vspace{.5cm} \noindent
\textit{The permutation contribution from $\{\mathcal{T}\}$.} It reads
\begin{equation}
X = 
\textcolor{Red}{4}\ 
    \tr \left[ 
\vcenter{\hbox{
\begin{tikzpicture}[scale=0.3]
    \draw[thick] (-5,2) -- ++ (3,0);\draw[thick,-<] (-5,2) -- ++ (1.5,0);
    \draw[thick] (-5,1.5) -- ++ (3,0);\draw[thick,->] (-5,1.5) -- ++ (1.5,0);
    \draw[thick] (-5,1.5) to[out=180,in=180] (-5,2);
    \draw[fill=Blue!50!white] (-3,1.5) circle (0.2);
    \draw[thick,-<] (-6,.5) -- ++ (1.5,0);\draw[thick] (-6,.5) -- ++ (4,0);
    \draw[thick,->] (-6,0) -- ++ (1.5,0);\draw[thick] (-6,0) -- ++ (4,0);
    \draw[thick] (5,2) -- ++ (-3,0);\draw[thick,->] (5,2) -- ++ (-1.5,0);
    \draw[thick] (5,1.5) -- ++ (-3,0);\draw[thick,-<] (5,1.5) -- ++ (-1.5,0);
    \draw[thick] (5,1.5) to[out=0,in=0] (5,2);
    \draw[fill=Blue!50!white] (3,1.5) circle (0.2);
    \draw[thick,->] (6,.5) -- ++ (-1.5,0);\draw[thick] (6,.5) -- ++ (-4,0);
    \draw[thick,-<] (6,0) -- ++ (-1.5,0);\draw[thick] (6,0) -- ++ (-4,0);
    \node at (-6,-.75) {\scriptsize $-\infty$};
    \node at (+6,-.75) {\scriptsize $-\infty$};
    \node at (-2,-.75) {\scriptsize $+\infty$};
    \node at (+2,-.75) {\scriptsize $+\infty$};
    \node at (-7,.75) {\scriptsize $\xvec$};
    \node at (7,.75) {\scriptsize $\yvec$};
    \draw[thick,Red] (-1.8,2) to[out=0,in=90] (-1.5,1.75) to[out=-90,in=0] (-1.8,1.5);
    \draw[thick,Red] (-1.8,.5) to[out=0,in=90] (-1.5,.25) to[out=-90,in=0] (-1.8,0);
    \draw[thick,Red] (1.8,2) to[out=180,in=90] (1.5,1.75) to[out=-90,in=180] (1.8,1.5);
    \draw[thick,Red] (1.8,.5) to[out=180,in=90] (1.5,.25) to[out=-90,in=180] (1.8,0);
    \draw[gluon,Red] (-1.5,1.75) to[out=0,in=180] (1.5,.25);
    \draw[gluon,Red] (-1.5,.25) to[out=0,in=180] (1.5,1.75);
\end{tikzpicture}}}
\,\right] = \left( - \hcalf^{(5)}_{gg}\right)
\end{equation}
Again, the action of the derivative enforces the system to be in the adjoint, thus we can replace the two gluon lines by quark lines, and we simply recover the operator in the rhs.
Having access to the identity and the permutation, we can already split the vector space between symmetric and anti-symmetric irreps at $+ \infty$.

\vspace{.5cm} \noindent
\textit{The anti-symmetric adjoint contribution.} From the $8 \otimes 8$ irrep decomposition, we write the projection according to
\begin{equation}
F = 
\textcolor{Red}{\frac{4}{N_c}}\ 
    \tr \left[ 
\vcenter{\hbox{
\begin{tikzpicture}[scale=0.3]
    \draw[thick] (-5,2) -- ++ (3,0);\draw[thick,-<] (-5,2) -- ++ (1.5,0);
    \draw[thick] (-5,1.5) -- ++ (3,0);\draw[thick,->] (-5,1.5) -- ++ (1.5,0);
    \draw[thick] (-5,1.5) to[out=180,in=180] (-5,2);
    \draw[fill=Blue!50!white] (-3,1.5) circle (0.2);
    \draw[thick,-<] (-6,.5) -- ++ (1.5,0);\draw[thick] (-6,.5) -- ++ (4,0);
    \draw[thick,->] (-6,0) -- ++ (1.5,0);\draw[thick] (-6,0) -- ++ (4,0);
    \draw[thick] (5,2) -- ++ (-3,0);\draw[thick,->] (5,2) -- ++ (-1.5,0);
    \draw[thick] (5,1.5) -- ++ (-3,0);\draw[thick,-<] (5,1.5) -- ++ (-1.5,0);
    \draw[thick] (5,1.5) to[out=0,in=0] (5,2);
    \draw[fill=Blue!50!white] (3,1.5) circle (0.2);
    \draw[thick,->] (6,.5) -- ++ (-1.5,0);\draw[thick] (6,.5) -- ++ (-4,0);
    \draw[thick,-<] (6,0) -- ++ (-1.5,0);\draw[thick] (6,0) -- ++ (-4,0);
    \node at (-6,-.75) {\scriptsize $-\infty$};
    \node at (+6,-.75) {\scriptsize $-\infty$};
    \node at (-2,-.75) {\scriptsize $+\infty$};
    \node at (+2,-.75) {\scriptsize $+\infty$};
    \node at (-7,.75) {\scriptsize $\xvec$};
    \node at (7,.75) {\scriptsize $\yvec$};
    \draw[thick,Red] (-1.8,2) to[out=0,in=90] (-1.5,1.75) to[out=-90,in=0] (-1.8,1.5);
    \draw[thick,Red] (-1.8,.5) to[out=0,in=90] (-1.5,.25) to[out=-90,in=0] (-1.8,0);
    \draw[thick,Red] (1.8,2) to[out=180,in=90] (1.5,1.75) to[out=-90,in=180] (1.8,1.5);
    \draw[thick,Red] (1.8,.5) to[out=180,in=90] (1.5,.25) to[out=-90,in=180] (1.8,0);
    \draw[gluon,Red] (-1.5,1.75) -- (-1,1);
    \draw[gluon,Red] (-1.5,.25) -- (-1,1);
    \draw[gluon,Red] (1,1) -- (1.5,1.75);
    \draw[gluon,Red] (1,1) -- (1.5,.25);
    \draw[gluon,Red] (-1,1) -- (1,1);
\end{tikzpicture}}}
\,\right]
\end{equation}
Let us recall the trace definition for the structure constant $if^{abc}$ and the fully symmmetric vertex $d^{abc}$, in birdtracks we write:
\begin{equation}
\left\{
\vcenter{\hbox{
\begin{tikzpicture}[scale=0.5]
    \draw[gluon] (0,0) -- (0,1);
    \draw[gluon] (0,0) -- (1,-1);
    \draw[gluon] (0,0) -- (-1,-1);
\end{tikzpicture}}}, 
\vcenter{\hbox{
\begin{tikzpicture}[scale=0.5]
    \draw[gluon] (0,0) -- (0,1);
    \draw[gluon] (0,0) -- (1,-1);
    \draw[gluon] (0,0) -- (-1,-1);
    \%
    \node[dabc] at (0,0) {};
\end{tikzpicture}}}
\right\}
= 2 \left[ 
\vcenter{\hbox{
\begin{tikzpicture}[scale=0.5]
    \draw[gluon] (0,0) -- (0,1);
    \draw[gluon] (0,0) -- (1,-1);
    \draw[gluon] (0,0) -- (-1,-1);
    \draw[fill=white] (0,0) circle (0.3);
    \draw[->] (.1,-.3) -- ++(.001,0);
\end{tikzpicture}}}
\pm
\vcenter{\hbox{
\begin{tikzpicture}[scale=0.5]
    \draw[gluon] (0,0) -- (0,1);
    \draw[gluon] (0,0) -- (1,-1);
    \draw[gluon] (0,0) -- (-1,-1);
    \draw[fill=white] (0,0) circle (0.3);
    \draw[-<] (.1,-.3) -- ++(.001,0);
\end{tikzpicture}}}
\right]
\end{equation}
Using the definition of the structure constant as the trace of generators and the Fierz identity, one finds:
\begin{align}
\hat{\Omega}_{gg}^{8_a} = F &= 
\textcolor{Red}{\frac{1}{2 N_c}}\ 
    \tr \left[ 
\vcenter{\hbox{
\begin{tikzpicture}[scale=0.3]
    \draw[thick] (-5,2) -- ++ (3,0);\draw[thick,-<] (-5,2) -- ++ (1.5,0);
    \draw[thick] (-5,1.5) -- ++ (3,0);\draw[thick,->] (-5,1.5) -- ++ (1.5,0);
    \draw[thick] (-5,1.5) to[out=180,in=180] (-5,2);
    \draw[fill=Blue!50!white] (-3,1.5) circle (0.2);
    \draw[thick,-<] (-6,.5) -- ++ (1.5,0);\draw[thick] (-6,.5) -- ++ (4,0);
    \draw[thick,->] (-6,0) -- ++ (1.5,0);\draw[thick] (-6,0) -- ++ (4,0);
    \draw[thick] (5,2) -- ++ (-3,0);\draw[thick,->] (5,2) -- ++ (-1.5,0);
    \draw[thick] (5,1.5) -- ++ (-3,0);\draw[thick,-<] (5,1.5) -- ++ (-1.5,0);
    \draw[thick] (5,1.5) to[out=0,in=0] (5,2);
    \draw[fill=Blue!50!white] (3,1.5) circle (0.2);
    \draw[thick,->] (6,.5) -- ++ (-1.5,0);\draw[thick] (6,.5) -- ++ (-4,0);
    \draw[thick,-<] (6,0) -- ++ (-1.5,0);\draw[thick] (6,0) -- ++ (-4,0);
    \draw[thick,Red] (-1.8,1.5) to[out=0,in=0,looseness=2] (-1.8,.5);
    \draw[thick,Red] (1.8,1.5) to[out=180,in=180,looseness=2] (1.8,.5);
    \draw[thick,Red] (-1.8,2) -- (1.8,2);
    \draw[thick,Red] (-1.8,0) -- (1.8,0);
\end{tikzpicture}}}
+
\vcenter{\hbox{
\begin{tikzpicture}[scale=0.3]
    \draw[thick] (-5,2) -- ++ (3,0);\draw[thick,-<] (-5,2) -- ++ (1.5,0);
    \draw[thick] (-5,1.5) -- ++ (3,0);\draw[thick,->] (-5,1.5) -- ++ (1.5,0);
    \draw[thick] (-5,1.5) to[out=180,in=180] (-5,2);
    \draw[fill=Blue!50!white] (-3,1.5) circle (0.2);
    \draw[thick,-<] (-6,.5) -- ++ (1.5,0);\draw[thick] (-6,.5) -- ++ (4,0);
    \draw[thick,->] (-6,0) -- ++ (1.5,0);\draw[thick] (-6,0) -- ++ (4,0);
    \draw[thick] (5,2) -- ++ (-3,0);\draw[thick,->] (5,2) -- ++ (-1.5,0);
    \draw[thick] (5,1.5) -- ++ (-3,0);\draw[thick,-<] (5,1.5) -- ++ (-1.5,0);
    \draw[thick] (5,1.5) to[out=0,in=0] (5,2);
    \draw[fill=Blue!50!white] (3,1.5) circle (0.2);
    \draw[thick,->] (6,.5) -- ++ (-1.5,0);\draw[thick] (6,.5) -- ++ (-4,0);
    \draw[thick,-<] (6,0) -- ++ (-1.5,0);\draw[thick] (6,0) -- ++ (-4,0);
    \draw[thick,Red] (-1.8,2) to[out=0,in=0,looseness=2] (-1.8,0);
    \draw[thick,Red] (1.8,2) to[out=180,in=180,looseness=2] (1.8,0);
    \draw[thick,Red] (-1.8,1.5) -- (1.8,1.5);
    \draw[thick,Red] (-1.8,0.5) -- (1.8,0.5);
\end{tikzpicture}}} \right. \notag \\[2ex]
&\qquad \qquad -\left.
\vcenter{\hbox{
\begin{tikzpicture}[scale=0.3]
    \draw[thick] (-5,2) -- ++ (3,0);\draw[thick,-<] (-5,2) -- ++ (1.5,0);
    \draw[thick] (-5,1.5) -- ++ (3,0);\draw[thick,->] (-5,1.5) -- ++ (1.5,0);
    \draw[thick] (-5,1.5) to[out=180,in=180] (-5,2);
    \draw[fill=Blue!50!white] (-3,1.5) circle (0.2);
    \draw[thick,-<] (-6,.5) -- ++ (1.5,0);\draw[thick] (-6,.5) -- ++ (4,0);
    \draw[thick,->] (-6,0) -- ++ (1.5,0);\draw[thick] (-6,0) -- ++ (4,0);
    \draw[thick] (5,2) -- ++ (-3,0);\draw[thick,->] (5,2) -- ++ (-1.5,0);
    \draw[thick] (5,1.5) -- ++ (-3,0);\draw[thick,-<] (5,1.5) -- ++ (-1.5,0);
    \draw[thick] (5,1.5) to[out=0,in=0] (5,2);
    \draw[fill=Blue!50!white] (3,1.5) circle (0.2);
    \draw[thick,->] (6,.5) -- ++ (-1.5,0);\draw[thick] (6,.5) -- ++ (-4,0);
    \draw[thick,-<] (6,0) -- ++ (-1.5,0);\draw[thick] (6,0) -- ++ (-4,0);
    \draw[thick,Red] (-1.8,1.5) to[out=0,in=0,looseness=2] (-1.8,.5);
    \draw[thick,Red] (1.8,2) to[out=180,in=180,looseness=2] (1.8,0);
    \draw[thick,Red] (-1.8,2) to[out=0,in=180,looseness=2] (1.8,0.5);
    \draw[thick,Red] (-1.8,0) to[out=0,in=180,looseness=2] (1.8,1.5);
\end{tikzpicture}}}
-
\vcenter{\hbox{
\begin{tikzpicture}[scale=0.3]
    \draw[thick] (-5,2) -- ++ (3,0);\draw[thick,-<] (-5,2) -- ++ (1.5,0);
    \draw[thick] (-5,1.5) -- ++ (3,0);\draw[thick,->] (-5,1.5) -- ++ (1.5,0);
    \draw[thick] (-5,1.5) to[out=180,in=180] (-5,2);
    \draw[fill=Blue!50!white] (-3,1.5) circle (0.2);
    \draw[thick,-<] (-6,.5) -- ++ (1.5,0);\draw[thick] (-6,.5) -- ++ (4,0);
    \draw[thick,->] (-6,0) -- ++ (1.5,0);\draw[thick] (-6,0) -- ++ (4,0);
    \draw[thick] (5,2) -- ++ (-3,0);\draw[thick,->] (5,2) -- ++ (-1.5,0);
    \draw[thick] (5,1.5) -- ++ (-3,0);\draw[thick,-<] (5,1.5) -- ++ (-1.5,0);
    \draw[thick] (5,1.5) to[out=0,in=0] (5,2);
    \draw[fill=Blue!50!white] (3,1.5) circle (0.2);
    \draw[thick,->] (6,.5) -- ++ (-1.5,0);\draw[thick] (6,.5) -- ++ (-4,0);
    \draw[thick,-<] (6,0) -- ++ (-1.5,0);\draw[thick] (6,0) -- ++ (-4,0);
    \draw[thick,Red] (-1.8,2) to[out=0,in=0,looseness=2] (-1.8,0);
    \draw[thick,Red] (1.8,1.5) to[out=180,in=180,looseness=2] (1.8,0.5);
    \draw[thick,Red] (-1.8,0.5) to[out=0,in=180,looseness=2] (1.8,2);
    \draw[thick,Red] (-1.8,1.5) to[out=0,in=180,looseness=2] (1.8,0);
\end{tikzpicture}}}
\,\right] \notag \\
&= \frac{1}{2N_c} \left[ (-N_c \hcalf_{gg}^{(1)})  + (-N_c \hcalf_{gg}^{(1)})^* - (-N_c\hcalf_{gg}^{(2)}) - (-N_c\hcalf_{gg}^{(2)})^* \right] \\
&=  - \hcalf_{gg}^{(1)} + \hcalf_{gg}^{(2)} \label{eq:gg_antisym_adj}
\end{align}
where we used in the last line the fact that the operator is real.

\vspace{.5cm} \noindent
\textit{The two decuplet contributions.}
The two decuplets have the same quadratic Casimir and the same eigenvalue under permutation, thus we combine them. The expression for the decuplet can be recovered from the previous results. Use the identity and the permutation to project onto the anti-symmetric subspace, then remove the anti-symmetric adjoint projector (e.g., appendix C of \cite{Cougoulic:2017ust}). One is left with the projection into the subspace corresponding to the sum of the two decuplets.
\begin{equation} \label{eq:gg_deculet}
\hat{\Omega}_{gg}^{10+\overline{10}} = 
    \frac{1}{2}\left[ I - X \right] - F = -\frac{1}{2} \left[(N_c^2 \hcalf^{(6)}_{gg} - \hcalf^{(3)}_{gg})- \hcalf^{(5)}_{gg}\right] - \left[ \hcalf_{gg}^{(2)} - \hcalf_{gg}^{(1)} \right]
\end{equation}
Note that those two irreps do not appear for $SU(2)$, thus the corresponding linear combination of TMD operators vanishes, which we can check on a per-realization basis in numerical simulations.

\vspace{.5cm} \noindent
\textit{The symmetric adjoint contribution.}
Let us introduce the Clebsch-Gordon coefficient relevant for the symmetric adjoint representation within the $8 \otimes 8$ decomposition:
\begin{align}
c_{8s \leftarrow 8\otimes 8} &= 
\vcenter{\hbox{
\begin{tikzpicture}[scale=0.3]
    \draw[thick] (-4,2) -- ++ (2,0);\draw[thick,-<] (-4,2) -- ++ (1.5,0);
    \draw[thick] (-4,1.5) -- ++ (2,0);\draw[thick,->] (-4,1.5) -- ++ (1.5,0);
    \draw[thick] (-4,.5) -- ++ (2,0);\draw[thick,-<] (-4,.5) -- ++ (1.5,0);
    \draw[thick] (-4,0) -- ++ (2,0);\draw[thick,->] (-4,0) -- ++ (1.5,0);
    \draw[thick] (-2,2) to[out=0,in=90] (-1,1.75) to[out=-90,in=0] (-2,1.5);
    \draw[thick] (-2,.5) to[out=0,in=90] (-1,.25) to[out=-90,in=0] (-2,0);
    \draw[gluon] (-1,1.75) to[out=0,in=135] (1,1);
    \draw[gluon] (-1,.25) to[out=0,in=-135] (1,1);
    \draw[gluon] (1,1) -- ++(2,0);
    \node[dabc] at (1,1) {};
\end{tikzpicture}}} \notag \\[2ex]
&= \left[ 
\frac{1}{2}\left(
\vcenter{\hbox{
\begin{tikzpicture}[scale=0.3]
    \draw[thick] (-4,2) -- ++ (2,0);\draw[thick,-<] (-4,2) -- ++ (1.5,0);
    \draw[thick] (-4,1.5) -- ++ (2,0);\draw[thick,->] (-4,1.5) -- ++ (1.5,0);
    \draw[thick] (-4,.5) -- ++ (2,0);\draw[thick,-<] (-4,.5) -- ++ (1.5,0);
    \draw[thick] (-4,0) -- ++ (2,0);\draw[thick,->] (-4,0) -- ++ (1.5,0);
    \draw[thick] (-2,2) to[out=0,in=90] (-1,1) to[out=-90,in=0] (-2,0);
    \draw[thick] (-2,1.5) to[out=0,in=90] (-1.5,1) to[out=-90,in=0] (-2,0.5);
    \draw[gluon] (-1,1) -- ++(2,0);
\end{tikzpicture}}}
-
\vcenter{\hbox{
\begin{tikzpicture}[scale=0.3]
    \draw[thick] (-4,2) -- ++ (2,0);\draw[thick,-<] (-4,2) -- ++ (1.5,0);
    \draw[thick] (-4,1.5) -- ++ (2,0);\draw[thick,->] (-4,1.5) -- ++ (1.5,0);
    \draw[thick] (-4,.5) -- ++ (2,0);\draw[thick,-<] (-4,.5) -- ++ (1.5,0);
    \draw[thick] (-4,0) -- ++ (2,0);\draw[thick,->] (-4,0) -- ++ (1.5,0);
    \draw[thick] (-2,2) to[out=0,in=90] (-1,1) to[out=-90,in=0] (-2,0);
    \draw[thick] (-2,1.5) to[out=0,in=90] (-1.5,1) to[out=-90,in=0] (-2,0.5);
    \draw[gluon] (-1.5,1) -- ++(2.5,0);
\end{tikzpicture}}}
\right)
-\frac{1}{N_c}
\vcenter{\hbox{
\begin{tikzpicture}[scale=0.3]
    \draw[thick] (-4,2) -- ++ (2,0);\draw[thick,-<] (-4,2) -- ++ (1.5,0);
    \draw[thick] (-4,1.5) -- ++ (2,0);\draw[thick,->] (-4,1.5) -- ++ (1.5,0);
    \draw[thick] (-4,.5) -- ++ (2,0);\draw[thick,-<] (-4,.5) -- ++ (1.5,0);
    \draw[thick] (-4,0) -- ++ (2,0);\draw[thick,->] (-4,0) -- ++ (1.5,0);
    \draw[thick] (-2,2) to[out=0,in=90] (-1,1.75) to[out=-90,in=0] (-2,1.5);
    \draw[thick] (-2,.5) to[out=0,in=90] (-1,.25) to[out=-90,in=0] (-2,0);
    \draw[gluon] (-1,1.75) to[out=0,in=180] (1,1);
\end{tikzpicture}}}
\right]
\end{align}
and the projection operator is\footnote{Note that symbolic operators are read right-to-left while birdtracks are read left-to-right. This peculiar issue is often irrelevant as we focus only on hermitian version of projectors and expectation values for the operators of interest are real.}:
\begin{equation}
    \calp_{8_s} = \frac{4\,N_c}{N_c^2-4} \ \left( c_{8s \leftarrow 8\otimes 8} \right)^\dagger c_{8s \leftarrow 8\otimes 8}
\end{equation}

\begin{equation}
D = \textcolor{Red}{\frac{4N_c}{N_c^2-4}}
\tr \left[ 
\vcenter{\hbox{
\begin{tikzpicture}[scale=0.3]
    \draw[thick] (-5,2) -- ++ (3,0);\draw[thick,-<] (-5,2) -- ++ (1.5,0);
    \draw[thick] (-5,1.5) -- ++ (3,0);\draw[thick,->] (-5,1.5) -- ++ (1.5,0);
    \draw[thick] (-5,1.5) to[out=180,in=180] (-5,2);
    \draw[fill=Blue!50!white] (-3,1.5) circle (0.2);
    \draw[thick,-<] (-6,.5) -- ++ (1.5,0);\draw[thick] (-6,.5) -- ++ (4,0);
    \draw[thick,->] (-6,0) -- ++ (1.5,0);\draw[thick] (-6,0) -- ++ (4,0);
    \draw[thick] (5,2) -- ++ (-3,0);\draw[thick,->] (5,2) -- ++ (-1.5,0);
    \draw[thick] (5,1.5) -- ++ (-3,0);\draw[thick,-<] (5,1.5) -- ++ (-1.5,0);
    \draw[thick] (5,1.5) to[out=0,in=0] (5,2);
    \draw[fill=Blue!50!white] (3,1.5) circle (0.2);
    \draw[thick,->] (6,.5) -- ++ (-1.5,0);\draw[thick] (6,.5) -- ++ (-4,0);
    \draw[thick,-<] (6,0) -- ++ (-1.5,0);\draw[thick] (6,0) -- ++ (-4,0);
    \draw[thick,Red] (-1.8,2) to[out=0,in=90] (-1.5,1.75) to[out=-90,in=0] (-1.8,1.5);
    \draw[thick,Red] (-1.8,.5) to[out=0,in=90] (-1.5,.25) to[out=-90,in=0] (-1.8,0);
    \draw[thick,Red] (1.8,2) to[out=180,in=90] (1.5,1.75) to[out=-90,in=180] (1.8,1.5);
    \draw[thick,Red] (1.8,.5) to[out=180,in=90] (1.5,.25) to[out=-90,in=180] (1.8,0);
    \draw[gluon,Red] (-1.5,1.75) -- (-1,1);
    \draw[gluon,Red] (-1.5,.25) -- (-1,1);
    \draw[gluon,Red] (1,1) -- (1.5,1.75);
    \draw[gluon,Red] (1,1) -- (1.5,.25);
    \draw[gluon,Red] (-1,1) -- (1,1);
    \node[dabc,Red,scale=.8] at (-1,1) {};
    \node[dabc,Red,scale=.8] at (1,1) {};
\end{tikzpicture}}}
\,\right]
\equiv \frac{4N_c}{N_c^2-4} \widetilde{D}
\end{equation}
In order to remove the internal gluons within $\widetilde{D}$, we use the trace representation for the three-gluon symmetric vertex $d^{abc}$, then recursively use the Fierz identity to remove the intermediate gluons. After some simplification, one gets the intermediate result:
\begin{align}
\widetilde{D} &=
\frac{1}{8}\tr \left[ 
\vcenter{\hbox{
\begin{tikzpicture}[scale=0.3]
    \draw[thick] (-5,2) -- ++ (3,0);\draw[thick,-<] (-5,2) -- ++ (1.5,0);
    \draw[thick] (-5,1.5) -- ++ (3,0);\draw[thick,->] (-5,1.5) -- ++ (1.5,0);
    \draw[thick] (-5,1.5) to[out=180,in=180] (-5,2);
    \draw[fill=Blue!50!white] (-3,1.5) circle (0.2);
    \draw[thick,-<] (-6,.5) -- ++ (1.5,0);\draw[thick] (-6,.5) -- ++ (4,0);
    \draw[thick,->] (-6,0) -- ++ (1.5,0);\draw[thick] (-6,0) -- ++ (4,0);
    \draw[thick] (5,2) -- ++ (-3,0);\draw[thick,->] (5,2) -- ++ (-1.5,0);
    \draw[thick] (5,1.5) -- ++ (-3,0);\draw[thick,-<] (5,1.5) -- ++ (-1.5,0);
    \draw[thick] (5,1.5) to[out=0,in=0] (5,2);
    \draw[fill=Blue!50!white] (3,1.5) circle (0.2);
    \draw[thick,->] (6,.5) -- ++ (-1.5,0);\draw[thick] (6,.5) -- ++ (-4,0);
    \draw[thick,-<] (6,0) -- ++ (-1.5,0);\draw[thick] (6,0) -- ++ (-4,0);
    \draw[thick,Red] (-1.8,2) -- (1.8,2);
    \draw[thick,Red] (-1.8,1.5) to[out=0,in=0,looseness=2] (-1.8,.5);
    \draw[thick,Red] (1.8,1.5) to[out=180,in=180,looseness=2] (1.8,.5);
    \draw[thick,Red] (-1.8,0) -- (1.8,0);
\end{tikzpicture}}}
+
\vcenter{\hbox{
\begin{tikzpicture}[scale=0.3]
    \draw[thick] (-5,2) -- ++ (3,0);\draw[thick,-<] (-5,2) -- ++ (1.5,0);
    \draw[thick] (-5,1.5) -- ++ (3,0);\draw[thick,->] (-5,1.5) -- ++ (1.5,0);
    \draw[thick] (-5,1.5) to[out=180,in=180] (-5,2);
    \draw[fill=Blue!50!white] (-3,1.5) circle (0.2);
    \draw[thick,-<] (-6,.5) -- ++ (1.5,0);\draw[thick] (-6,.5) -- ++ (4,0);
    \draw[thick,->] (-6,0) -- ++ (1.5,0);\draw[thick] (-6,0) -- ++ (4,0);
    \draw[thick] (5,2) -- ++ (-3,0);\draw[thick,->] (5,2) -- ++ (-1.5,0);
    \draw[thick] (5,1.5) -- ++ (-3,0);\draw[thick,-<] (5,1.5) -- ++ (-1.5,0);
    \draw[thick] (5,1.5) to[out=0,in=0] (5,2);
    \draw[fill=Blue!50!white] (3,1.5) circle (0.2);
    \draw[thick,->] (6,.5) -- ++ (-1.5,0);\draw[thick] (6,.5) -- ++ (-4,0);
    \draw[thick,-<] (6,0) -- ++ (-1.5,0);\draw[thick] (6,0) -- ++ (-4,0);
    \draw[thick,Red] (-1.8,1.5) -- (1.8,1.5);
    \draw[thick,Red] (-1.8,2) to[out=0,in=0,looseness=2] (-1.8,0);
    \draw[thick,Red] (1.8,2) to[out=180,in=180,looseness=2] (1.8,0);
    \draw[thick,Red] (-1.8,.5) -- (1.8,.5);
\end{tikzpicture}}}
\right] \notag \\[2ex]
&+\frac{1}{8}\tr \left[ 
\vcenter{\hbox{
\begin{tikzpicture}[scale=0.3]
    \draw[thick] (-5,2) -- ++ (3,0);\draw[thick,-<] (-5,2) -- ++ (1.5,0);
    \draw[thick] (-5,1.5) -- ++ (3,0);\draw[thick,->] (-5,1.5) -- ++ (1.5,0);
    \draw[thick] (-5,1.5) to[out=180,in=180] (-5,2);
    \draw[fill=Blue!50!white] (-3,1.5) circle (0.2);
    \draw[thick,-<] (-6,.5) -- ++ (1.5,0);\draw[thick] (-6,.5) -- ++ (4,0);
    \draw[thick,->] (-6,0) -- ++ (1.5,0);\draw[thick] (-6,0) -- ++ (4,0);
    \draw[thick] (5,2) -- ++ (-3,0);\draw[thick,->] (5,2) -- ++ (-1.5,0);
    \draw[thick] (5,1.5) -- ++ (-3,0);\draw[thick,-<] (5,1.5) -- ++ (-1.5,0);
    \draw[thick] (5,1.5) to[out=0,in=0] (5,2);
    \draw[fill=Blue!50!white] (3,1.5) circle (0.2);
    \draw[thick,->] (6,.5) -- ++ (-1.5,0);\draw[thick] (6,.5) -- ++ (-4,0);
    \draw[thick,-<] (6,0) -- ++ (-1.5,0);\draw[thick] (6,0) -- ++ (-4,0);
    \draw[thick,Red] (-1.8,2) to[out=0,in=0,looseness=2] (-1.8,0);
    \draw[thick,Red] (1.8,1.5) to[out=180,in=180,looseness=2] (1.8,0.5);
    \draw[thick,Red] (-1.8,0.5) to[out=0,in=180,looseness=2] (1.8,2);
    \draw[thick,Red] (-1.8,1.5) to[out=0,in=180,looseness=2] (1.8,0);
\end{tikzpicture}}}
+
\vcenter{\hbox{
\begin{tikzpicture}[scale=0.3]
    \draw[thick] (-5,2) -- ++ (3,0);\draw[thick,-<] (-5,2) -- ++ (1.5,0);
    \draw[thick] (-5,1.5) -- ++ (3,0);\draw[thick,->] (-5,1.5) -- ++ (1.5,0);
    \draw[thick] (-5,1.5) to[out=180,in=180] (-5,2);
    \draw[fill=Blue!50!white] (-3,1.5) circle (0.2);
    \draw[thick,-<] (-6,.5) -- ++ (1.5,0);\draw[thick] (-6,.5) -- ++ (4,0);
    \draw[thick,->] (-6,0) -- ++ (1.5,0);\draw[thick] (-6,0) -- ++ (4,0);
    \draw[thick] (5,2) -- ++ (-3,0);\draw[thick,->] (5,2) -- ++ (-1.5,0);
    \draw[thick] (5,1.5) -- ++ (-3,0);\draw[thick,-<] (5,1.5) -- ++ (-1.5,0);
    \draw[thick] (5,1.5) to[out=0,in=0] (5,2);
    \draw[fill=Blue!50!white] (3,1.5) circle (0.2);
    \draw[thick,->] (6,.5) -- ++ (-1.5,0);\draw[thick] (6,.5) -- ++ (-4,0);
    \draw[thick,-<] (6,0) -- ++ (-1.5,0);\draw[thick] (6,0) -- ++ (-4,0);
    \draw[thick,Red] (-1.8,2) to[out=0,in=0,looseness=2] (-1.8,0);
    \draw[thick,Red] (1.8,1.5) to[out=180,in=180,looseness=2] (1.8,0.5);
    \draw[thick,Red] (-1.8,0.5) to[out=0,in=180,looseness=2] (1.8,2);
    \draw[thick,Red] (-1.8,1.5) to[out=0,in=180,looseness=2] (1.8,0);
\end{tikzpicture}}}
\right] \notag \\[2ex]
&-\frac{1}{2N_c}
\tr \left[
\vcenter{\hbox{
\begin{tikzpicture}[scale=0.3]
    \draw[thick] (-5,2) -- ++ (3,0);\draw[thick,-<] (-5,2) -- ++ (1.5,0);
    \draw[thick] (-5,1.5) -- ++ (3,0);\draw[thick,->] (-5,1.5) -- ++ (1.5,0);
    \draw[thick] (-5,1.5) to[out=180,in=180] (-5,2);
    \draw[fill=Blue!50!white] (-3,1.5) circle (0.2);
    \draw[thick,-<] (-6,.5) -- ++ (1.5,0);\draw[thick] (-6,.5) -- ++ (4,0);
    \draw[thick,->] (-6,0) -- ++ (1.5,0);\draw[thick] (-6,0) -- ++ (4,0);
    \draw[thick] (5,2) -- ++ (-3,0);\draw[thick,->] (5,2) -- ++ (-1.5,0);
    \draw[thick] (5,1.5) -- ++ (-3,0);\draw[thick,-<] (5,1.5) -- ++ (-1.5,0);
    \draw[thick] (5,1.5) to[out=0,in=0] (5,2);
    \draw[fill=Blue!50!white] (3,1.5) circle (0.2);
    \draw[thick,->] (6,.5) -- ++ (-1.5,0);\draw[thick] (6,.5) -- ++ (-4,0);
    \draw[thick,-<] (6,0) -- ++ (-1.5,0);\draw[thick] (6,0) -- ++ (-4,0);
    \draw[thick,Red] (-1.8,2) to[out=0,in=0,looseness=2] (-1.8,0);
    \draw[thick,Red] (-1.8,1.5) to[out=0,in=0,looseness=2] (-1.8,.5);
    \draw[thick,Red] (1.8,2) to[out=180,in=180,looseness=2] (1.8,0);
    \draw[thick,Red] (1.8,1.5) to[out=180,in=180,looseness=2] (1.8,0.5);
\end{tikzpicture}}}
+ \frac{1}{N_c}
\vcenter{\hbox{
\begin{tikzpicture}[scale=0.3]
    \draw[thick] (-5,2) -- ++ (3,0);\draw[thick,-<] (-5,2) -- ++ (1.5,0);
    \draw[thick] (-5,1.5) -- ++ (3,0);\draw[thick,->] (-5,1.5) -- ++ (1.5,0);
    \draw[thick] (-5,1.5) to[out=180,in=180] (-5,2);
    \draw[fill=Blue!50!white] (-3,1.5) circle (0.2);
    \draw[thick,-<] (-6,.5) -- ++ (1.5,0);\draw[thick] (-6,.5) -- ++ (4,0);
    \draw[thick,->] (-6,0) -- ++ (1.5,0);\draw[thick] (-6,0) -- ++ (4,0);
    \draw[thick] (5,2) -- ++ (-3,0);\draw[thick,->] (5,2) -- ++ (-1.5,0);
    \draw[thick] (5,1.5) -- ++ (-3,0);\draw[thick,-<] (5,1.5) -- ++ (-1.5,0);
    \draw[thick] (5,1.5) to[out=0,in=0] (5,2);
    \draw[fill=Blue!50!white] (3,1.5) circle (0.2);
    \draw[thick,->] (6,.5) -- ++ (-1.5,0);\draw[thick] (6,.5) -- ++ (-4,0);
    \draw[thick,-<] (6,0) -- ++ (-1.5,0);\draw[thick] (6,0) -- ++ (-4,0);
    \draw[thick,Red] (-1.8,2) -- (1.8,2);
    \draw[thick,Red] (-1.8,1.5) -- (1.8,1.5);
    \draw[thick,Red] (-1.8,.5) to[out=0,in=0,looseness=2] (-1.8,0);
    \draw[thick,Red] (1.8,.5) to[out=180,in=180,looseness=2] (1.8,0);
\end{tikzpicture}}}
\right] \notag \\[2ex]
&-\frac{1}{4N_c} \tr \left[ 
\vcenter{\hbox{
\begin{tikzpicture}[scale=0.3]
    \draw[thick] (-5,2) -- ++ (3,0);\draw[thick,-<] (-5,2) -- ++ (1.5,0);
    \draw[thick] (-5,1.5) -- ++ (3,0);\draw[thick,->] (-5,1.5) -- ++ (1.5,0);
    \draw[thick] (-5,1.5) to[out=180,in=180] (-5,2);
    \draw[fill=Blue!50!white] (-3,1.5) circle (0.2);
    \draw[thick,-<] (-6,.5) -- ++ (1.5,0);\draw[thick] (-6,.5) -- ++ (4,0);
    \draw[thick,->] (-6,0) -- ++ (1.5,0);\draw[thick] (-6,0) -- ++ (4,0);
    \draw[thick] (5,2) -- ++ (-3,0);\draw[thick,->] (5,2) -- ++ (-1.5,0);
    \draw[thick] (5,1.5) -- ++ (-3,0);\draw[thick,-<] (5,1.5) -- ++ (-1.5,0);
    \draw[thick] (5,1.5) to[out=0,in=0] (5,2);
    \draw[fill=Blue!50!white] (3,1.5) circle (0.2);
    \draw[thick,->] (6,.5) -- ++ (-1.5,0);\draw[thick] (6,.5) -- ++ (-4,0);
    \draw[thick,-<] (6,0) -- ++ (-1.5,0);\draw[thick] (6,0) -- ++ (-4,0);
    \draw[thick,Red] (-1.8,2) -- (1.8,2);
    \draw[thick,Red] (-1.8,1.5) to[out=0,in=180,looseness=2] (1.8,0);
    \draw[thick,Red] (-1.8,.5) to[out=0,in=0,looseness=2] (-1.8,0);
    \draw[thick,Red] (1.8,.5) to[out=180,in=180,looseness=2] (1.8,1.5);
\end{tikzpicture}}}
+
\vcenter{\hbox{
\begin{tikzpicture}[scale=0.3]
    \draw[thick] (-5,2) -- ++ (3,0);\draw[thick,-<] (-5,2) -- ++ (1.5,0);
    \draw[thick] (-5,1.5) -- ++ (3,0);\draw[thick,->] (-5,1.5) -- ++ (1.5,0);
    \draw[thick] (-5,1.5) to[out=180,in=180] (-5,2);
    \draw[fill=Blue!50!white] (-3,1.5) circle (0.2);
    \draw[thick,-<] (-6,.5) -- ++ (1.5,0);\draw[thick] (-6,.5) -- ++ (4,0);
    \draw[thick,->] (-6,0) -- ++ (1.5,0);\draw[thick] (-6,0) -- ++ (4,0);
    \draw[thick] (5,2) -- ++ (-3,0);\draw[thick,->] (5,2) -- ++ (-1.5,0);
    \draw[thick] (5,1.5) -- ++ (-3,0);\draw[thick,-<] (5,1.5) -- ++ (-1.5,0);
    \draw[thick] (5,1.5) to[out=0,in=0] (5,2);
    \draw[fill=Blue!50!white] (3,1.5) circle (0.2);
    \draw[thick,->] (6,.5) -- ++ (-1.5,0);\draw[thick] (6,.5) -- ++ (-4,0);
    \draw[thick,-<] (6,0) -- ++ (-1.5,0);\draw[thick] (6,0) -- ++ (-4,0);
    \draw[thick,Red] (-1.8,2) -- (1.8,2);
    \draw[thick,Red] (-1.8,0) to[out=0,in=180,looseness=2] (1.8,1.5);
    \draw[thick,Red] (-1.8,.5) to[out=0,in=0,looseness=2] (-1.8,1.5);
    \draw[thick,Red] (1.8,.5) to[out=180,in=180,looseness=2] (1.8,0);
\end{tikzpicture}}}
\right] \notag \\[2ex]
&-\frac{1}{4N_c} \tr \left[
\vcenter{\hbox{
\begin{tikzpicture}[scale=0.3]
    \draw[thick] (-5,2) -- ++ (3,0);\draw[thick,-<] (-5,2) -- ++ (1.5,0);
    \draw[thick] (-5,1.5) -- ++ (3,0);\draw[thick,->] (-5,1.5) -- ++ (1.5,0);
    \draw[thick] (-5,1.5) to[out=180,in=180] (-5,2);
    \draw[fill=Blue!50!white] (-3,1.5) circle (0.2);
    \draw[thick,-<] (-6,.5) -- ++ (1.5,0);\draw[thick] (-6,.5) -- ++ (4,0);
    \draw[thick,->] (-6,0) -- ++ (1.5,0);\draw[thick] (-6,0) -- ++ (4,0);
    \draw[thick] (5,2) -- ++ (-3,0);\draw[thick,->] (5,2) -- ++ (-1.5,0);
    \draw[thick] (5,1.5) -- ++ (-3,0);\draw[thick,-<] (5,1.5) -- ++ (-1.5,0);
    \draw[thick] (5,1.5) to[out=0,in=0] (5,2);
    \draw[fill=Blue!50!white] (3,1.5) circle (0.2);
    \draw[thick,->] (6,.5) -- ++ (-1.5,0);\draw[thick] (6,.5) -- ++ (-4,0);
    \draw[thick,-<] (6,0) -- ++ (-1.5,0);\draw[thick] (6,0) -- ++ (-4,0);
    \draw[thick,Red] (-1.8,1.5) -- (1.8,1.5);
    \draw[thick,Red] (-1.8,2) to[out=0,in=180,looseness=2] (1.8,.5);
    \draw[thick,Red] (-1.8,.5) to[out=0,in=0,looseness=2] (-1.8,0);
    \draw[thick,Red] (1.8,0) to[out=180,in=180,looseness=1.5] (1.8,2);
\end{tikzpicture}}}
+
\vcenter{\hbox{
\begin{tikzpicture}[scale=0.3]
    \draw[thick] (-5,2) -- ++ (3,0);\draw[thick,-<] (-5,2) -- ++ (1.5,0);
    \draw[thick] (-5,1.5) -- ++ (3,0);\draw[thick,->] (-5,1.5) -- ++ (1.5,0);
    \draw[thick] (-5,1.5) to[out=180,in=180] (-5,2);
    \draw[fill=Blue!50!white] (-3,1.5) circle (0.2);
    \draw[thick,-<] (-6,.5) -- ++ (1.5,0);\draw[thick] (-6,.5) -- ++ (4,0);
    \draw[thick,->] (-6,0) -- ++ (1.5,0);\draw[thick] (-6,0) -- ++ (4,0);
    \draw[thick] (5,2) -- ++ (-3,0);\draw[thick,->] (5,2) -- ++ (-1.5,0);
    \draw[thick] (5,1.5) -- ++ (-3,0);\draw[thick,-<] (5,1.5) -- ++ (-1.5,0);
    \draw[thick] (5,1.5) to[out=0,in=0] (5,2);
    \draw[fill=Blue!50!white] (3,1.5) circle (0.2);
    \draw[thick,->] (6,.5) -- ++ (-1.5,0);\draw[thick] (6,.5) -- ++ (-4,0);
    \draw[thick,-<] (6,0) -- ++ (-1.5,0);\draw[thick] (6,0) -- ++ (-4,0);
    \draw[thick,Red] (-1.8,1.5) -- (1.8,1.5);
    \draw[thick,Red] (-1.8,.5) to[out=0,in=180,looseness=2] (1.8,2);
    \draw[thick,Red] (-1.8,0) to[out=0,in=0,looseness=1.5] (-1.8,2);
    \draw[thick,Red] (1.8,.5) to[out=180,in=180,looseness=2] (1.8,0);
\end{tikzpicture}}}
\right]
\end{align}
One finally expresses those operators in $\widetilde{D}$ in terms of TMD operators, where we used again the fact that the operators are real:
\begin{equation}\label{eq:gg_sym_adj}
    \hat{\Omega}^{8_s} = D = \frac{4N_c}{N_c^2-4} \left[\frac{1}{4}(-N_c \hcalf^{(1+2)}_{gg}) +\frac{1}{2N_c} \hcalf^{(3+4)}_{gg}\right]
\end{equation}
Note that this irrep does not appear for $SU(2)$, thus the linear combination of TMD operators, within the square bracket, vanishes. This can be checked explicitly within the numerical simulations on a per-realization basis.

\vspace{.5cm} \noindent
\textit{The remaining two symmetric representations}
One can easily access the subspace corresponding to the irreps $0 \oplus 27$ by using the identity and the permutation to single out the symmetric subspace, then removing the previously computed symmetric irreps $1 \oplus 8_s$. Namely
\begin{equation}
    \frac{1}{2} \left[ I + X \right] - D - S
\end{equation}
To separate the two remaining irreps, we need to consider the following contributions (e.g., Appendix A.1 of \cite{Dokshitzer:2005ig}):
\begin{equation}
    \calp_{27/0} = \pi^{\pm} - \frac{N_c \mp 2}{2N_c} \calp_{8_s} - \frac{N_c \mp 1}{2N_c} \calp_{1}
\end{equation}
with
\begin{equation}
    \pi^{\pm} =
    \vcenter{\hbox{
\begin{tikzpicture}[scale=.3,decoration={markings, mark=at position 0.5 with {\arrow{>}}}]
    \draw[thick,postaction={decorate}] (-2,1) -- (-4,1);
    \draw[thick,postaction={decorate}] (-4,.5) -- (-2,.5);
    \draw[thick,postaction={decorate}] (-2,-.5) -- (-4,-.5);
    \draw[thick,postaction={decorate}] (-4,-1) -- (-2,-1);
    \draw[thick,postaction={decorate}] (4,1) -- (2,1);
    \draw[thick,postaction={decorate}] (2,.5) -- (4,.5);
    \draw[thick,postaction={decorate}] (4,-.5) -- (2,-.5);
    \draw[thick,postaction={decorate}] (2,-1) -- (4,-1);
    \draw[thick] (-2,1) to (0,1) to (2,1);
    \draw[thick] (-2,-.5) to[out=0,in=180] (0,.5) to[out=0,in=180] (2,-.5);
    \draw[thick] (-2,.5) to[out=0,in=180] (0,-.5) to[out=0,in=180] (2,.5);
    \draw[thick] (-2,-1) to (0,-1) to (2,-1);
    \draw[fill=black!30!white] (-.3,1.3) rectangle (.3,.2);
    \draw[fill=black!30!white] (-.3,-1.3) rectangle (.3,-.2);
    \draw[thick] (-4,1) to (-4.5,.75) to (-4,.5);
    \draw[gluon] (-7,.75) -- (-4.5,.75);
    \draw[thick] (-4,-1) to (-4.5,-.75) to (-4,-.5);
    \draw[gluon] (-7,-.75) -- (-4.5,-.75);
    \draw[thick] (4,1) to (4.5,.75) to (4,.5);
    \draw[gluon] (4.5,.75) -- (7,.75);
    \draw[thick] (4,-1) to (4.5,-.75) to (4,-.5);
    \draw[gluon] (4.5,-.75) -- (7,-.75);
\end{tikzpicture}}}
\end{equation}
and the shaded rectangles denote either the symmetrizer or the anti-symmetrizer, depending on the sign $\pm$ used for $\pi^\pm$.
Placed into the usual trace, the contribution which is \underline{not} proportional to the identity $I$ and the permutation $X$ reads:
\begin{align}
W &= \textcolor{Red}{4}\ \tr \left[ 
    \vcenter{\hbox{
\begin{tikzpicture}[scale=0.3]
    \draw[thick] (-7,2) -- ++ (3,0);\draw[thick,-<] (-7,2) -- ++ (1.5,0);
    \draw[thick] (-7,1.5) -- ++ (3,0);\draw[thick,->] (-7,1.5) -- ++ (1.5,0);
    \draw[thick] (-7,1.5) to[out=180,in=180] (-7,2);
    \draw[fill=Blue!50!white] (-5,1.5) circle (0.2);
    \draw[thick,-<] (-8,.5) -- ++ (1.5,0);\draw[thick] (-8,.5) -- ++ (4,0);
    \draw[thick,->] (-8,0) -- ++ (1.5,0);\draw[thick] (-8,0) -- ++ (4,0);
    \draw[thick] (7,2) -- ++ (-3,0);\draw[thick,->] (7,2) -- ++ (-1.5,0);
    \draw[thick] (7,1.5) -- ++ (-3,0);\draw[thick,-<] (7,1.5) -- ++ (-1.5,0);
    \draw[thick] (7,1.5) to[out=0,in=0] (7,2);
    \draw[fill=Blue!50!white] (5,1.5) circle (0.2);
    \draw[thick,->] (8,.5) -- ++ (-1.5,0);\draw[thick] (8,.5) -- ++ (-4,0);
    \draw[thick,-<] (8,0) -- ++ (-1.5,0);\draw[thick] (8,0) -- ++ (-4,0);
    \draw[Red,thick] (-1,2) -- (1,2);
    \draw[Red,thick,-<] (-1,2) -- (0,2);
    \draw[Red,thick] (-1,.5) -- (1,.5);
    \draw[Red,thick] (-1,1.5) to[out=0,in=180] (1,0);
    \draw[Red,thick] (-1,0) to[out=0,in=180] (1,1.5);
    \draw[Orange,thick] (-1,2) -- (-1.5,1.75) -- (-1,1.5);
    \draw[Orange,thick] (-1,0) -- (-1.5,.25) -- (-1,.5);
    \draw[Orange,gluon] (-3,1.75) -- (-1.5,1.75);
    \draw[Orange,thick] (-3.5,2) -- (-3,1.75) -- (-3.5,1.5);
    \draw[Orange,gluon] (-3,.25) -- (-1.5,.25);
    \draw[Orange,thick] (-3.5,.5) -- (-3,.25) -- (-3.5,0);
    \draw[Orange,thick] (1,2) -- (1.5,1.75) -- (1,1.5);
    \draw[Orange,thick] (1,0) -- (1.5,.25) -- (1,.5);
    \draw[Orange,lgluon] (3,1.75) -- (1.5,1.75);
    \draw[Orange,thick] (3.5,2) -- (3,1.75) -- (3.5,1.5);
    \draw[Orange,lgluon] (3,.25) -- (1.5,.25);
    \draw[Orange,thick] (3.5,.5) -- (3,.25) -- (3.5,0);
\end{tikzpicture}}} + \text{c.c.}
\right] \notag \\[2ex]
&= \tr \left[ 
    \vcenter{\hbox{
\begin{tikzpicture}[scale=0.3]
    \draw[thick] (-7,2) -- ++ (3,0);\draw[thick,-<] (-7,2) -- ++ (1.5,0);
    \draw[thick] (-7,1.5) -- ++ (3,0);\draw[thick,->] (-7,1.5) -- ++ (1.5,0);
    \draw[thick] (-7,1.5) to[out=180,in=180] (-7,2);
    \draw[fill=Blue!50!white] (-5,1.5) circle (0.2);
    \draw[thick,-<] (-8,.5) -- ++ (1.5,0);\draw[thick] (-8,.5) -- ++ (4,0);
    \draw[thick,->] (-8,0) -- ++ (1.5,0);\draw[thick] (-8,0) -- ++ (4,0);
    \draw[thick] (7,2) -- ++ (-3,0);\draw[thick,->] (7,2) -- ++ (-1.5,0);
    \draw[thick] (7,1.5) -- ++ (-3,0);\draw[thick,-<] (7,1.5) -- ++ (-1.5,0);
    \draw[thick] (7,1.5) to[out=0,in=0] (7,2);
    \draw[fill=Blue!50!white] (5,1.5) circle (0.2);
    \draw[thick,->] (8,.5) -- ++ (-1.5,0);\draw[thick] (8,.5) -- ++ (-4,0);
    \draw[thick,-<] (8,0) -- ++ (-1.5,0);\draw[thick] (8,0) -- ++ (-4,0);
    \draw[Red,thick] (-1,2) -- (1,2);
    \draw[Red,thick,-<] (-1,2) -- (0,2);
    \draw[Red,thick] (-1,.5) -- (1,.5);
    \draw[Red,thick] (-1,1.5) to[out=0,in=180] (1,0);
    \draw[Red,thick] (-1,0) to[out=0,in=180] (1,1.5);
    \draw[thick,Orange] (-3.5,2) -- (-1,2);
    \draw[thick,Orange] (-3.5,1.5) -- (-1,1.5);
    \draw[Orange,thick] (-1,0) -- (-1.5,.25) -- (-1,.5);
    \draw[Orange,gluon] (-3,.25) -- (-1.5,.25);
    \draw[Orange,thick] (-3.5,.5) -- (-3,.25) -- (-3.5,0);
    \draw[thick,Orange] (3.5,2) -- (1,2);
    \draw[thick,Orange] (3.5,1.5) -- (1,1.5);
    \draw[Orange,thick] (1,0) -- (1.5,.25) -- (1,.5);
    \draw[Orange,lgluon] (3,.25) -- (1.5,.25);
    \draw[Orange,thick] (3.5,.5) -- (3,.25) -- (3.5,0);
\end{tikzpicture}}} + \text{c.c.}
\right]
\end{align}
At this point, it is worth noticing that upon using the Fierz identity to remove the last two gluons, the singlet contribution will be orthogonal to the adjoint contribution within the trace. Thus, we simply add the two contributions incoherently:
\begin{align}
    W &= \frac{1}{4}
    \tr \left[ 
    \vcenter{\hbox{
\begin{tikzpicture}[scale=0.3]
    \draw[thick] (-7,2) -- ++ (3,0);\draw[thick,-<] (-7,2) -- ++ (1.5,0);
    \draw[thick] (-7,1.5) -- ++ (3,0);\draw[thick,->] (-7,1.5) -- ++ (1.5,0);
    \draw[thick] (-7,1.5) to[out=180,in=180] (-7,2);
    \draw[fill=Blue!50!white] (-5,1.5) circle (0.2);
    \draw[thick,-<] (-8,.5) -- ++ (1.5,0);\draw[thick] (-8,.5) -- ++ (4,0);
    \draw[thick,->] (-8,0) -- ++ (1.5,0);\draw[thick] (-8,0) -- ++ (4,0);
    \draw[thick] (7,2) -- ++ (-3,0);\draw[thick,->] (7,2) -- ++ (-1.5,0);
    \draw[thick] (7,1.5) -- ++ (-3,0);\draw[thick,-<] (7,1.5) -- ++ (-1.5,0);
    \draw[thick] (7,1.5) to[out=0,in=0] (7,2);
    \draw[fill=Blue!50!white] (5,1.5) circle (0.2);
    \draw[thick,->] (8,.5) -- ++ (-1.5,0);\draw[thick] (8,.5) -- ++ (-4,0);
    \draw[thick,-<] (8,0) -- ++ (-1.5,0);\draw[thick] (8,0) -- ++ (-4,0);
    \draw[Red,thick] (-1,2) -- (1,2);
    \draw[Red,thick,-<] (-1,2) -- (0,2);
    \draw[Red,thick] (-1,.5) -- (1,.5);
    \draw[Red,thick] (-1,1.5) to[out=0,in=180] (1,0);
    \draw[Red,thick] (-1,0) to[out=0,in=180] (1,1.5);
    \draw[thick,Orange] (-3.5,2) -- (-1,2);
    \draw[thick,Orange] (-3.5,1.5) -- (-1,1.5);
    \draw[thick,Orange] (-3.5,.5) -- (-1,.5);
    \draw[thick,Orange] (-3.5,0) -- (-1,0);
    \draw[thick,Orange] (3.5,2) -- (1,2);
    \draw[thick,Orange] (3.5,1.5) -- (1,1.5);
    \draw[thick,Orange] (3.5,.5) -- (1,.5);
    \draw[thick,Orange] (3.5,0) -- (1,0);
\end{tikzpicture}}} + \text{c.c.}
\right] \notag \\[2ex]
&-\frac{1}{4N_c^2}
    \tr \left[ 
    \vcenter{\hbox{
\begin{tikzpicture}[scale=0.3]
    \draw[thick] (-7,2) -- ++ (3,0);\draw[thick,-<] (-7,2) -- ++ (1.5,0);
    \draw[thick] (-7,1.5) -- ++ (3,0);\draw[thick,->] (-7,1.5) -- ++ (1.5,0);
    \draw[thick] (-7,1.5) to[out=180,in=180] (-7,2);
    \draw[fill=Blue!50!white] (-5,1.5) circle (0.2);
    \draw[thick,-<] (-8,.5) -- ++ (1.5,0);\draw[thick] (-8,.5) -- ++ (4,0);
    \draw[thick,->] (-8,0) -- ++ (1.5,0);\draw[thick] (-8,0) -- ++ (4,0);
    \draw[thick] (7,2) -- ++ (-3,0);\draw[thick,->] (7,2) -- ++ (-1.5,0);
    \draw[thick] (7,1.5) -- ++ (-3,0);\draw[thick,-<] (7,1.5) -- ++ (-1.5,0);
    \draw[thick] (7,1.5) to[out=0,in=0] (7,2);
    \draw[fill=Blue!50!white] (5,1.5) circle (0.2);
    \draw[thick,->] (8,.5) -- ++ (-1.5,0);\draw[thick] (8,.5) -- ++ (-4,0);
    \draw[thick,-<] (8,0) -- ++ (-1.5,0);\draw[thick] (8,0) -- ++ (-4,0);
    \draw[Red,thick] (-1,2) -- (1,2);
    \draw[Red,thick,-<] (-1,2) -- (0,2);
    \draw[Red,thick] (-1,1.5) -- (1,1.5);
    \draw[thick,Orange] (-3.5,2) -- (-1,2);
    \draw[thick,Orange] (-3.5,1.5) -- (-1,1.5);
    \draw[Orange,thick] (-3.5,.5) -- (-3,.25) -- (-3.5,0);
    \draw[thick,Orange] (3.5,2) -- (1,2);
    \draw[thick,Orange] (3.5,1.5) -- (1,1.5);
    \draw[Orange,thick] (3.5,.5) -- (3,.25) -- (3.5,0);
\end{tikzpicture}}} + \text{c.c.}
\right] \notag \\
&=  \frac{1}{4} \left( -N_c \hcalf^{(7)}_{gg} + c.c.\right)
-\frac{1}{4N_c^2} \left(-N_c \hcalf^{(3)}_{gg} + c.c.\right)
\end{align}
Finally, the contributions of interest are (for $\eta = +$ that is irrep $27$, and for $\eta = -1$ that is irrep $0$)
\begin{align}
\hat{\Omega}_{gg}^{27/0}
    &=\frac{1}{4}[I+X] + \eta W - \frac{N_c -\eta \times 2}{2N_c} D - \frac{N_c -\eta \times 1}{2 N_c} S \notag \\
    &= 
    \frac{1}{4}\left[(-  N_c^2 \hcalf^{(6)}_{gg} + \hcalf^{(3)}_{gg}) + (-\hcalf^{(5)}_{gg})\right] 
    + \frac{\eta}{2} \left[  ( -N_c \hcalf^{(7)}_{gg}) -\frac{1}{N_c^2} (-N_c \hcalf^{(3)}_{gg})\right] \notag \\
    &- \frac{N_c -\eta \times 2}{2N_c} \frac{N_c}{N_c^2-4} \left[(-N_c \hcalf^{(1+2)}_{gg}) +\frac{2}{N_c} \hcalf^{(3+4)}_{gg}\right] \notag \\
    &- \frac{N_c -\eta \times 1}{2 N_c} \frac{1}{N_c^2-1} \left( - \hcalf^{(4)}_{gg}\right) \label{eq:gg_27_and_0}
\end{align}
For $N_c=2$, we already know that the symmetric adjoint vanishes; we could focus on the first and last lines only.
For $N_c =2,3$ the total combination with $\eta=-1$ vanishes, which gives a relation between the seven TMD operators, see \req{eq:sum_rule_gg_su3} for $SU(3)$. 

This completes the full decomposition into irreps at $+\infty$ and their relation to the seven gluon-gluon TMD operators. In the following section, we evaluate the expectation value of those operators in the Gaussian approximation.

\section{Gaussian approximation for the irreps}
\label{Sec:Gaussian_approx}

In order the evaluate the expectation value of an operator $\hcalo$ made of several WLs, we rely on the following construction. The fields are sourced by $\rho$ which we assume to have local correlation in both longitudinal and transverse variables:
\begin{equation}\label{eq:rhorhoCorrelation}
    \left\langle \rho^a(t,\xvec) \rho^b(t',\yvec) \right\rangle_\rho \propto \delta^{ab} \delta(t-t') \delta^2(\xvec-\yvec).
\end{equation}
The field $\alpha$ is obtained from the sources $\rho$ by the use of a Green function \calg, and we have
\begin{equation}\label{eq:AlphaAlphaCorrelation}
    \left\langle \alpha^a(t,\xvec) \alpha^b(t',\yvec)\right\rangle \propto \delta^{ab} \delta(t-t') \overline{\calv}(|\xvec-\yvec|).
\end{equation}
The interaction potential, denoted by $\overline{\calv}$, results from the convolution of two Green functions:
\begin{equation}
    \overline{\calv}(|\xvec-\yvec|) \propto \int\dd\wvec\ \calg(\xvec-\wvec) \calg(\yvec-\wvec),
\end{equation}
where we assume \underline{translation invariance}.
In the following, we denote the angle bracket as the average over the field realizations:
\begin{equation}
    \langle \hcalo \rangle = \int \cald\alpha\ \calo[\alpha]\ \calw[\alpha],
\end{equation}
while the average over the source realizations will be specified by introducing the subscript on the angle brackets $\langle \cdots \rangle_\rho$.
Under those approximations, it becomes simple to evaluate the expectation value of an arbitrary even number of fields using Wick contractions. For example, in the case of four insertions, we write:
\begin{equation}
    \langle \alpha_1\alpha_2\alpha_3\alpha_4 \rangle
    = \langle \alpha_1 \alpha_2\rangle \langle \alpha_3\alpha_4\rangle
    + \langle \alpha_1 \alpha_3\rangle \langle \alpha_2\alpha_4\rangle
    + \langle \alpha_1 \alpha_4\rangle \langle \alpha_2\alpha_3\rangle,
\end{equation}
where each expectation value $\langle \alpha\alpha\rangle$ is given by \req{eq:AlphaAlphaCorrelation}

In the following sections, we will first introduce the Gaussian approximation for the expectation value of the dipole. Building on this foundation, we then compute the Gaussian approximation for the $\hat{\Omega}$ operator.

\subsection{Introduction: the expectation value of the dipole}
\label{Sec:intro_ev_dipole}

In order to illustrate the basics of the Gaussian approximation, let us apply it to the dipole operator.
Introduce the length $L$, probed by the dipole, to be the support of the background field $\alpha$. 
We write the dipole as
\begin{equation}
    \hat{S}_{\xvec\yvec} = \tr\, P \exp \left\{ig\int_{-L/2}^{L/2} dt_x \alpha(t_x,\xvec) \right\} 
    P \exp \left\{-ig\int_{-L/2}^{L/2} dt_x \alpha(t_x,\xvec) \right\}.
\end{equation}
Let us discretize the path with $\ell = L / N$. The path exponential reads:
\begin{equation}
    V_\xvec \sim e^{ig\,\alpha_{N\ell-L/2}\left(\xvec\right)} e^{ig\,\alpha_{(N-1)\ell-L/2}\left(\xvec\right)} \cdots
    e^{ig\,\alpha_{2\ell-L/2}\left(\xvec\right)}
    e^{ig\,\alpha_{\ell-L/2}\left(,\xvec\right)},
\end{equation}
which is explicitly a $SU(\Nc)$ matrix. The quark-antiquark system being an overall color singlet at every step of the time evolution (interactions are mediated by Glauber gluons, which are instantaneous), it is possible to project out the singlet component using
\begin{equation}
    \underbrace{\delta^{j}_{\ i}}_{\text{quark}} \underbrace{\delta^{\ m}_{n}}_{\text{antiquark}}
    \underset{\text{singlet}}{\longrightarrow}
    \frac{1}{\Nc} \delta^j_n\delta^m_n,
\end{equation}
between the discretized times.
Thus, the dipole operator to be evaluated is
\begin{align}
     \hat{S}_{\xvec\yvec} &= 
    \frac{1}{\Nc} \tr \left[e^{ig\, \alpha_{N\ell-L/2}(\xvec)} e^{-ig\,\alpha_{N\ell-L/2}(\yvec)} \right] \notag \\
    &\times
    \frac{1}{\Nc} \tr \left[e^{ig\, \alpha_{(N-1)\ell-L/2}(\xvec)} e^{-ig\,\alpha_{(N-1)\ell-L/2}(\yvec)} \right] \notag \\
    &\cdots \notag\\
    &\times 
    \frac{1}{\Nc} \tr \left[e^{ig\, \alpha_{\ell-L/2}(\xvec)} e^{-ig\,\alpha_{\ell-L/2}(\yvec)} \right].
\end{align}
Using the discretized version of the two-point expectation
\begin{equation}
    g^2 \langle \alpha_{t_i}(\uvec) \alpha_{t_j}(\wvec) \rangle = \ell\, \delta_{t_i,t_j} \overline{\calv}^{(t_i)}(\uvec-\wvec),
\end{equation}
and Wick contractions,
\begin{align}
     \langle\hat{S}_{\xvec\yvec}\rangle &= 
    \frac{1}{\Nc} \langle\tr \left[e^{ig\, \alpha_{N\ell-L/2}(\xvec)} e^{-ig\,\alpha_{N\ell-L/2}(\yvec)} \right] \rangle \notag \\
    &\times
    \frac{1}{\Nc}  \langle\tr \left[e^{ig\, \alpha_{(N-1)\ell-L/2}(\xvec)} e^{-ig\,\alpha_{(N-1)\ell-L/2}(\yvec)} \right] \rangle \notag \\
    &\cdots \notag\\
    &\times 
    \frac{1}{\Nc}  \langle\tr \left[e^{ig\, \alpha_{\ell-L/2}(\xvec)} e^{-ig\,\alpha_{\ell-L/2}(\yvec)} \right]\rangle.
\end{align}
We have
\begin{align}
    \langle \hat{S}_{\xvec\yvec}\rangle &= 
    \prod_{k=1}^{N} \exp\left\{-C_F\ell \left[ \overline{\calv}^{(k)}(0) - \overline{\calv}^{(k)}(\xvec-\yvec) \right]\right\} 
    = \prod_{k=1}^{N} \exp\left\{-C_F\ell\ \overline{\Gamma}^{(k)}(\xvec-\yvec) \right\},
\end{align}
where the label $(k)$ is a shorthand for the time $ t = k\, \ell -L/2$.
As we assumed translation invariance (in the longitudinal direction) of the interaction potential, $\overline{\Gamma}^{(k)} \rightarrow \overline{\Gamma}$, the average of this operator yields
\begin{equation}\label{eq:DipoleGaussianExpectation}
    S(r=|\xvec-\yvec|) = \langle \hat{S}_{\xvec\yvec} \rangle = e^{-C_F L \overline{\Gamma}_r} = e^{-C_F \Gamma_r}.
\end{equation}
The function $\Gamma_r$ is related to the interaction potential $\mathcal{V}$ by
\begin{equation}
    \Gamma_r = \mathcal{V}(0) - \mathcal{V}(r).
\end{equation}
Relevant to us is the result of the MV model \cite{McLerran:1993ni,McLerran:1993ka,McLerran:1994vd} approximations, where the logarithm of the dipole reads
\begin{equation}
    \Gamma_r \propto r^2 \ln (\tfrac{1}{\lambda r} + e),
\end{equation}
or the GBW model \cite{Golec-Biernat:1998zce,Golec-Biernat:1999qor} where we have
\begin{equation}
    \Gamma_r \propto r^2.
\end{equation}
Both models fulfill color transparency:
\begin{equation}
    \lim_{r\rightarrow 0} \Gamma_r = 0.
\end{equation}

\noindent
Note that we explicitly stripped the color factor $C_F$ in \req{eq:DipoleGaussianExpectation} when defining $\Gamma_r$, this will make the Casimir scaling properties explicit in the following sections.

\subsection{Derivation for \texorpdfstring{$\Omega$}{Omega}s}
\label{Sec:derivation_omega}
\paragraph{Quark - Gluon:}
Let us factor out the common part of the three quark-gluon TMD operators, and project along the final state cut at $+\infty$, according to the decomposition
\begin{equation}
   (\overline{3}\otimes 3 ) \otimes \overline{3} \quad \supset \quad  8 \otimes \overline{3} = \overline{3} \oplus 6 \oplus \overline{15}.
\end{equation}
In section~\ref{sec_birdtrack}, we related three $\hcalf_{qg}$ to the following birdtrack
\begin{align}
\hat{\Omega}_{qg} = 
{2} \times 
\vcenter{\hbox{
\begin{tikzpicture}[scale=0.3]
    \draw[thick] (-5,1.5) -- ++ (10,0);\draw[thick,->] (-5,1.5) -- ++ (3,0);
    \draw[thick] (-5,2) -- ++ (10,0);\draw[thick,-<] (-5,2) -- ++ (3,0);
    \draw[fill=Blue!50!white] (1,1.5) circle (0.2);
    \draw[thick] (-5,1) -- ++ (10,0);\draw[thick,-<] (-5,1) -- ++ (4,0);
    \draw[thick] (-5,-1) -- ++ (10,0);\draw[thick,->] (-5,-1) -- ++ (6,0);
    \draw[thick] (-5,-1.5) -- ++ (10,0);\draw[thick,-<] (-5,-1.5) -- ++ (7,0);
    \draw[thick] (-5,-2) -- ++ (10,0);\draw[thick,->] (-5,-2) -- ++ (7,0);
    \draw[fill=Blue!50!white] (-1,-2) circle (0.2);
    \draw[thick] (-5,2) to[out=180,in=180] (-5,1.5);
    \draw[thick] (-5,-2) to[out=180,in=180] (-5,-1.5);
    \draw[thick] (-5,1) to[out=180,in=180] (-5,-1);
    \draw[thick] (5.5,2) to[out=0,in=90,looseness=1.5] (6,1.75) to [out=-90,in=0, looseness=1.5] (5.5,1.5);
    \draw[thick] (5.5,-2) to[out=0,in=-90,looseness=1.5] (6,-1.75) to [out=90,in=0, looseness=1.5] (5.5,-1.5);
    \draw[gluon] (6,1.75) to[out=0,in=0,looseness=2] (6,-1.75);
    \draw[thick] (5.5,1) -- (6,1) to[out=0,in=0,looseness=2] (6,-1) -- (5.5,-1);
    \draw[fill=white] (7.5,0) ellipse (1.2 and .5) node {\scriptsize $\omega$};
\end{tikzpicture}}},
\end{align}
which we now evaluate in the Gaussian approximation.
The symbol $\omega$ denotes the irreps of $8 \otimes \overline{3}$, while the blob is the corresponding projection. The factor $2$ follows from the normalization $\tr(t^at^b) = \thalf \delta^{ab}$.
We first note that at $-\infty$ the WLs have been projected in a peculiar way, which corresponds to the irrep $\overline{3}$ obtained from
\begin{equation}
    (\overline{3}\otimes 3 ) \otimes \overline{3} \quad \supset \quad  1 \otimes \overline{3} = \overline{3} .
\end{equation}
We first introduce the derivative of a WL as 
\begin{equation}
    \partial_\xvec^j V_\xvec = \int dt\ V_\xvec[\infty,t] \left( ig\partial^j_\xvec \alpha(t,\xvec) \right) V_\xvec[t,-\infty].
\end{equation}
Then we simplify the operator $\calb^j_\xvec$:
\begin{align}
    \calb^j_\xvec \equiv \left(\partial_\xvec^j V_\xvec \right) V_\xvec^\dagger 
    &= \int dt\ V_\xvec[\infty,t] \left( ig\partial^j_\xvec \alpha(t,\xvec) \right) V_\xvec[t,-\infty] V_\xvec^\dagger \notag \\
    &= \int dt\ V_\xvec[\infty,t] \left( ig \partial^j_\xvec \alpha(t,\xvec) \right) V^\dagger_\xvec[t,+\infty] 
\end{align}
From the last line, we see that there will be no interaction from $-\infty$ to $t$ associated with this operator. The first interaction involving an operator $\calb^j$ is $\partial^j \alpha$, which effectively rotates the pointlike system to an adjoint representation that will then interact from $t$ to $+\infty$ as such. As such, we can write
\begin{align}
\Omega_{qg} \equiv  
\langle \hat{\Omega}_{qg} \rangle^{\text{(Gaussian)}}  = 
{2} \times 
\vcenter{\hbox{
\begin{tikzpicture}[scale=0.3]
    \draw[Green,fill=Green!30!white] (1.5,2.5) rectangle ++(1,-5);
    \node[below] at (2,-2.5) {\scriptsize \textcolor{Green}{$t_x$}};
    \draw[Green,fill=Green!30!white] (-2.5,2.5) rectangle ++(1,-5);
    \node[below] at (-2,-2.5) {\scriptsize \textcolor{Green}{$t_y$}};
    \draw[Orange,fill=Orange!30!white] (-5,2.5) rectangle ++(2,-5);
    \draw[Orange,fill=Orange!30!white] (-1,2.5) rectangle ++(2,-5);
    \draw[Orange,fill=Orange!30!white] (3,2.5) rectangle ++(2,-5);
    \draw[thick] (5,1.5) -- (2,1.5) to[out=180,in=180,looseness=2] (2,2) -- (5,2);
    \draw[fill=Blue!50!white] (2,1.5) circle (0.2);
    \draw[thick] (-5,1) -- (5,1);
    \draw[thick] (5,-1.5) -- (-2,-1.5) to[out=180,in=180,looseness=2] (-2,-2) -- (5,-2);
    \draw[thick] (-5,-1) -- (5,-1);    
    \draw[fill=Blue!50!white] (-2,-2) circle (0.2);
    \draw[thick] (-5,1) to[out=180,in=180,looseness=2] (-5,-1);
    \draw[thick] (5.5,2) to[out=0,in=90,looseness=1.5] (6,1.75) to [out=-90,in=0, looseness=1.5] (5.5,1.5);
    \draw[thick] (5.5,-2) to[out=0,in=-90,looseness=1.5] (6,-1.75) to [out=90,in=0, looseness=1.5] (5.5,-1.5);
    \draw[gluon] (6,1.75) to[out=0,in=0,looseness=2] (6,-1.75);
    \draw[thick] (5.5,1) -- (6,1) to[out=0,in=0,looseness=2] (6,-1) -- (5.5,-1);
    \draw[fill=white] (7.5,0) ellipse (1.2 and .5) node {\scriptsize $\omega$};
\end{tikzpicture}}}.
\end{align}
We denote in orange shaded rectangles the domain $\cald_{eik}$, with $t_\gtrless = \substack{\text{max} \\ \text{min}}(t_x,t_y)$, that contains eikonal interaction between the WLs:
\begin{equation}
    \cald_{eik} = 
    (-\infty, t_< )
    \ \cup\ 
    ( t_<, t_> )
    \ \cup\ 
    (t_>, +\infty)
\end{equation}
One has to distinguish two cases: (i) the irrep $\omega$ at $+\infty$ is the same as the irrep $\overline{3}$ at $-\infty$, and (ii) the irrep $\omega$ is distinct from the irrep at $-\infty$.
In case (i), there is no need for an irrep-rotation to occur, in particular within the domain $(t_<,t_>)$, we can form a singlet out of the system at $\xvec$ and the system at $\yvec$. This implies that in the integrals over $t_x$ and $t_y$, the full range contributes, and we distinguish two types of correlations
\begin{equation}
    \langle \alpha (\partial^i \alpha) \rangle \qquad \text{and} \qquad
    \langle (\partial^i \alpha) (\partial^i \alpha) \rangle,
\end{equation}
in addition to the usual eikonal correlations $\langle \alpha \alpha \rangle$.
In the case (ii), the region $(t_<,t_>)$ cannot occur since there is no way to construct a singlet out of the system at $\xvec$ with the system at $\yvec$. One has to rotate the whole system, which enforces $t_x = t_y$. 

\vspace{.5cm} \noindent
\textit{Case (a): $t_x \neq t_y$.}
It involves the correlations:
\begin{equation}
    \left\langle (ig \alpha^b)_\yvec (ig \partial^j\alpha^a)_\xvec \right\rangle_{t_x} 
    \left\langle (ig \partial^j \alpha^d)_\yvec (ig \alpha^c)_\xvec \right\rangle_{t_y}
\end{equation}
Into the definition of $\Omega_{qg}$, we have (recall that there are two orderings):
\begin{align}
\Omega_{qg}^{(a)} \Big|_{t_x > t_y} = 
{2} \times 
\vcenter{\hbox{
\begin{tikzpicture}[scale=0.3]
    \draw[Green,fill=Green!30!white] (1.5,2.5) rectangle ++(1,-5);
    \node[below] at (2,-2.5) {\scriptsize \textcolor{Green}{$t_x$}};
    \draw[Green,fill=Green!30!white] (-2.5,2.5) rectangle ++(1,-5);
    \node[below] at (-2,-2.5) {\scriptsize \textcolor{Green}{$t_y$}};
    \draw[Orange,fill=Orange!30!white] (-5,2.5) rectangle ++(2,-5);
    \draw[Orange,fill=Orange!30!white] (-1,2.5) rectangle ++(2,-5);
    \draw[Orange,fill=Orange!30!white] (3,2.5) rectangle ++(2,-5);
    \draw[thick] (5,1.5) -- (2,1.5) to[out=180,in=180,looseness=2] (2,2) -- (5,2);
    \draw[fill=Blue!50!white] (2,1.5) circle (0.2);
    \draw[thick] (-5,1) -- (5,1);
    \draw[thick] (5,-1.5) -- (-2,-1.5) to[out=180,in=180,looseness=2] (-2,-2) -- (5,-2);
    \draw[thick] (-5,-1) -- (5,-1);    
    \draw[fill=Blue!50!white] (-2,-2) circle (0.2);
    \draw[thick] (-5,1) to[out=180,in=180,looseness=2] (-5,-1);
    \draw[thick] (5.5,2) to[out=0,in=90,looseness=1.5] (6,1.75) to [out=-90,in=0, looseness=1.5] (5.5,1.5);
    \draw[thick] (5.5,-2) to[out=0,in=-90,looseness=1.5] (6,-1.75) to [out=90,in=0, looseness=1.5] (5.5,-1.5);
    \draw[gluon] (6,1.75) to[out=0,in=0,looseness=2] (6,-1.75);
    \draw[thick] (5.5,1) -- (6,1) to[out=0,in=0,looseness=2] (6,-1) -- (5.5,-1);
    \draw[fill=white] (7.5,0) ellipse (1.2 and .5) node {\scriptsize $\omega$};
\end{tikzpicture}}},
\end{align}
The requirement that we can form a singlet between the system at $\xvec$ and the system at $\yvec$ implies that the irrep $\omega$ is the dual-fundamental $\overline{3}$. Thus we can simply write
\begin{align}
\Omega_{qg}^{(a)} \Big|_{t_x > t_y} &= 
\frac{2}{C_F} \times 
\vcenter{\hbox{
\begin{tikzpicture}[scale=0.3]
    \draw[Orange,fill=Orange!30!white] (-1,2.5) rectangle ++(2,-5);
    \draw[thick] (-1,1) -- (1,1) to[out=0,in=0] (1,-1) -- (-1,-1) to[out=180,in=180] cycle; 
    \node at (-3,0) {$\frac{1}{N_c}$};
    \node at (3,0) {$\frac{1}{N_c}$};
    \node at (-10,0) {$\frac{1}{N_c}$};
    \node at (10,0) {$\frac{1}{N_c}$};
    \draw[Green,fill=Green!30!white] (-7.5,2.5) rectangle ++(1,-5);
    \draw[thick] (-8,1) -- (-5,1) to[out=0,in=0] (-5,-1) -- (-8,-1) to[out=180,in=180] cycle;
    \draw[thick] (-7,-2) -- ++(.5,0) to[out=0,in=-90] ++(.5,.25) to[out=90,in=0] ++(-.5,.25) -- ++(-1,0) to[out=180,in=90] ++(-.5,-.25) to[out=-90,in=180] ++(.5,-.25) --cycle;
    \draw[fill=Blue!50!white] (-7,-2) circle (0.2);
    \draw[lgluon] (-6,-1.75) to[out=0,in=-90] (-5,-1);
    \draw[Orange,fill=Orange!30!white] (-14,2.5) rectangle ++(2,-5);
    \draw[thick] (-14,1) -- (-12,1) to[out=0,in=0] (-12,-1) -- (-14,-1) to[out=180,in=180] cycle; 
    \draw[Green,fill=Green!30!white] (5.5,2.5) rectangle ++(1,-5);
    \draw[thick] (5,1) -- (8,1) to[out=0,in=0] (8,-1) -- (5,-1) to[out=180,in=180] cycle;
    \draw[thick] (6,1.5) -- ++(.5,0) to[out=0,in=-90] ++(.5,.25) to[out=90,in=0] ++(-.5,.25) -- ++(-1,0) to[out=180,in=90] ++(-.5,-.25) to[out=-90,in=180] ++(.5,-.25) --cycle;
    \draw[fill=Blue!50!white] (6,1.5) circle (0.2);
    \draw[gluon] (7,1.75) to[out=0,in=90] (8,1);
    \draw[Orange,fill=Orange!30!white] (12,2.5) rectangle ++(2,-5);
    \draw[thick] (12,1) -- (14,1) to[out=0,in=0] (14,-1) -- (12,-1) to[out=180,in=180] cycle; 
\end{tikzpicture}}} \notag \\[2ex]
&=  \frac{2}{N_c\, C_F} e^{-C_F \Gamma_r} \times \frac{L^2}{2} \left(\frac{1}{2}C_F N_c \nabla^j\overline{\Gamma}_r\right) \left(\frac{1}{2}C_F N_c \nabla^j\overline{\Gamma}_r\right) \\
&=  \frac{1}{4} N_c C_F (\nabla \Gamma_r)^2 e^{-C_F \Gamma_r}
\end{align}
Introducing the other ordering simply doubles this result, and we find
\begin{equation}\label{eq:Omega_qg_A}
    \Omega_{qg}^{(a)} = \frac{N_c}{2} C_F (\nabla \Gamma_r)^2\, e^{-C_F \Gamma_r}\ \delta_{\omega = \overline{3}}
\end{equation}
Note the relation to a contribution from the Laplacian acting on the fundamental dipole.
The second term of the Laplacian will be recovered in the next paragraph.

\vspace{.5cm} \noindent
\textit{Case (b): $t_x = t_y$.} It involves the correlation:
\begin{equation}
    \left\langle(ig \partial^j \alpha)_\xvec (ig\partial^j \alpha)_\yvec \right\rangle_{t_x=t_y}
\end{equation}
Into the definition of $\Omega_{qg}$, it reads\footnote{We use used $\text{sinhc}(x) \equiv \frac{sinh(x)}{x}$, and $\text{sinh}(x) \equiv \thalf \left( e^x - e^{-x}\right)$. In particular, we have $\text{sinhc}(0) = 1$.}
\begin{align}
\Omega_{qg}^{(b)} &= 
{2} \times 
\vcenter{\hbox{
\begin{tikzpicture}[scale=0.3]
    \draw[Green,fill=Green!30!white] (-.5,2.5) rectangle ++(1,-5);
    \node[below] at (0,-2.5) {\scriptsize \textcolor{Green}{$t_x = t_y$}};
    \draw[Orange,fill=Orange!30!white] (-5,2.5) rectangle ++(2,-5);
    \draw[Orange,fill=Orange!30!white] (3,2.5) rectangle ++(2,-5);
    \draw[thick] (5,1.5) -- (0,1.5) to[out=180,in=180,looseness=2] (0,2) -- (5,2);
    \draw[fill=Blue!50!white] (0,1.5) circle (0.2);
    \draw[thick] (-5,1) -- (5,1);
    \draw[thick] (5,-1.5) -- (0,-1.5) to[out=180,in=180,looseness=2] (0,-2) -- (5,-2);
    \draw[thick] (-5,-1) -- (5,-1);    
    \draw[fill=Blue!50!white] (0,-2) circle (0.2);
    \draw[thick] (-5,1) to[out=180,in=180,looseness=2] (-5,-1);
    \draw[thick] (5.5,2) to[out=0,in=90,looseness=1.5] (6,1.75) to [out=-90,in=0, looseness=1.5] (5.5,1.5);
    \draw[thick] (5.5,-2) to[out=0,in=-90,looseness=1.5] (6,-1.75) to [out=90,in=0, looseness=1.5] (5.5,-1.5);
    \draw[gluon] (6,1.75) to[out=0,in=0,looseness=2] (6,-1.75);
    \draw[thick] (5.5,1) -- (6,1) to[out=0,in=0,looseness=2] (6,-1) -- (5.5,-1);
    \draw[fill=white] (7.5,0) ellipse (1.2 and .5) node {\scriptsize $\omega$};
\end{tikzpicture}}} \notag \\
&= 2 \times 
\vcenter{\hbox{
\begin{tikzpicture}[scale=0.3]
    \draw[Green,fill=Green!30!white] (-.5,2.5) rectangle ++(1,-5);
    \node[below] at (0,-2.5) {\scriptsize \textcolor{Green}{$t_x = t_y$}};
    \draw[Orange,fill=Orange!30!white] (9,2.5) rectangle ++(2,-5);
    \draw[Orange,fill=Orange!30!white] (-8,2.5) rectangle ++(2,-5);
    \draw[thick] (.5,1.5) -- (0,1.5) to[out=180,in=180,looseness=2] (0,2) -- (.5,2) to[out=0,in=90,looseness=1.5] (1,1.75) to [out=-90,in=0, looseness=1.5] cycle;
    \draw[fill=Blue!50!white] (0,1.5) circle (0.2);
    \draw[thick] (.5,-1.5) -- (0,-1.5) to[out=180,in=180,looseness=2] (0,-2) -- (.5,-2) to[out=0,in=-90,looseness=1.5] (1,-1.75) to [out=90,in=0, looseness=1.5] cycle;  
    \draw[fill=Blue!50!white] (0,-2) circle (0.2);
    \draw[gluon] (1,1.75) to[out=0,in=0,looseness=2] (1,-1.75);
    \draw[thick] (-1,1) -- (1,1) to[out=0,in=0,looseness=2] (1,-1) -- (-1,-1) to[out=180,in=180,looseness=2] cycle;
    \draw[fill=white] (2.5,0) ellipse (1.2 and .5) node {\scriptsize $\omega$};
    \node at (5,0) {$\frac{1}{K_\omega}$};
    \draw[gluon] (9,-1.75) to[out=180,in=180,looseness=2] (9,1.75) -- (11,1.75) to[out=0,in=0,looseness=2] (11,-1.75) -- cycle;
    \draw[thick] (9,1) to[out=180,in=180,looseness=2] (9,-1) -- (11,-1) to[out=0,in=0,looseness=2] (11,1) -- cycle;
    \draw[fill=white] (7.5,0) ellipse (1.2 and .5) node {\scriptsize $\omega$};
    \draw[fill=white] (12.5,0) ellipse (1.2 and .5) node {\scriptsize $\omega$};
    \node at (-3.5,0) {$\frac{1}{N_c}$};
    \draw[thick] (-6,1) -- (-8,1) to[out=180,in=180,looseness=2] (-8,-1) -- (-6,-1) to[out=0,in=0,looseness=2] cycle;
\end{tikzpicture}}} \\
&= -\frac{K_\omega}{2}(\nabla^2 {\Gamma}_r )\int\limits_{-L/2}^{L/2} dt\ e^{-C_\omega (\frac{L}{2} - t) \overline{\Gamma}_r}  e^{-C_F(t+\frac{L}{2})} \\
&= -\frac{K_\omega}{2}(\nabla^2 {\Gamma}_r )\, e^{-\frac{C_F + C_\omega}{2}\Gamma_r}\ \text{sinhc} \left[ \frac{1}{2}(C_\omega - C_F) \Gamma_r\right] \label{eq:Omega_qg_B}
\end{align}
with 
\begin{subequations}\label{CK_coeff_values_qg}
\begin{align}
&C_{\overline{3}} = C_F = \frac{N_c^2-1}{2N_c}, \qquad &K_{\overline{3}} = K_F = N_c, \\
&C_{\overline{15}/6} = \Nc \pm 1 + C_F,\quad &K_{\overline{15}/6} = \thalf(\Nc \pm2)\Nc(\Nc\mp 1)
\end{align}
\end{subequations}
such that 
\begin{equation}\label{eq:Omega_qg}
\Omega_{qg}^\omega \equiv \Omega_{qg}^{(a)} + \Omega_{qg}^{(b)}
\end{equation}
\noindent
The following form is convenient to emphasize the relation to the Laplacian of the fundamental dipole when $\omega = \overline{3}$:
\begin{subequations}
\begin{align}
    \frac{2C_F}{N_c} \times \Omega_{qg}^{(a)} &= (-C_F\nabla_r \Gamma_r)^2\, e^{-C_F \Gamma_r}\ \delta_{\omega = \overline{3}},\\
    \forall \omega \in \overline{3}\otimes8,\quad \frac{2C_F}{K_\omega} \times \Omega_{qg}^{(b)} &=(-C_F\nabla^2_r {\Gamma}_r )\, e^{-\frac{C_F + C_\omega}{2}\Gamma_r}\ \text{sinhc} \left[ \frac{1}{2}(C_\omega - C_F) \Gamma_r\right].
\end{align}
\end{subequations}

\paragraph{Gluon - Gluon:}
In section~\ref{sec_birdtrack}, we related six $\hcalf_{gg}$ to the operator:
\begin{equation}
\hat{\Omega}_{gg} = {4} \times
\vcenter{\hbox{
\begin{tikzpicture}[scale=0.3]
    \draw[thick] (-5,2) -- ++ (10,0);\draw[thick,->] (-5,2) -- ++ (3,0);
    \draw[thick] (-5,2.5) -- ++ (10,0);\draw[thick,-<] (-5,2.5) -- ++ (3,0);
    \draw[fill=Blue!50!white] (1,2) circle (0.2);
    \draw[thick] (-5,1.5) -- ++ (10,0);\draw[thick,-<] (-5,1.5) -- ++ (4,0);
    \draw[thick] (-5,1) -- ++(10,0);\draw[thick,->] (-5,1) -- ++(4,0);
    \draw[thick] (-5,-1) -- ++(10,0);\draw[thick,-<] (-5,-1) -- ++(6,0);
    \draw[thick] (-5,-1.5) -- ++ (10,0);\draw[thick,->] (-5,-1.5) -- ++ (6,0);
    \draw[thick] (-5,-2) -- ++ (10,0);\draw[thick,-<] (-5,-2) -- ++ (7,0);
    \draw[thick] (-5,-2.5) -- ++ (10,0);\draw[thick,->] (-5,-2.5) -- ++ (7,0);
    \draw[fill=Blue!50!white] (-1,-2.5) circle (0.2);
    \draw[thick] (-5,2) to[out=180,in=180] (-5,2.5);
    \draw[thick] (-5,-2) to[out=180,in=180] (-5,-2.5);
    \draw[thick] (-5,1) to[out=180,in=180] (-5,-1);
    \draw[thick] (-5,1.5) to[out=180,in=180] (-5,-1.5);
    \draw[thick] (5.5,2.5) to[out=0,in=90] ++(.5,-.25) to[out=-90,in=0] ++(-.5,-.25);
    \draw[thick] (5.5,1.5) to[out=0,in=90] ++(.5,-.25) to[out=-90,in=0] ++(-.5,-.25);
    \draw[thick] (5.5,-1) to[out=0,in=90] ++(.5,-.25) to[out=-90,in=0] ++(-.5,-.25);
    \draw[thick] (5.5,-2) to[out=0,in=90] ++(.5,-.25) to[out=-90,in=0] ++(-.5,-.25);
    \draw[gluon] (6,2.25) to[out=0,in=0,looseness=2] (6,-2.25);
    \draw[gluon] (6,1.25) to[out=0,in=0,looseness=2] (6,-1.25);
    \draw[fill=white] (8,0) ellipse (1.2 and .5) node {\scriptsize $\omega$};
\end{tikzpicture}}}
\end{equation}
We now evaluate it using the Gaussian approximation for $\omega \in 8 \otimes 8$. Due to the generators being traceless, we can replace the trace at $-\infty$  between $\xvec$ and $\yvec$ over the quark anti-quark indices to the corresponding adjoint irrep using the Fierz identity.
\begin{align}
\Omega_{gg} \equiv \langle \hat{\Omega}_{gg}\rangle^{\text{(Gaussian)}}= 
{2} \times 
\vcenter{\hbox{
\begin{tikzpicture}[scale=0.3]
    \draw[Green,fill=Green!30!white] (1.5,2.5) rectangle ++(1,-5);
    \node[below] at (2,-2.5) {\scriptsize \textcolor{Green}{$t_x$}};
    \draw[Green,fill=Green!30!white] (-2.5,2.5) rectangle ++(1,-5);
    \node[below] at (-2,-2.5) {\scriptsize \textcolor{Green}{$t_y$}};
    \draw[Orange,fill=Orange!30!white] (-5,2.5) rectangle ++(2,-5);
    \draw[Orange,fill=Orange!30!white] (-1,2.5) rectangle ++(2,-5);
    \draw[Orange,fill=Orange!30!white] (3,2.5) rectangle ++(2,-5);
    \draw[thick] (5,1.5) -- (2,1.5) to[out=180,in=180,looseness=2] (2,2) -- (5,2);
    \draw[fill=Blue!50!white] (2,1.5) circle (0.2);
    \draw[gluon] (5,1) -- (-5,1) to[out=180,in=180,looseness=2] (-5,-1) -- (5,-1);
    \draw[thick] (5,-1.5) -- (-2,-1.5) to[out=180,in=180,looseness=2] (-2,-2) -- (5,-2);    
    \draw[fill=Blue!50!white] (-2,-2) circle (0.2);
    \draw[thick] (5.5,2) to[out=0,in=90,looseness=1.5] (6,1.75) to [out=-90,in=0, looseness=1.5] (5.5,1.5);
    \draw[thick] (5.5,-2) to[out=0,in=-90,looseness=1.5] (6,-1.75) to [out=90,in=0, looseness=1.5] (5.5,-1.5);
    \draw[gluon] (6,1.75) to[out=0,in=0,looseness=2] (6,-1.75);
    \draw[gluon] (5.5,1) -- (6,1) to[out=0,in=0,looseness=2] (6,-1) -- (5.5,-1);
    \draw[fill=white] (7.5,0) ellipse (1.2 and .5) node {\scriptsize $\omega$};
\end{tikzpicture}}}
\end{align}
The topology being exactly the same as in the quark-gluon case, we can directly conclude the result to be:\footnote{We mention that at first inspection the contribution $(a)$ could have been non-vanishing for $\omega = 8_s$. This is not the case since at $t_x$ or $t_y$, it would involve a scalar from a birdtrack with a single insertion of the symmetric three gluon vertex, which can only vanish due to symmetry: $f^{abc} d^{abc} = 0$. }
\begin{subequations}\label{eq:Omega_gg}
\begin{align}
\Omega_{gg}^\omega &\equiv \Omega_{gg}^{(a)} + \Omega_{gg}^{(b)} \\
\label{eq:Omega_gg_A}
\Omega_{gg}^{(a)}
&= \frac{(N_c^2-1)}{2} C_A (\nabla \Gamma_r)^2 e^{-C_A \Gamma_r}\  \delta_{\omega = 8_a}\\
\label{eq:Omega_gg_B}
\Omega_{gg}^{(b)}&= -\frac{K_\omega}{2}(\nabla^2 {\Gamma}_r )\, e^{-\frac{C_A + C_\omega}{2}\Gamma_r}\ \text{sinhc} \left[ \frac{1}{2}(C_\omega - C_A) \Gamma_r\right]
\end{align}
\end{subequations}
where we have the quadratic charges and dimensions given by\footnote{Take notice that the dimension of the subspace associated to $\omega = 10+\overline{10}$ is $K_{10 + \overline{10}} = 2 \times K_{10}$, which is $20$ for $SU(3)$ even through we make use of the short-hand notation ``$\omega = 10$'' in the following sections.}:
\begin{subequations}
\begin{align}
    &C_1 =0,\quad 
    C_{10} = C_{\overline{10}} = 2N_c,\quad 
    C_{8} = C_A = N_c,\quad 
    C_{27/0} = 2(N_c \pm 1).\\
    &K_1 = 1, \quad K_{8a/8s} = K_A = \Nc^2-1, \\
    &K_{10/\overline{10}} = \frac{(\Nc^2-1)(\Nc^2-4)}{4}, \quad
    K_{27/0} = \Nc^2 \frac{(\Nc \mp 1)(\Nc\pm3)}{4}.
\end{align}
\label{CK_coeff_values}
\end{subequations}
In a form similar to the quark-gluon sector, we write
\begin{align}
\frac{2C_A}{K_A} \times \Omega_{gg}^{(a)}
&= (-C_A\nabla_r \Gamma_r)^2 e^{-C_A \Gamma_r}\  \delta_{\omega = 8_a},\\
\forall \omega \in 8\otimes 8, \quad \frac{2C_A}{K_\omega} \times \Omega_{gg}^{(b)}&= (-C_A\nabla^2_r {\Gamma}_r ) e^{-\frac{C_A + C_\omega}{2}\Gamma_r}\ \text{sinhc} \left[ \frac{1}{2}(C_\omega - C_A) \Gamma_r\right].
\end{align}
which emphasizes relation to the Laplacian of the adjoint dipole obtained for $\omega = 8_a$

\paragraph{The last adjoint.}
Finally, one can select the singlet out of the trace at $-\infty$.
This contribution is directly proportional to the expectation value of $\hcalf^{(3)}_{gg}$ and reads
\begin{align}\label{eq:last_adj}
\langle \hcalf^{(3)}_{gg} \rangle  = - \left\langle
\vcenter{\hbox{
\begin{tikzpicture}[scale=0.3]
    \draw[thick] (-5,1) -- ++ (10,0);\draw[thick,->] (-5,1) -- ++ (3,0);
    \draw[thick] (-5,1.5) -- ++ (10,0);\draw[thick,-<] (-5,1.5) -- ++ (3,0);
    \draw[fill=Blue!50!white] (1,1) circle (0.2);
    \draw[thick] (-5,-1) -- ++ (10,0);\draw[thick,-<] (-5,-1) -- ++ (7,0);
    \draw[thick] (-5,-1.5) -- ++ (10,0);\draw[thick,->] (-5,-1.5) -- ++ (7,0);
    \draw[fill=Blue!50!white] (-1,-1.5) circle (0.2);
    \draw[thick] (5,-1.5) to[out=0,in=0,looseness=1] (5,1.5);
    \draw[thick] (5,-1) to[out=0,in=0,looseness=1] (5,1);
    \draw[thick] (-5,1) to[out=180,in=180] (-5,1.5);
    \draw[thick] (-5,-1) to[out=180,in=180] (-5,-1.5);
\end{tikzpicture}}}
\right\rangle
= C_F \left( \frac{\nabla^2 \Gamma_r}{\Gamma_r}\right) (1-e^{-N_c \Gamma_r})
\end{align}
This is the well-known Weizs{\"a}cker-Williams (WW) gluon distribution. The Gaussian approximation for this TMD distribution was obtained in Ref.~\cite{Dominguez:2011wm}, and \req{eq:last_adj} agrees with their Eq.~(A15).
Due to time reversal symmetry of the setup, we can take the expression for $\langle \hcalf^{(4)}_{gg} \rangle$ in the Gaussian approximation and check that we recover $\langle \hcalf^{(3)}_{gg} \rangle$ within the same approximation:
\begin{align}
    \langle \hcalf_{gg}^{(4)} \rangle
    &= -(N_c^2-1) \langle S \rangle = -(N_c^2-1) \Omega_{gg}^{1} 
    = \langle \hcalf^{(3)}_{gg} \rangle
    \label{F4_F3_omega_rel}
\end{align}

\section{Numerical results for the irreps  \texorpdfstring{$\hat{\Omega}^\omega$}{Omega}}\label{sec:num_result_omega}

In the previous section \ref{Sec:Gaussian_approx}, we computed the Gaussian approximation for the operators $\Omega^\omega$ in both the quark-gluon and gluon-gluon channels. Those operators have an advantageous interpretation in terms of color charges, for which the $N_c$-behavior is easily tractable, as shown by \req{eq:Omega_qg_A}, \req{eq:Omega_qg_B} and \req{eq:Omega_gg}.
The quantification of $N_c$ effects within the initial condition will be key to the analysis of observables computed after small-$x$ evolution, as we can single them out from the initial condition, and look for genuine $N_c$-dependent effects of the BK and JIMWLK evolutions.

In this section, we provide numerical results for the expectation values of $\hat{\Omega}^\omega$ obtained numerically with the (massive) MV initial condition of WLs, generated for $N_c \in \{2,3,4,5\}$.
Results are obtained by generating $N_{\textrm{real}}$ independent realizations from which we calculate the mean and the variance to estimate the value of the operator's average and its error. The number of realization vary from  $N_{\textrm{real}}=128$ for $N_c=3,4,5$ to  $N_{\textrm{real}}=512$ for $N_c=2$. This choice is dictated by the observation that statistical fluctuations decrease considerably with increasing $N_c$.
Details on the generation procedure of a realization can be found in the Appendix \ref{app:MV_impl}.
Let us mention that we expect very good agreement between the numerical evaluations using the MV-model and the analytic evaluations using the Gaussian approximation. While features of the Gaussian approximation are captured by the correlation between the fields $\langle \alpha \alpha \rangle$ given in \req{eq:AlphaAlphaCorrelation}, one can recover those features from the locality of the correlation between sources $\langle \rho \rho \rangle$, given in \req{eq:rhorhoCorrelation}, used in the MV-model by simply noting that the field relates to the sources through a Green function. On the one hand, the Gaussian approximation gives analytic expressions for the correlation function, while on the other hand, the numerical simulation using the MV-model gives us realizations of Wilson lines, which we use to build any correlation functions of interest.

In order not to overload the notation, we do not introduce any additional notation for the distinct ways of obtaining the expectation values of an operator $\calo \equiv \langle \hcalo \rangle$. We assume the context to be sufficient to distinguish between: the simulations of the MV model $\langle \hcalo \rangle^{(\textrm{MV-num})}$ where data points with error bars are shown; and the analytic results obtained in the Gaussian approximation $\langle \hcalo \rangle^{(\textrm{Gaussian})}$ where curves are shown.

\subsection{Dipole fit}\label{sec:dipole_fit}

\begin{figure}\centering
    \includegraphics[width=0.7\textwidth]{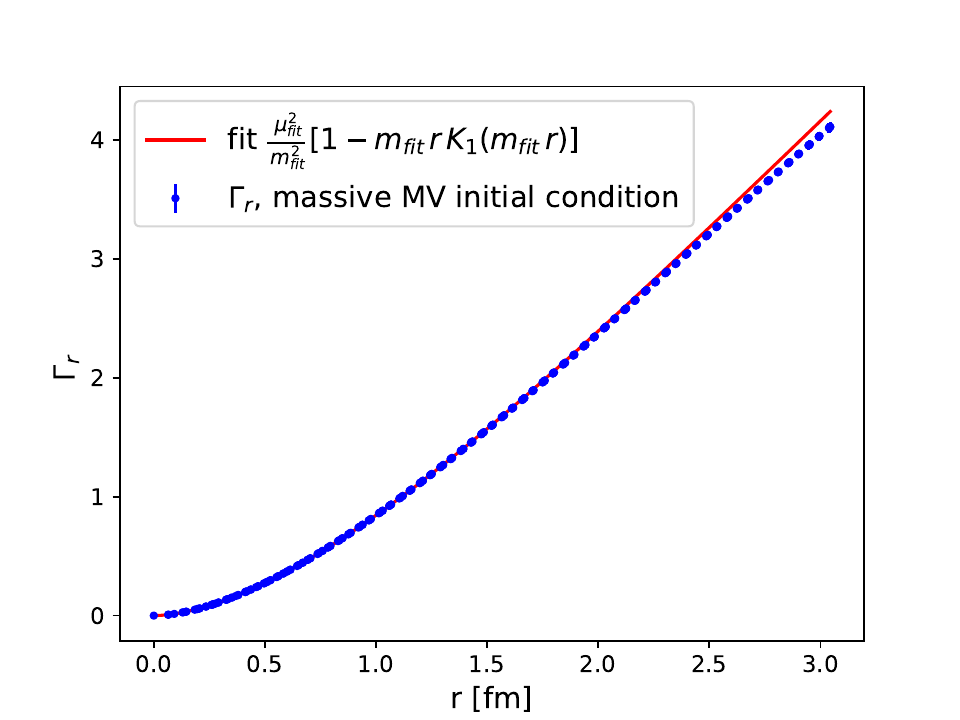}
    \caption{The logarithm of the (fundamental) dipole generated with $\mu = 0.7 \ \text{fm}^{-1},\ m=0.2 \ \text{fm}^{-1},\ L = 100 \ \text{fm}$ and $N_c = 3$ (blue points) and the fit obtained in the range $0<r \leq 3.04$~fm (red curve). 
    \label{fig:dip_fit_v2} }
\end{figure}

We related TMD operators to the formal operators $\hat{\Omega}^\omega$ in the section \ref{sec_birdtrack}. Those operators have been evaluated in the Gaussian approximation in the section \ref{Sec:derivation_omega}. However, in the Gaussian approximation, the expectation values $\langle \hat{\Omega}^\omega \rangle$ are functionals of the (logarithm of the) dipole, a consequence of the assumption that the interaction potential $\calv(\xvec-\yvec)$ is (light-cone)-time independent. Thus, we proceed to fit the dipole in this section. 

To mitigate the impact of the finite volume and ensure the requisite number of points for evaluating the derivative within the range of a few Fermi, we employ the Green function that solves the massive Poisson equation: 
\begin{equation}
    \left(\nabla^2 - m^2 \right) \mathcal{G}(\yvec, \xvec) = \delta^{2}(\yvec-\xvec)
\end{equation}
with the constrain that the mass fulfills $\frac{1}{mL} \ll 1$, where $L$ is the length in fermi of our simulation geometry. The result for the continuum\&massless MV model is recovered in the limit $L\rightarrow \infty$, $m \rightarrow 0$.
Employing the Green function for the Poisson equation, the fundamental dipole $S_r$ can be expressed as follows:
\begin{equation}\label{eq:fit_func_form}
    S_r = e^{-(4/3)\Gamma^{\rm{fit}}_r}, \qquad \text{with }
    \Gamma^{\rm{fit}}_r = \frac{\mu_{\rm{fit}}^2}{m^2_{\rm{fit}}}\left[ 1-m_{\rm{fit}}\, r\,K_1\left(m_{\rm{fit}}\, r\right)\right],
\end{equation}
{where the modified Bessel function of the second kind is denoted by $K_n$.}
The coefficient $4/3$ in \req{eq:fit_func_form} is simply the quadratic Casimir of the fundamental irrep when $N_c =3$, which serves as the reference to the study of $N_c$-scaling properties of TMD distributions in the following sections.

We fit the two parameters of $\Gamma_r^{\rm{fit}}$, namely $\mu_{\rm{fit}}$ and $m_{\rm{fit}}$, in the range $r \leq 3.04\, \text{fm}$.
The result of the fit can be seen in {Fig.~\ref{fig:dip_fit_v2}}, and the parameters extracted from the fit are:
\begin{equation}\label{eq:fit_param}
    \mu_{\rm{fit}} = 0.859 \textrm{fm}^{-1}, \qquad 
    m_{\rm{fit}} = 0.191\textrm{fm}^{-1}.
\end{equation}
Let us stress that the transverse charge density obtained from this fit $\mu_{\rm{fit}}^2\approx 0.74$~fm$^{-2}$ is distinct from the simulation parameter $\mu^2=0.49$~fm$^{-2}$ mentioned in Tab.~\ref{tab:qg_gg_parameters} (see Appendix \ref{params_choice}). 
This behavior is expected and has been illustrated in the literature (see e.g. \cite{Lappi:2007ku}).
We also notice that the fitted mass $m_{\rm{fit}}$ acquires a $\sim 5\%$ correction compared to the parameter used for $SU(3)$ in Tab.~\ref{tab:qg_gg_parameters}, which we assume to be a finite volume effect.

\subsection{\texorpdfstring{$N_c$}{Nc}-scaling of the dipole.}
\begin{figure}
\begin{center}
\includegraphics[width=0.49\textwidth]{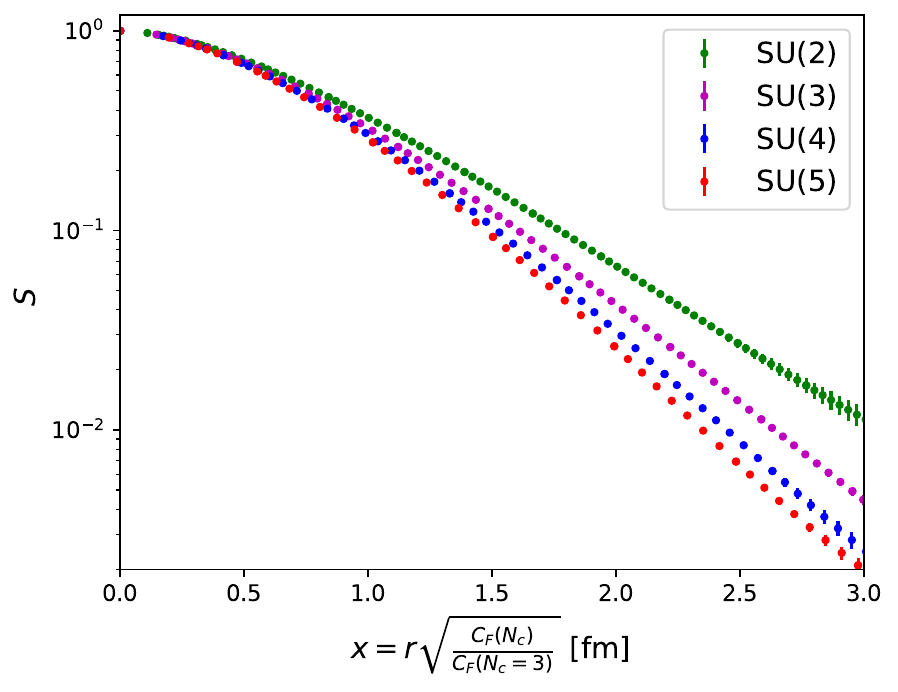}
\includegraphics[width=0.49\textwidth]{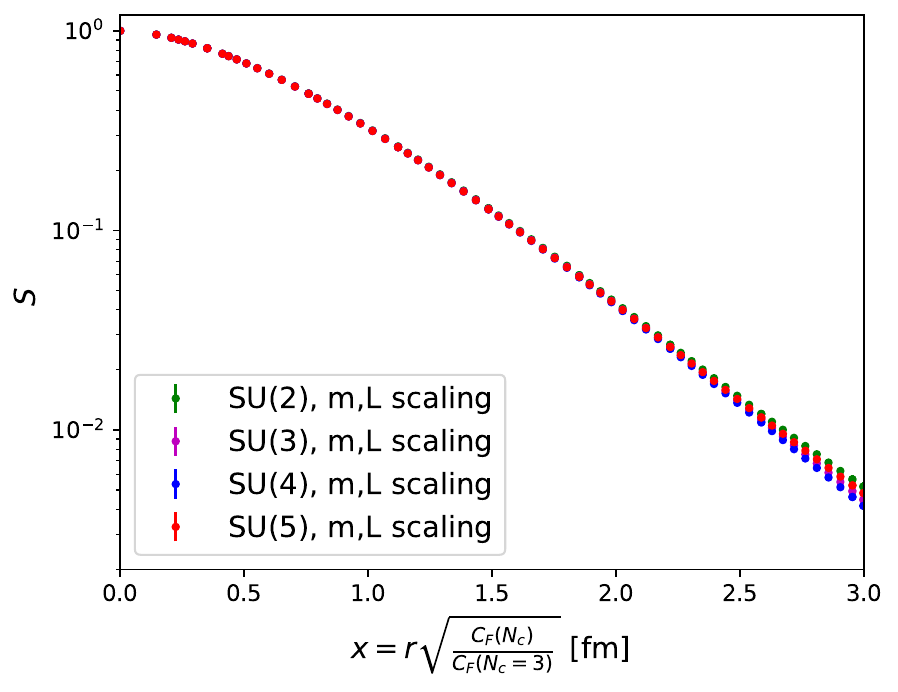}
\caption{Dipole in the MV model obtained with (left panel) and without (right panel) scaling of the mass and volume. The simulation parameters for $qg$ sector are given in Tab.~\ref{tab:qg_gg_parameters} of Appendix \ref{params_choice}. \label{fig. dipole scaling} 
}
\end{center}
\end{figure}

In the left panel of {Fig.~\ref{fig. dipole scaling}}, we show the dipole obtained from simulations for several values of $N_c \in \{2,3,4,5\}$  while all parameters of the simulation ($\mu, m, L$) are fixed. The dipole is plotted in terms of the $N_c$-scaled distance:
\begin{equation}
    \textrm{x}= r \sqrt{\frac{C_F(N_c)}{C_F(N_c = 3)}}, 
    \label{Nc_scaled_distance}
\end{equation}
which accounts for different values of the quadratic Casimir for different $N_c$. In the case of infinite $L$ and zero mass regulator $m=0$, this scaling would be exact. However, in our case, the breaking is observed at large distances, $\textrm{x} \gtrsim 0.5\, \text{fm}$ due to the non-zero mass.

We address this observation by promoting the infrared regulators $m$ and $L$ to depend on the value of $N_c$ within our simulation, as shown in Tab.~\ref{tab:qg_gg_parameters}.
While the scaling of the mass is required from the functional form defined in \req{eq:fit_func_form}, we also scale the length of the torus in the simulation in order to have a similar finite volume effect over the range in $N_c$:
\begin{subequations}\label{eq:rescale_dip}
    \begin{align} 
    r &\longrightarrow r / \sqrt{C_F(N_c) / C_F(N_c = 3)}, \\
    m &\longrightarrow m  \sqrt{C_F(N_c) / C_F(N_c = 3)},\\
    L &\longrightarrow L / \sqrt{C_F(N_c) / C_F(N_c = 3)}.
\end{align}
\end{subequations}
The resulting dipole for the values of $N_c = 2, \ldots 5$ is shown in the right panel of {Fig.~\ref{fig. dipole scaling}}. We clearly see an improvement with regard to the $N_c$-scaling property of the fundamental dipole since the data points agree with a single curve in the range in $\textrm{x} \in [0\, \text{fm},3\, \text{fm}]$ described by the $SU(3)$ reference \req{eq:fit_func_form} using the parameters in \req{eq:fit_param}.

Having the dipole $N_c$-scaling properly understood, we now focus on the $N_c$-scaling properties of TMD distributions in the following sections.

\paragraph{Remarks on the error evaluation in the dipole and following distributions.}

Subsequent to each realization, the distribution and its square are aggregated.
This allows us to calculate the mean value and the standard deviation, as shown in {Fig. \ref{fig. dipole scaling}} for the dipole. The fundamental premise underlying the extraction of the variance is that the data points, as a function of the separation $r$, are independent. 
This assumption is erroneous, as evidenced by the values of distribution on a per-realization basis. 
Nevertheless, this is an efficient approach that eliminates the need for storing voluminous data files while still providing insight into the fluctuations on a per-realization basis.
Therefore, it is imperative to acknowledge that certain error bars may be under/over-evaluated.

A more thorough investigation into the correlated errors would be essential for conducting a quantitative analysis; however, this falls outside the scope of the present manuscript. 
It is not anticipated that the conclusions derived from this analysis would be influenced by such an examination, as average quantities exhibit reduced sensitivity to outlier events (assuming their distribution does not have a heavy tail), and we see remarkable agreement with analytical results.

\subsection{\texorpdfstring{$N_c$}{Nc} scaling of \texorpdfstring{$\Omega$s}{Omegas}}
\label{N_c_scaling_omega_section}
In order to study the $N_c$-scaling features of TMD distributions, we first consider the study of the expectation values $\Omega^\omega_{qg}$ and $\Omega^\omega_{gg}$. 
The methodology employed in the following sections will be consistent with that of the fundamental dipole in {Sec.~\ref{sec:dipole_fit}}. The reference will be the operator expectation values for $SU(3)$.
For other values of $N_c$, the following will be considered.
Separations are scaled according to the reference - the $SU(3)$ case - by 
\begin{subequations}\label{eqs:Scaling_generic}
\begin{align}
    r &\longrightarrow r / \sqrt{C_R(N_c) / C_R(N_c = 3)},
\end{align} 
operators involving derivatives in the Gaussian approximation will use 
\begin{align}
    \nabla^i_r\nabla^j_r &\longrightarrow \frac{C_R(N_c)}{C_R(N_c=3)} \nabla^i_r\nabla^j_r.
\end{align}
The mass term in the fit of the dipole is scaled according to 
\begin{equation}
    m_{\rm{fit}} \longrightarrow m_{\rm{fit}} \sqrt{C_R(N_c) / C_R(N_c = 3)},
\end{equation}
and finally, the infrared regulators in the simulation are scaled according to 
\begin{align}
    m &\longrightarrow m  \sqrt{C_R(N_c) / C_R(N_c = 3)}, \\
    L &\longrightarrow L / \sqrt{C_R(N_c) / C_R(N_c = 3)},
\end{align}
\end{subequations}
with $R = \{F, A\}$ depending on the sector of interest, quark-quark (F) or gluon-gluon (A). The simulation parameters are summarized in Tab.~\ref{tab:qg_gg_parameters} of Appendix \ref{params_choice}.

\subsubsection{\texorpdfstring{$qg$}{qg} sector}

Let us quote again the result of the Gaussian approximation obtained in \req{eq:Omega_qg_A} and \req{eq:Omega_qg_B} for the quark-gluon sector:
\begin{equation}
    \frac{C_F}{N_c}\Omega_{qg}^\omega =  \thalf(C_F \nabla_r \Gamma_r)^2 e^{-C_F \Gamma_r}\ \delta_{\omega = \overline{3}}
    -\frac{K_\omega}{2N_c}(C_F\nabla^2_r {\Gamma}_r ) e^{-\frac{C_\omega + C_F}{2}\Gamma_r}\ \text{sinhc} \left[ \frac{1}{2}(C_\omega - C_F) \Gamma_r\right],
    \label{eq:recap_omega_qg}
\end{equation}
where the charges $C_\omega$ and dimensions $K_\omega$ are given in (\ref{CK_coeff_values_qg}).

\begin{figure}[h]
    \includegraphics[width=0.35\textwidth, angle=270]{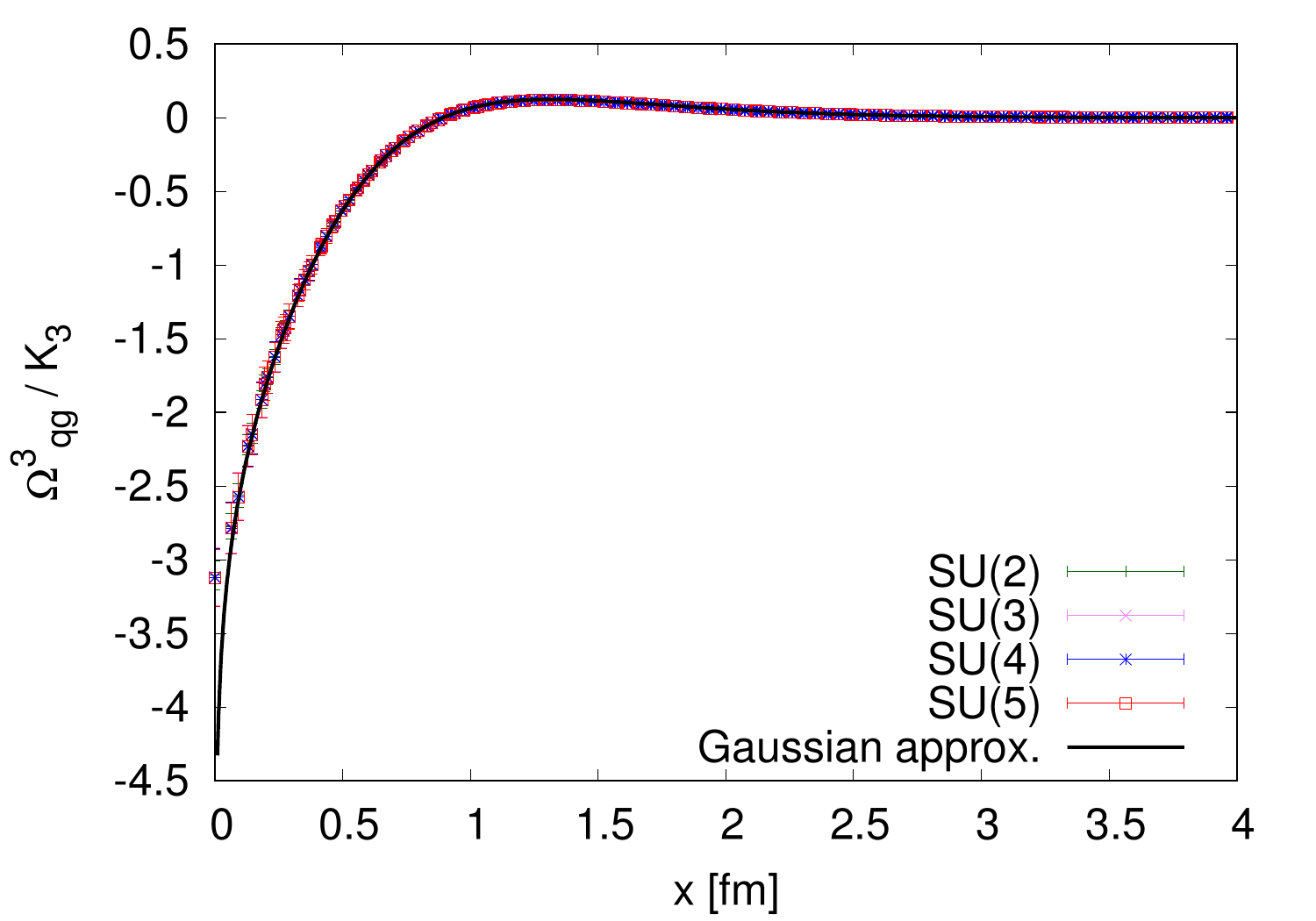}
    \includegraphics[width=0.35\textwidth, angle=270]{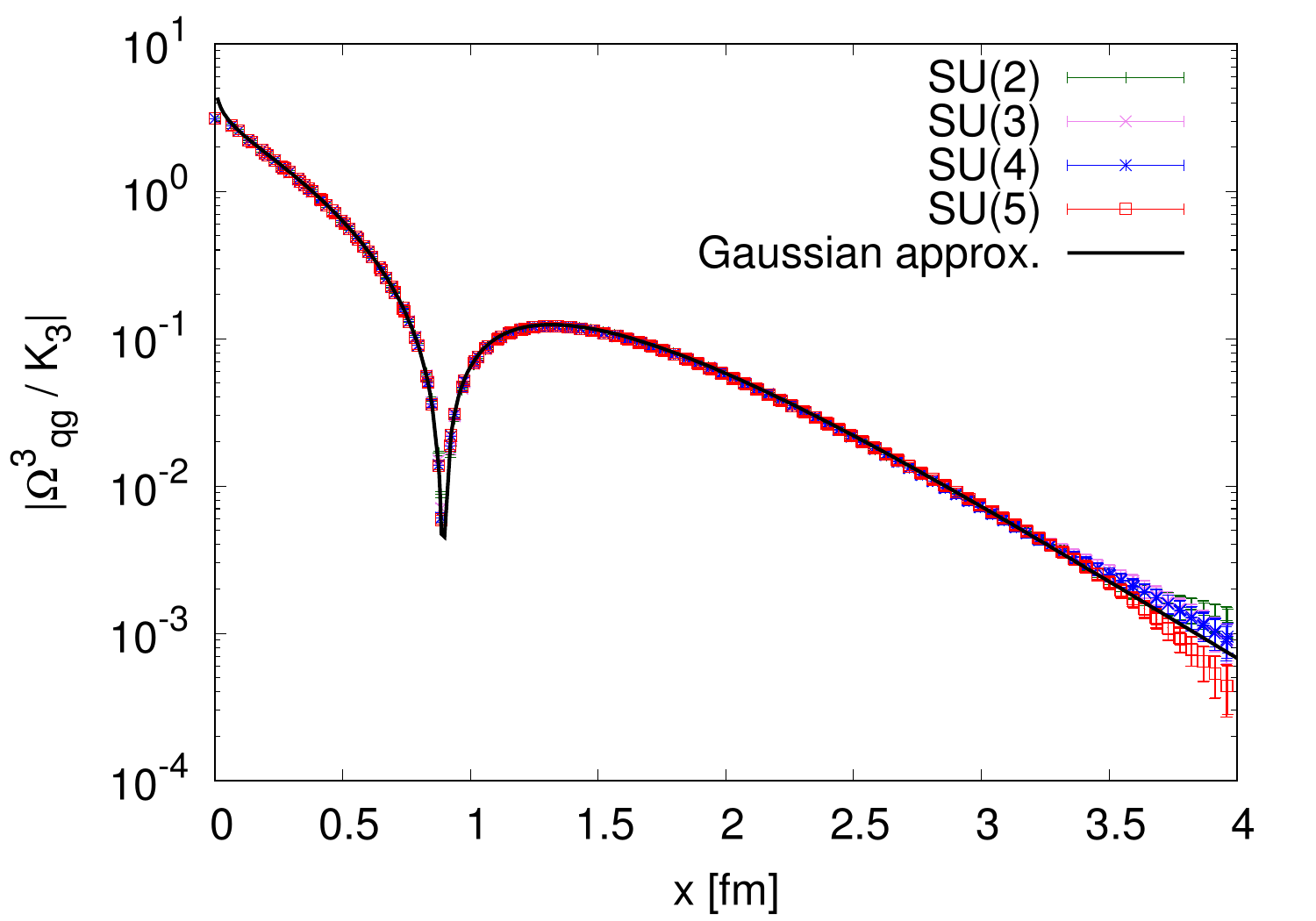}
    \includegraphics[width=0.35\textwidth, angle=270]{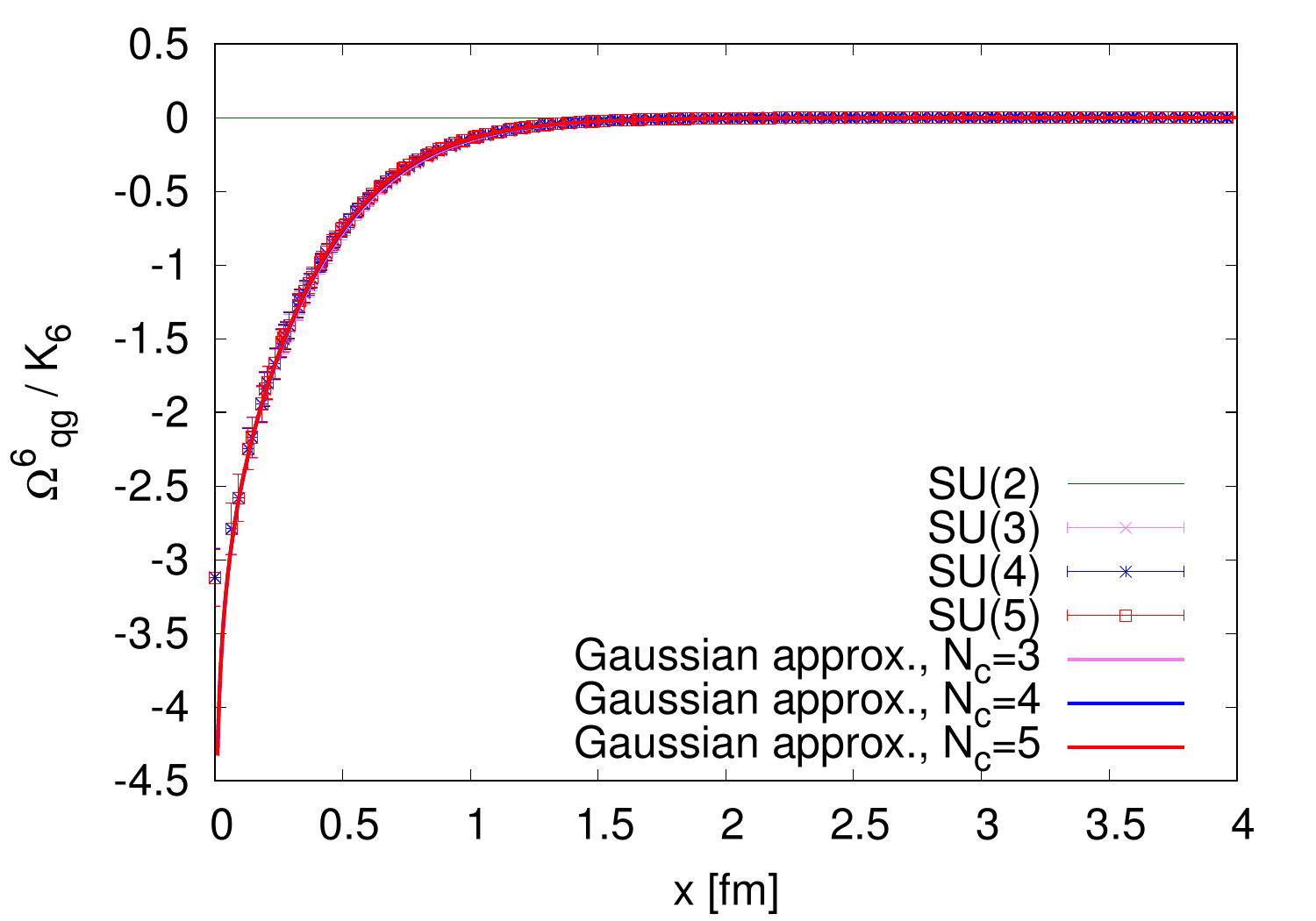}
    \includegraphics[width=0.35\textwidth, angle=270]{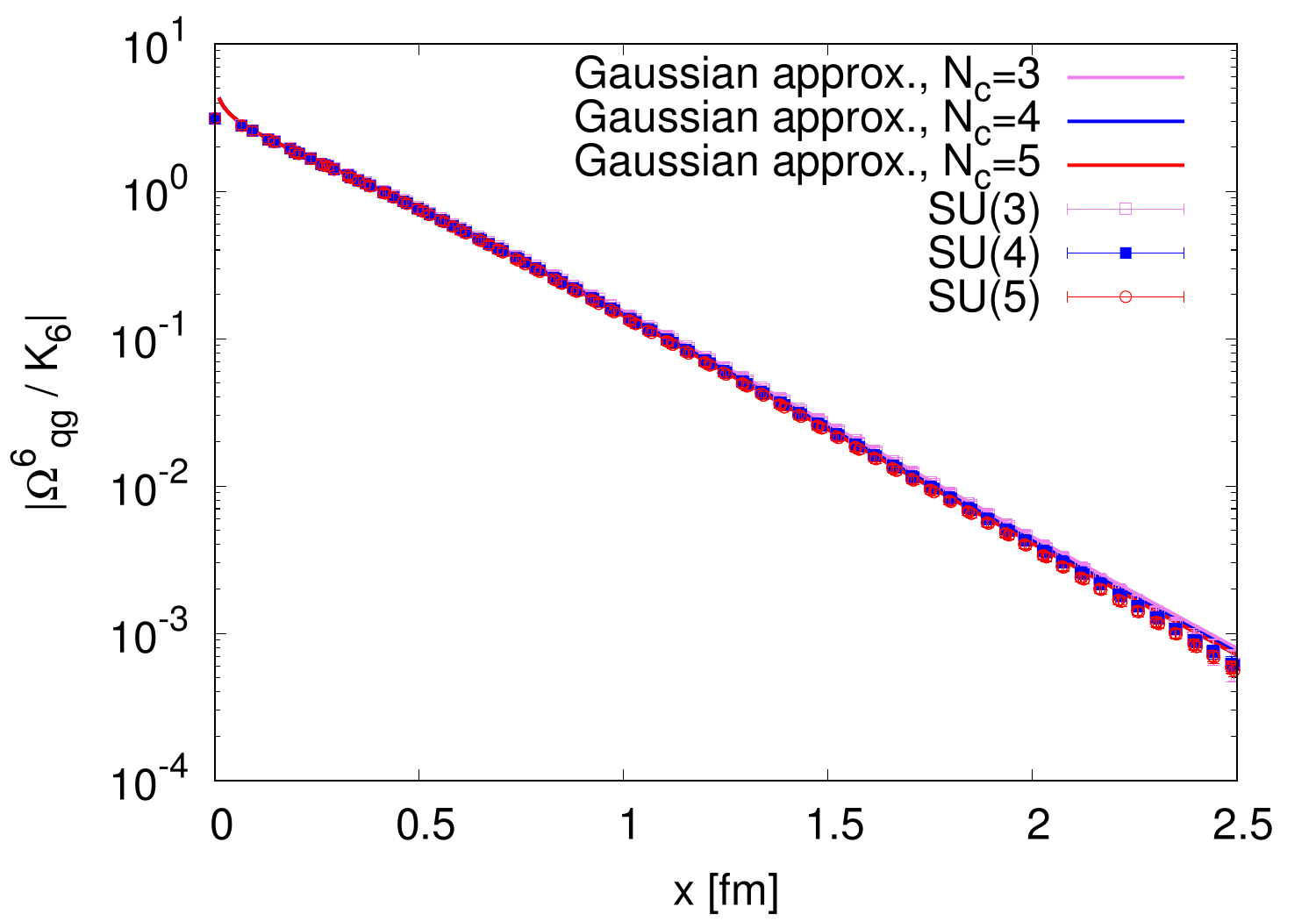}
    \includegraphics[width=0.35\textwidth, angle=270]{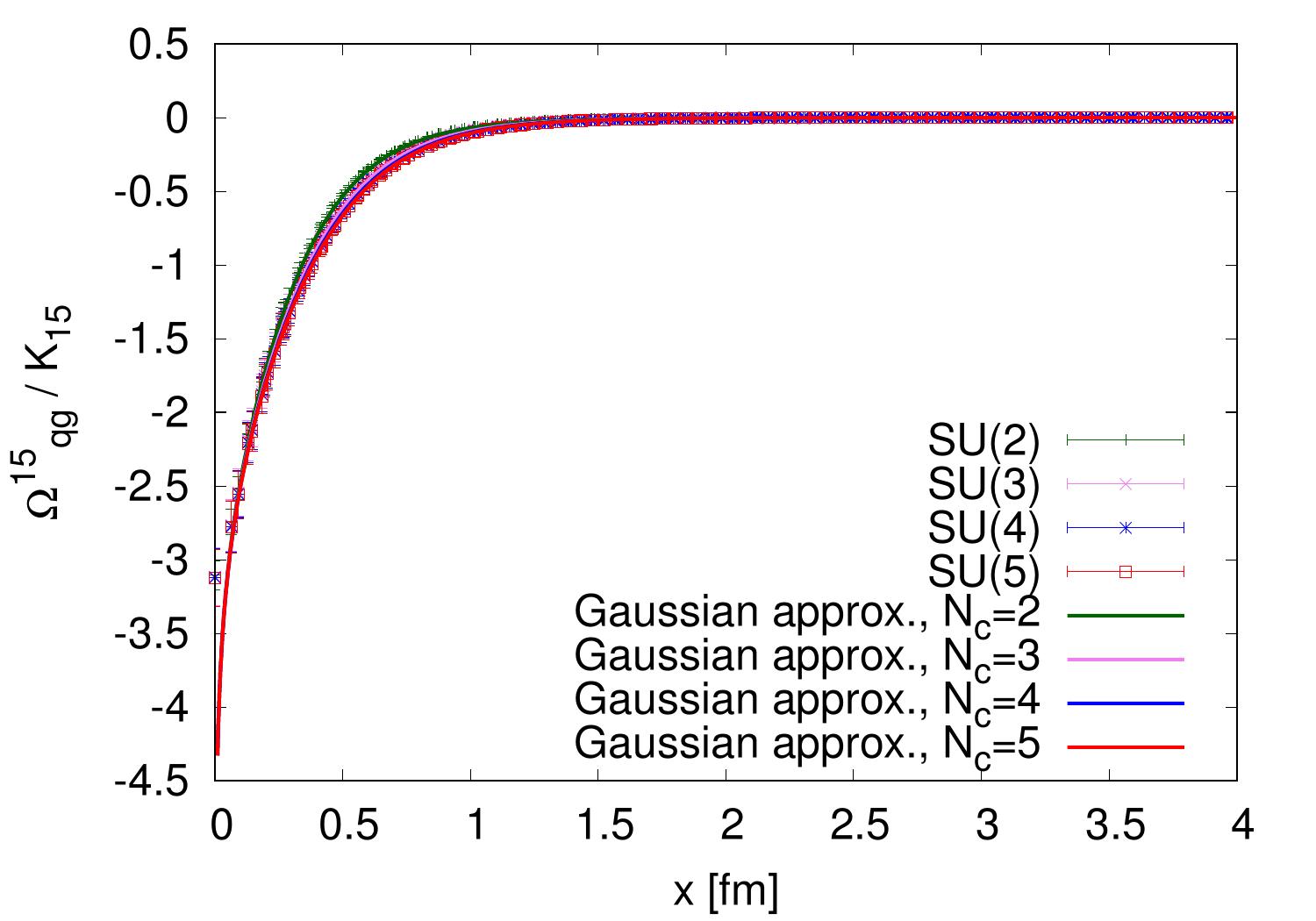}
    \includegraphics[width=0.35\textwidth, angle=270]{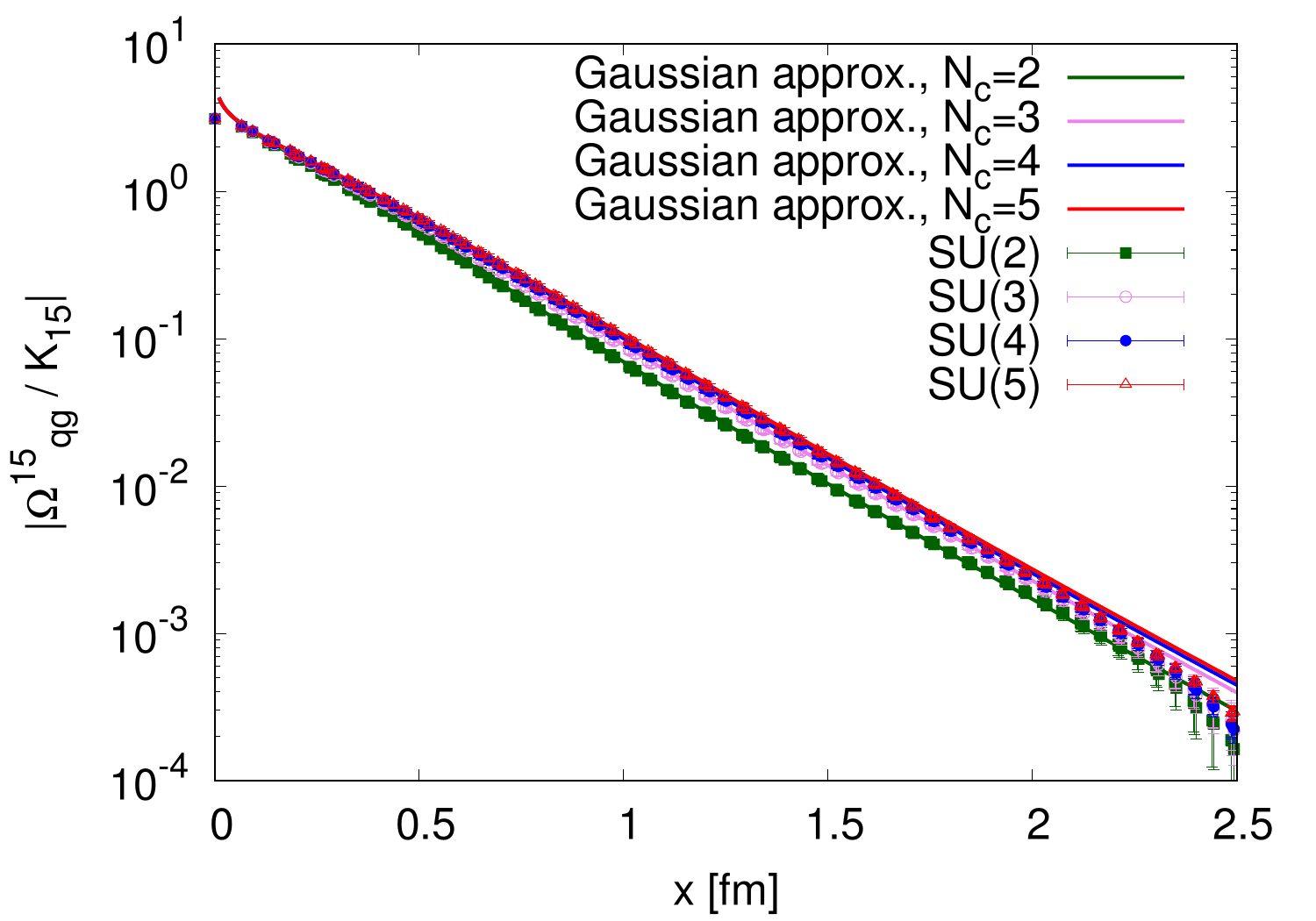}
    \caption{Dependence of $\Omega^{\omega}_{qg}/K_{\omega}$ on $N_c$. Additionally, the Gaussian approximation is shown for each $N_c$ with a continuous curve with color matching the color of the data. For $\overline{3}$, only one curve is shown in black because of the exact scaling. 
    \label{fig. omega qg}
    }
\end{figure}

\paragraph{Scaling.}
On the {r.h.s.} we identify for $\omega = \overline{3}$:
\begin{equation}\label{eq:omega_bar3_equation}
     \frac{\Omega_{qg}^{\overline{3}}}{K_{\overline{3}}}  = \frac{\nabla_r^2}{2C_F} e^{-C_F \Gamma_r}.
\end{equation}
This contribution is expected to scale with $C_F$, provided we perform the substitutions given in \req{eqs:Scaling_generic}. This behavior is verified by numerical simulations shown in the top row of {Fig.~\ref{fig. omega qg}}, where we show $\Omega_{qg}^{\overline{3}}/{K_{\overline{3}}}$ in linear scale (left panel) and $|\Omega_{qg}^{\overline{3}}/{K_{\overline{3}}}|$ on the log scale (right panel).

On top of the points obtained with the MV model, we have shown a black curve for the Gaussian approximation (\ref{eq:omega_bar3_equation}), which was scaled in a similar way.
This curve follows the points very closely. Although we could not fit the dipole at (\ref{eq:fit_func_form}) $\textrm{x} > 3$fm, for $\Omega_{qg}^{\overline{3}}$ the Gaussian approximation curve follows the points even above $3$fm. 

\paragraph{Lack of scaling.}  In \req{eq:Omega_qg_B}, or the second term of \req{eq:recap_omega_qg}, there are two charges involved for each of the two irreps $\{\overline{15},6\}$. Those charges read
\begin{equation}
(C_\omega \pm C_F) = 
\begin{cases}
    C_F(1 \pm 1) + N_c + 1, & \text{for } \omega = \overline{15} \\
    C_F(1 \pm 1) + N_c - 1, & \text{for } \omega = 6 
\end{cases}.
\end{equation}
They cannot be factorized without any remaining $N_c$ dependence left in the expression. Thus the resulting expression for $\Omega^\omega_{qg} / K_\omega$ with $\omega \in \{\overline{15},6\}$ are not expected to properly scale with $C_F$. 

In the middle row of {Fig.~\ref{fig. omega qg}} we show $\Omega^6_{qg} / K_6$ as function of $N_c$-scaled distance (\ref{Nc_scaled_distance}). For $SU(2)$, the $\Omega^6_{qg}$ vanishes; however, when $N_c \ge3$ the data points are compatible with each others. In this case, the breaking of scaling is smaller than the errors of the points. Such a small breaking of scaling can be observed only on the Gaussian approximation curves, where the difference between $N_c=3$ and $N_c=5$ is of the order of 10\% at $\textrm{x}=2.5$ fm.

For $\Omega^{\overline{15}}_{qg} / K_{\overline{15}}$ (bottom row of Fig.~\ref{fig. omega qg}), the breaking of $C_F$-scaling is also small, and only $N_c=2$ MV result is clearly distinguishable from the others on a logarithmic scale.
\begin{figure}
\begin{center}
\includegraphics[width=0.345\textwidth, angle=270]{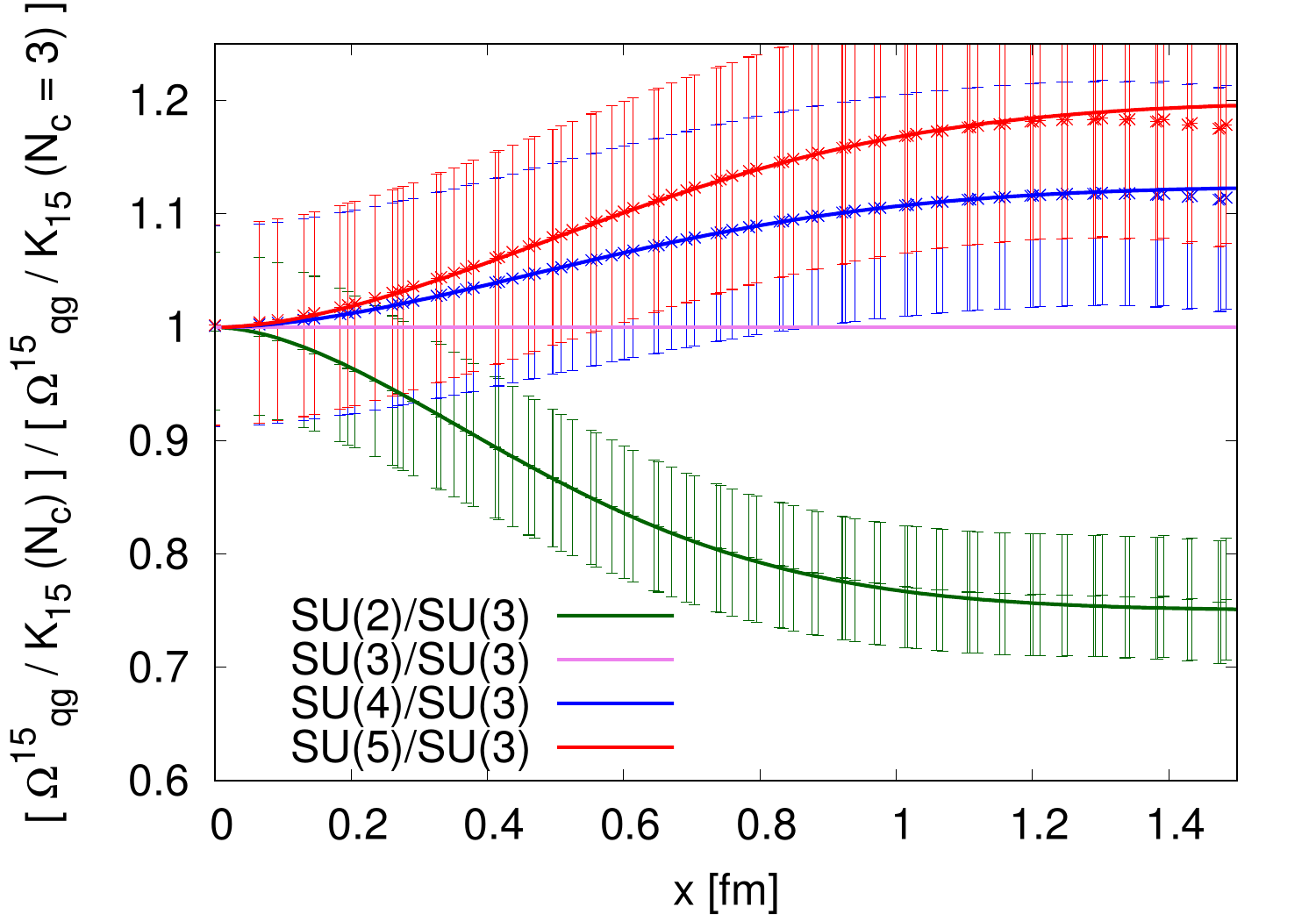}
\includegraphics[width=0.345\textwidth, angle=270]{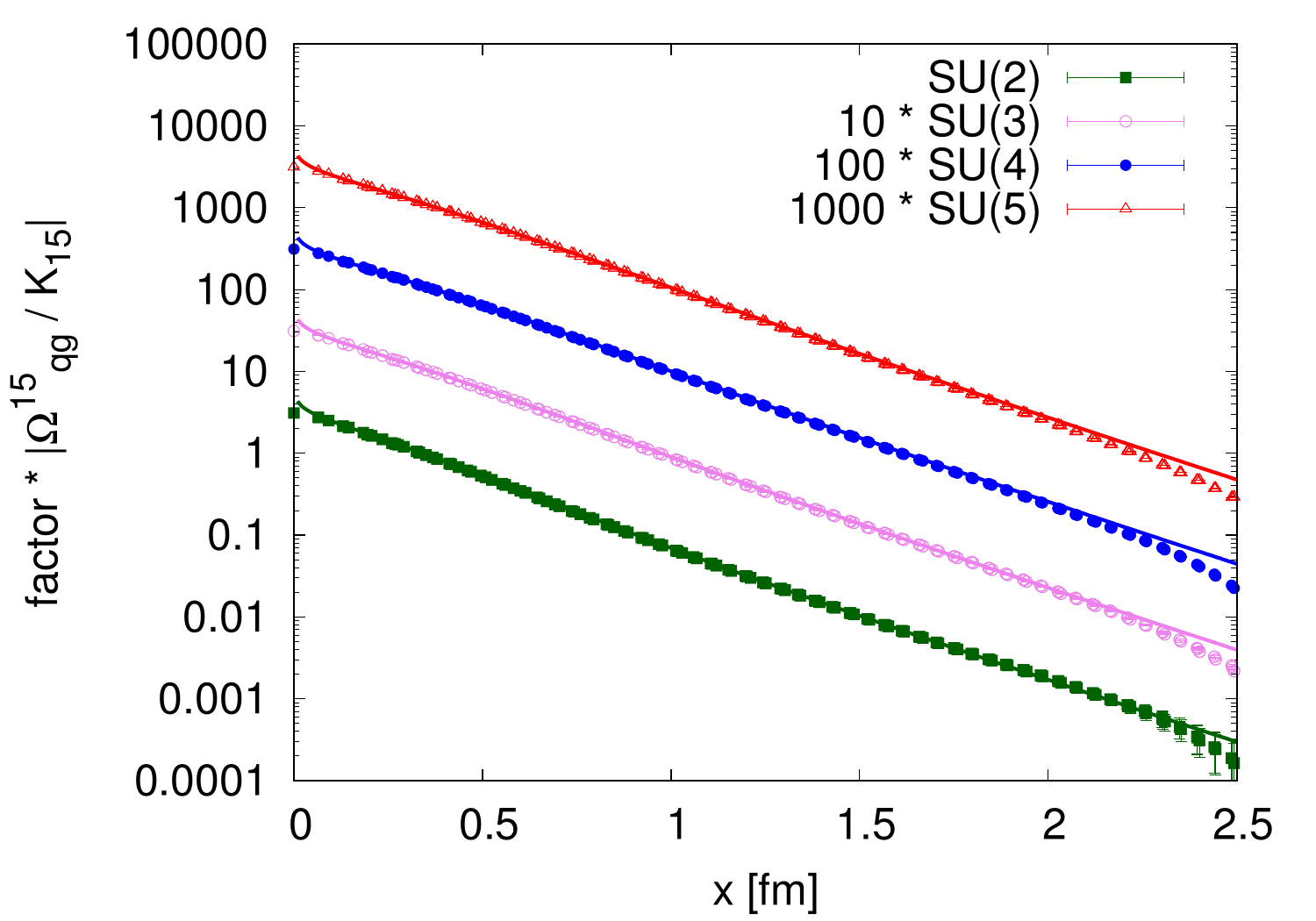}
\caption{
Left: ratio of $\Omega^{\overline{15}}_{qg}/K_{\overline{15}}$ for $N_c = \{2,3,4,5\}$ over $\Omega^{\overline{15}}_{qg}/K_{\overline{15}}$ for $N_c=3$. The Gaussian approximation (plain lines) accurately describes the numerical data for all $N_c$ values. The data points are displayed up to $1.5\,\text{fm}$ for which a relevant signal can be extracted. \\ 
Right: $\Omega^{\overline{15}}_{qg}/K_{\overline{15}}$ on a logarithmic scale, where the curves are shifted by changing the normalization by a factor $10^{N_c-2}$. We see the agreement between the simulated data and the Gaussian approximation in the range $N_c \in \{ 2,3,4,5 \}$. 
\label{fig. omega 15 ratio}
}
\end{center}
\end{figure}
To further highlight the dependence on $N_c$ of $\Omega_{qg}^{\overline{15}}$, we display in left panel of {Fig.~\ref{fig. omega 15 ratio}} the ratio of simulated data for $SU(N_c)$ over the simulated data for $SU(3)$. Due to the normalization choice, those curves agree in the limit $\rm{x} \rightarrow 0$. As we increase the separation $\rm{x}$, the degeneracy is clearly lifted, and we can see the ordering of those curves from top to bottom: $N_c = \{5,4,3,2\}$. The Gaussian approximation accurately captures this feature of the data, as shown by the plain lines of {Fig.~\ref{fig. omega 15 ratio}}. The general agreement is shown again in the right panel of {Fig.~\ref{fig. omega 15 ratio}} where we separated the data with different $N_c$ by an artificial factor.

\begin{figure}[h]
    \centering
    \includegraphics[width=0.65\linewidth]{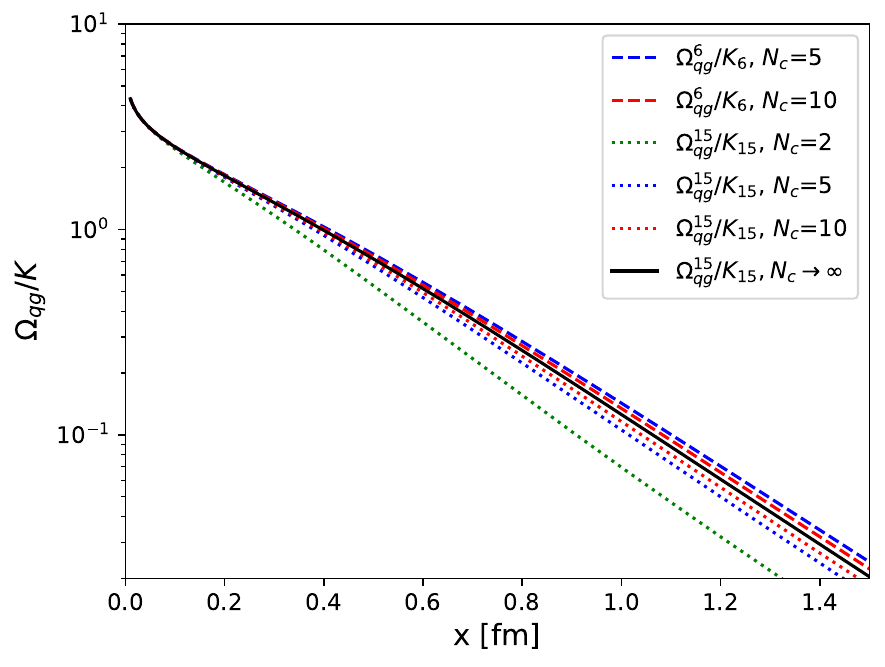}
    \caption{The comparison of $\Omega^\omega_{qg} / K_\omega$ for $\omega = \{\overline{15},6\}$ for various $N_c$ using the Gaussian approximation. Both sets of data converge to a common limit: $\lim\limits_{N_c \rightarrow \infty} {\Omega_{qg}^{\overline{15}}}/{K_{\overline{15}}} = \beta_{\overline{15}} = \lim\limits_{N_c \rightarrow \infty } {\Omega_{qg}^{6}}/{K_6} = \beta_6$. 
    }
    \label{fig:Omega_large_Nc_qg}
\end{figure}

For large values of $N_c$, the charges for $\omega = \{ \overline{15}, 6\}$ in \req{eq:Omega_qg_B} can be approximated as follows:
\begin{equation}
    \label{eq: scaling of 6 and 15}
    (C_\omega \pm C_F) \sim C_F (3\pm 1).
\end{equation}
In this approximation, we can now properly factorize $C_F$ and observe scaling. 
This is shown in {Fig.~\ref{fig:Omega_large_Nc_qg}} where we can read the asymptotic behavior of $\Omega^\omega_{qg} / K_\omega$ for $\omega = \{\overline{15},6\}$ approaching a common curve, $\beta_{\overline{15}}(\textrm{x})=\beta_6(\textrm{x})$, along the trajectory $N_c \in \{2,5,10\}$. 
This feature will be used in {sec.~\ref{sec:largeNc_of_Gaussian_approx}} when the large-$N_c$ limit of the TMD distributions will be extracted within the Gaussian approximation.

\paragraph{Sum rule.} An interesting feature obtained from $\Omega_{qg}^6 = 0$ within \req{eq:Omega_to_Fqg} for $SU(2)$, is that there are only two independent TMD distributions out of the three $\calf_{qg}$ for this specific value of $N_c = 2$:
\begin{align}
    \left[ 2 \hcalf^{(2)}_{qg} - \hcalf^{(3)}_{qg} -\hcalf^{(1)}_{qg} \right]_{N_c = 2} = 0
\end{align}
While $SU(2)$ is an academic case, we will see that this type of conclusion also appears in the gluon-gluon sector for the practical case of $SU(3)$.

\subsubsection{\texorpdfstring{$gg$}{gg} sector}

Let us quote again the result of the Gaussian approximation obtained in section \ref{Sec:Gaussian_approx} for the gluon-gluon sector, Eqs.~(\ref{eq:Omega_gg}), (\ref{F4_F3_omega_rel}):
\begin{align}
\Omega_{gg}^\omega &= \frac{K_A}{2} C_A (\nabla \Gamma_r)^2 e^{-C_A \Gamma_r}\  \delta_{\omega = 8_a}
-\frac{K_\omega}{2}(\nabla^2 {\Gamma}_r )\, e^{-\frac{C_A + C_\omega}{2}\Gamma_r}\ \text{sinhc} \left[ \frac{1}{2}(C_\omega - C_A) \Gamma_r\right], 
\label{omega_gg_recap}
\end{align}
for $ \omega \in \{1,8_a,8_s,10+\overline{10},27,0\}$ and
\begin{equation}
  \Omega_{gg}^{8\ell}= -(N_c^2-1) \Omega_{gg}^{1} .
  \label{omega_gg_last_recap}
  \end{equation}
The charges $C_\omega$ and dimensions $K_\omega$ are given in (\ref{CK_coeff_values}).

\paragraph{Scaling.}
In a similar fashion to the quark-gluon sector, for $\omega = 8_a$ we observe exact $N_c$-scaling for $\Omega_{gg}^{8_a}/K_{A}$. This contribution relates to the Laplacian of the adjoint dipole.

A novel feature appears in the gluon-gluon sector compared to the quark-gluon sector. 
For the irreps $\{1,10\}$, the contribution from \req{eq:Omega_gg_B} still contains two distinct charges given by
\begin{equation}
C_\omega \pm C_A =
\begin{cases}
    \pm C_A, & \text{for } \omega = 1\\
    C_A(2 \pm 1 ), & \text{for } \omega = 10
\end{cases}\ .
\end{equation}
However, we can still factorize $C_A = N_c$ such that the remaining factor is independent of $N_c$. 

\begin{figure}
    \includegraphics[width=0.35\textwidth, angle=270]{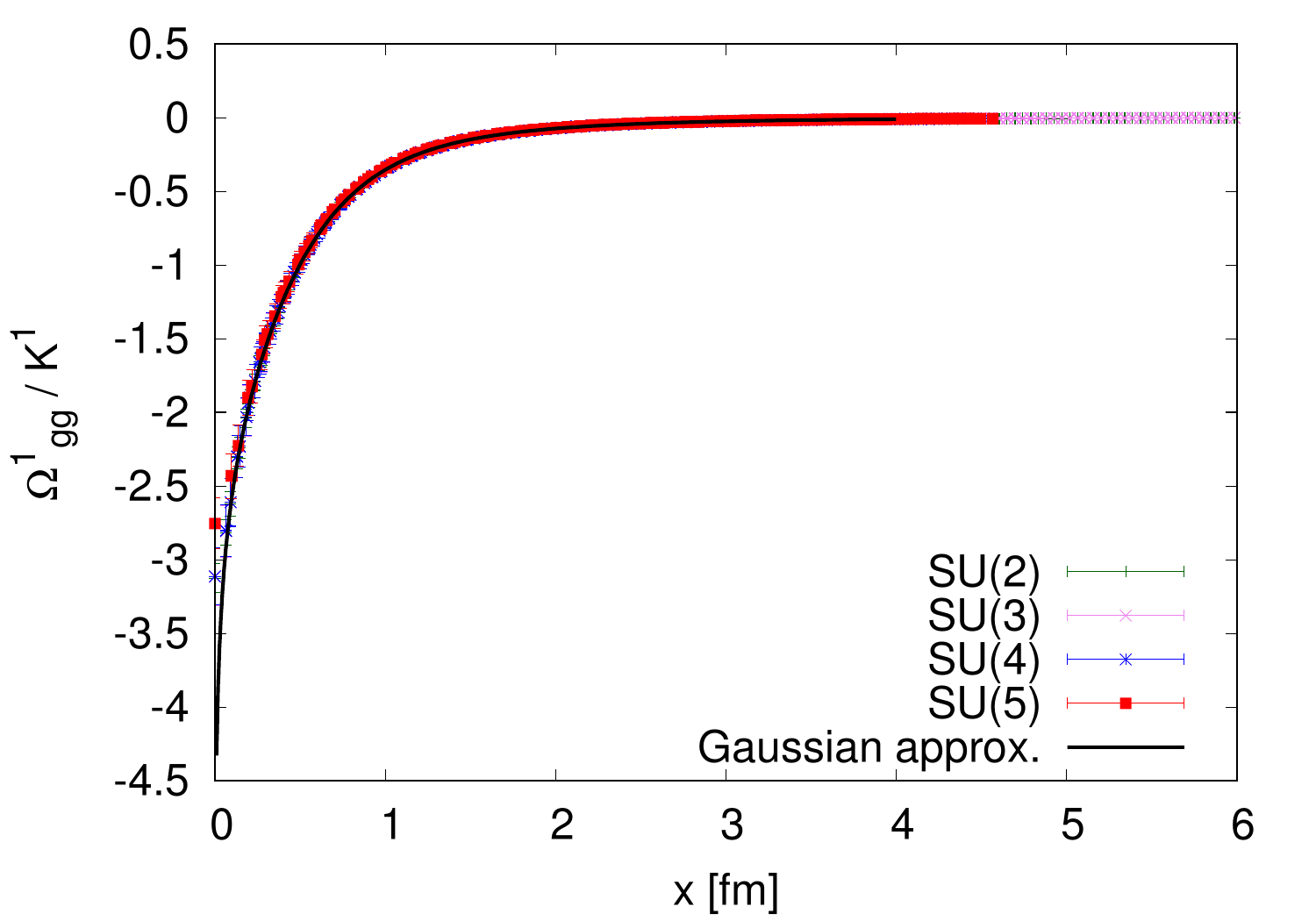}
    \includegraphics[width=0.35\textwidth, angle=270]{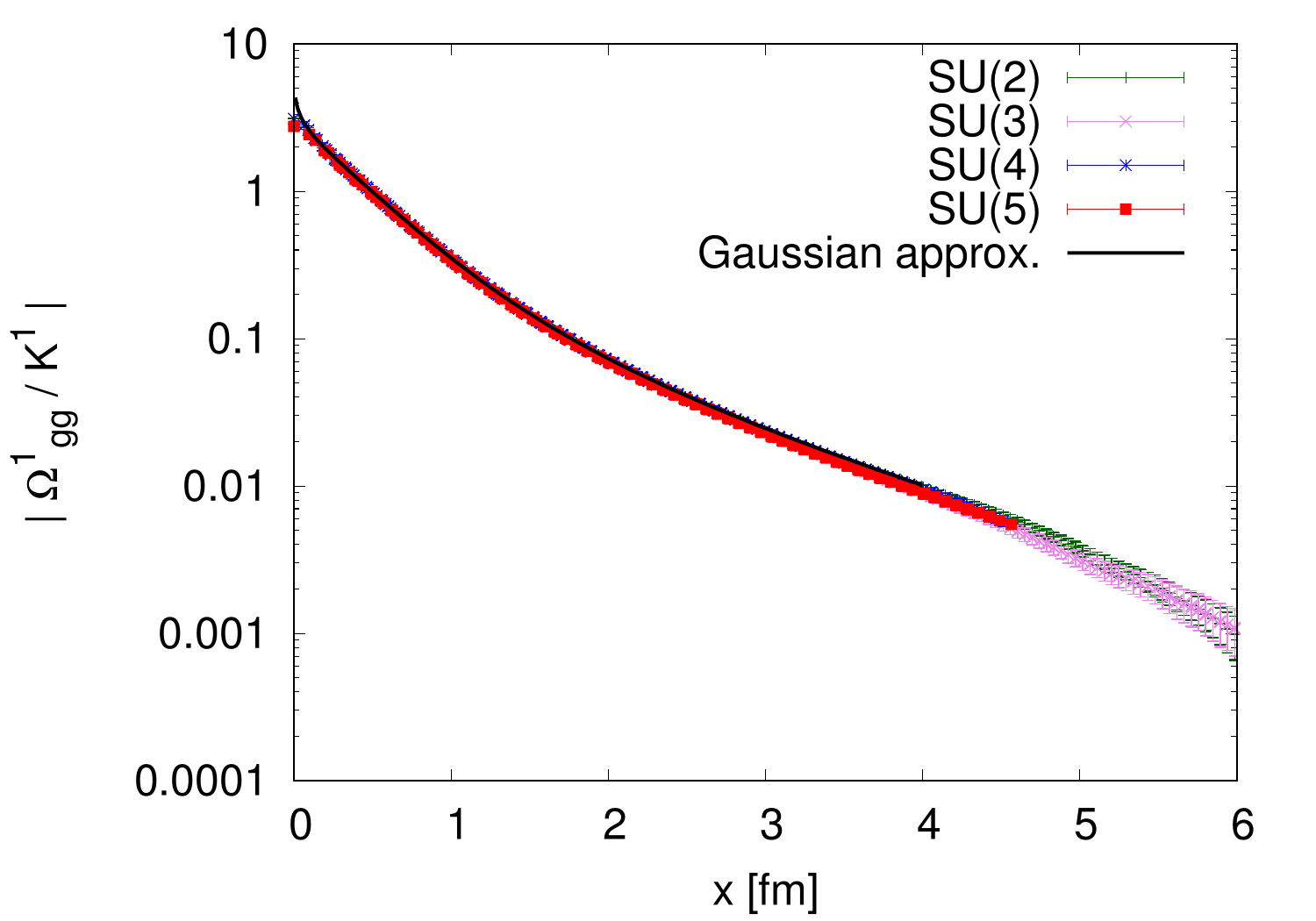}
    \includegraphics[width=0.35\textwidth, angle=270]{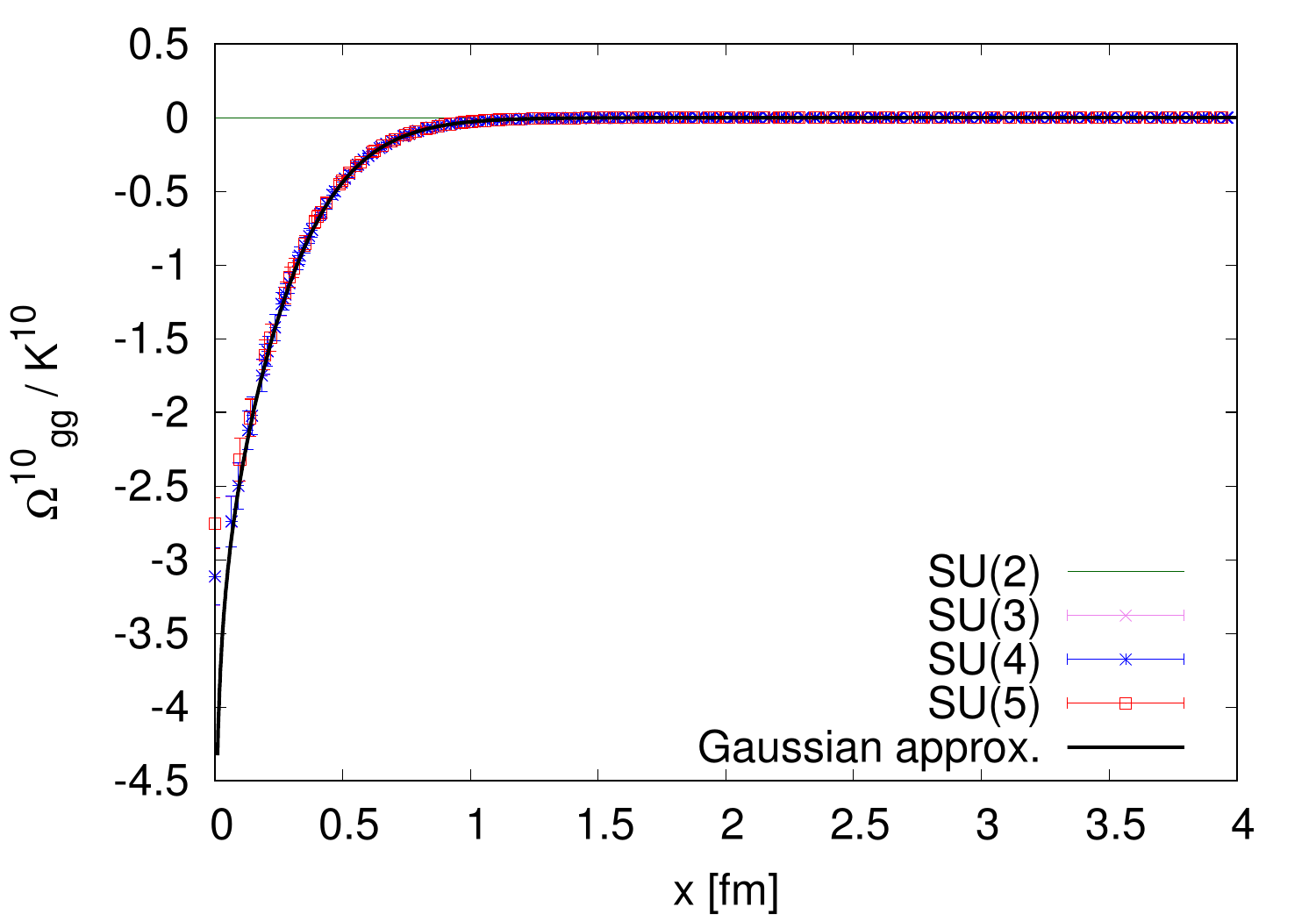}
    \includegraphics[width=0.35\textwidth, angle=270]{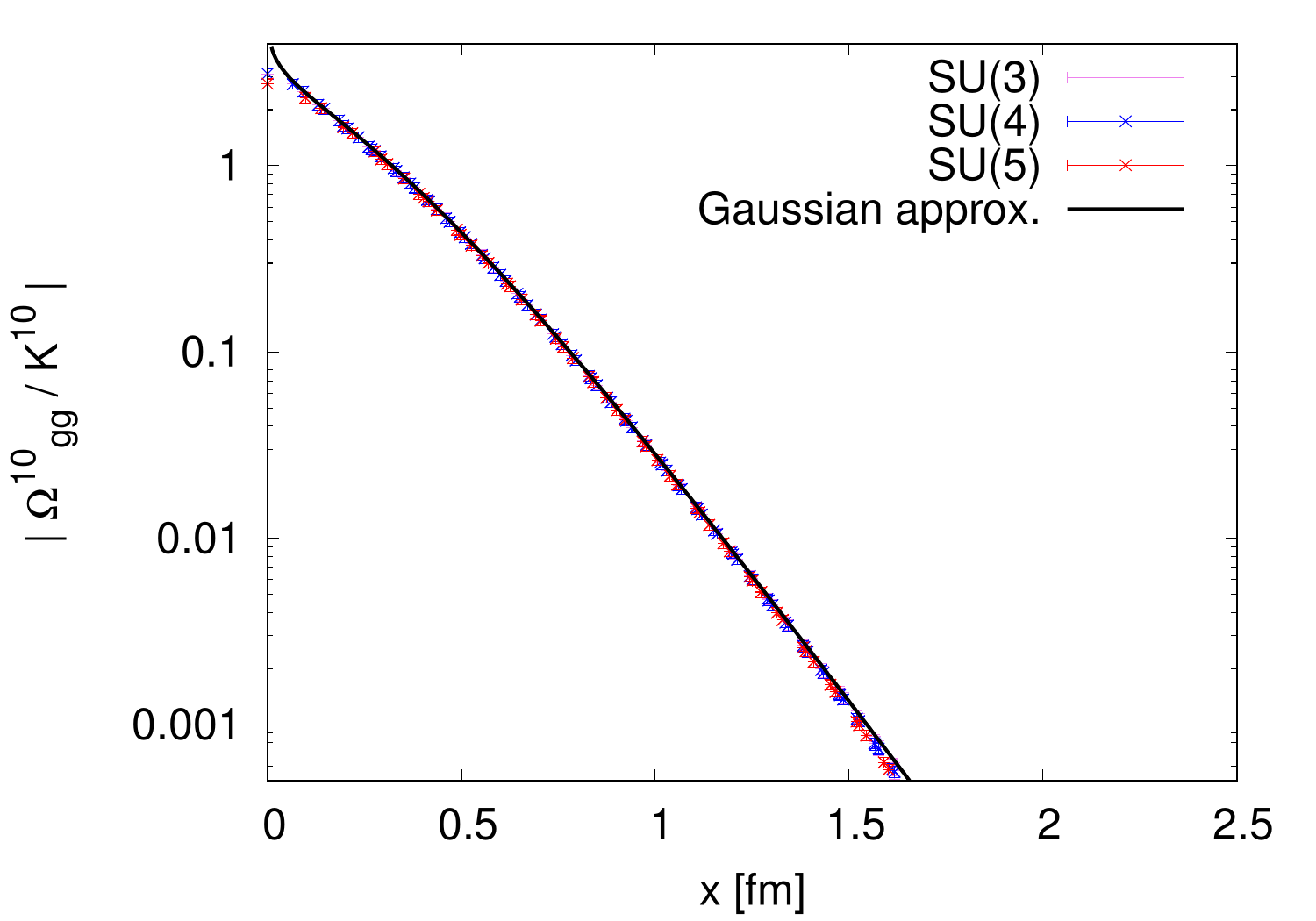}
    \includegraphics[width=0.35\textwidth, angle=270]{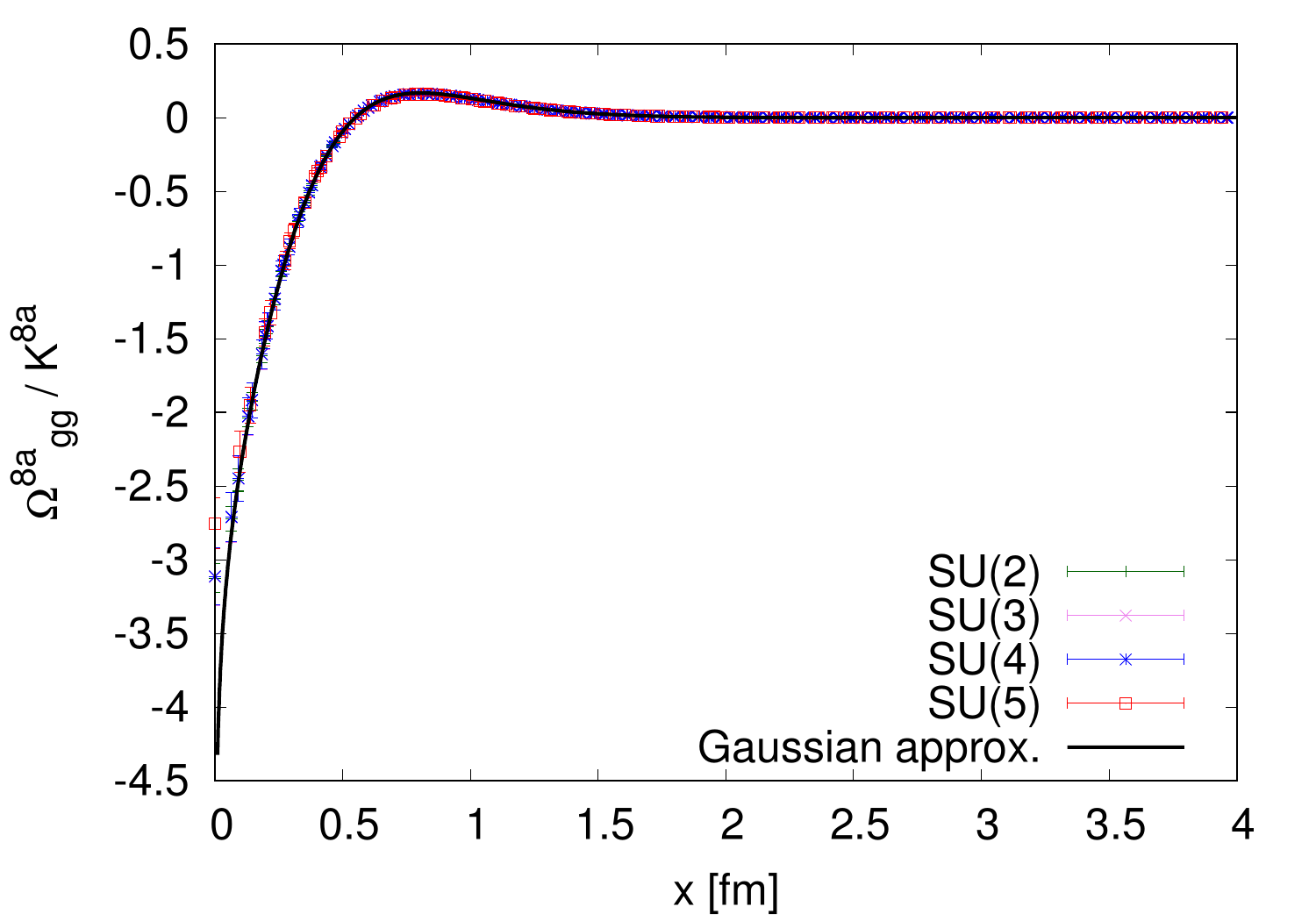}
    \includegraphics[width=0.35\textwidth, angle=270]{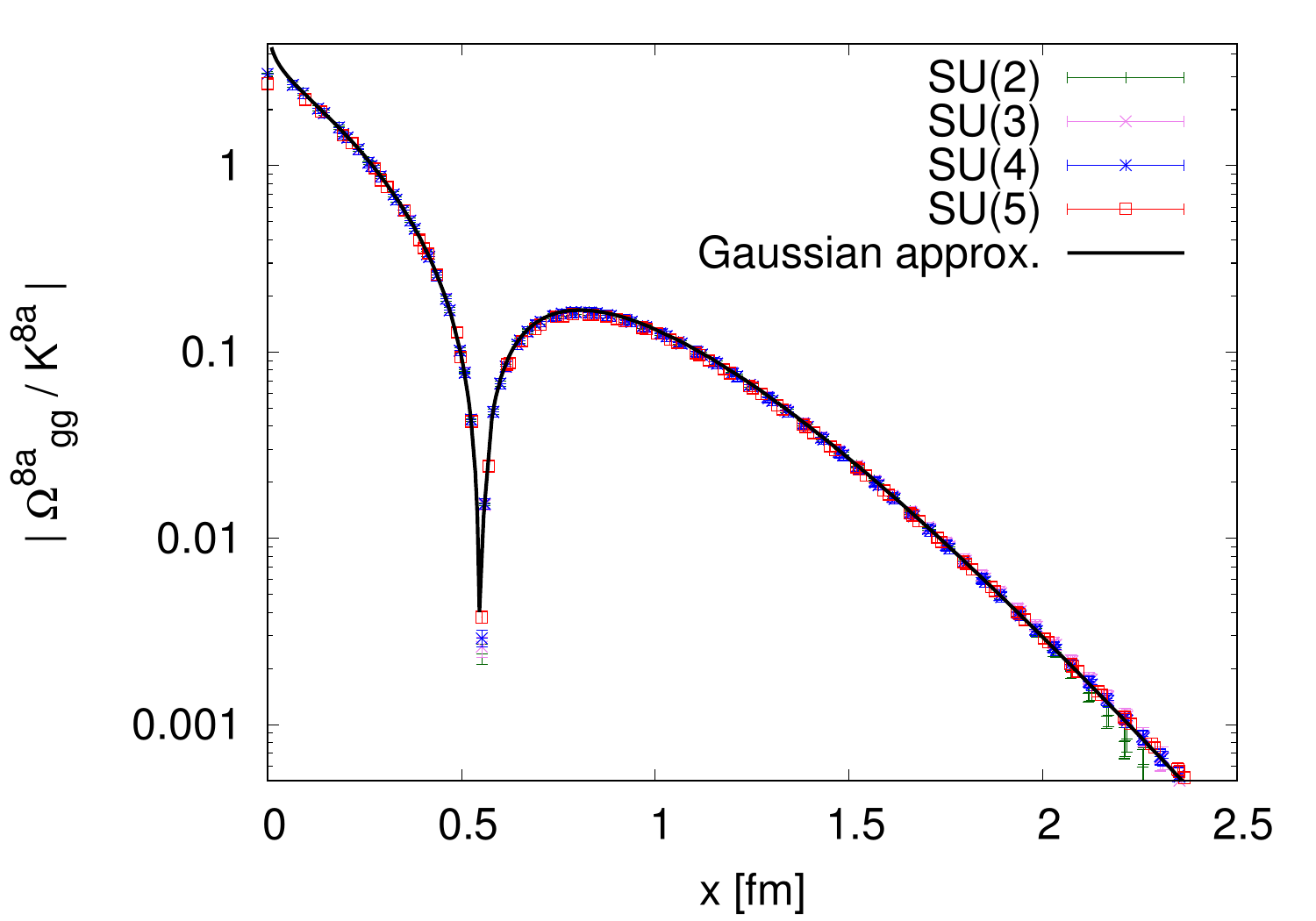}
    \includegraphics[width=0.35\textwidth, angle=270]{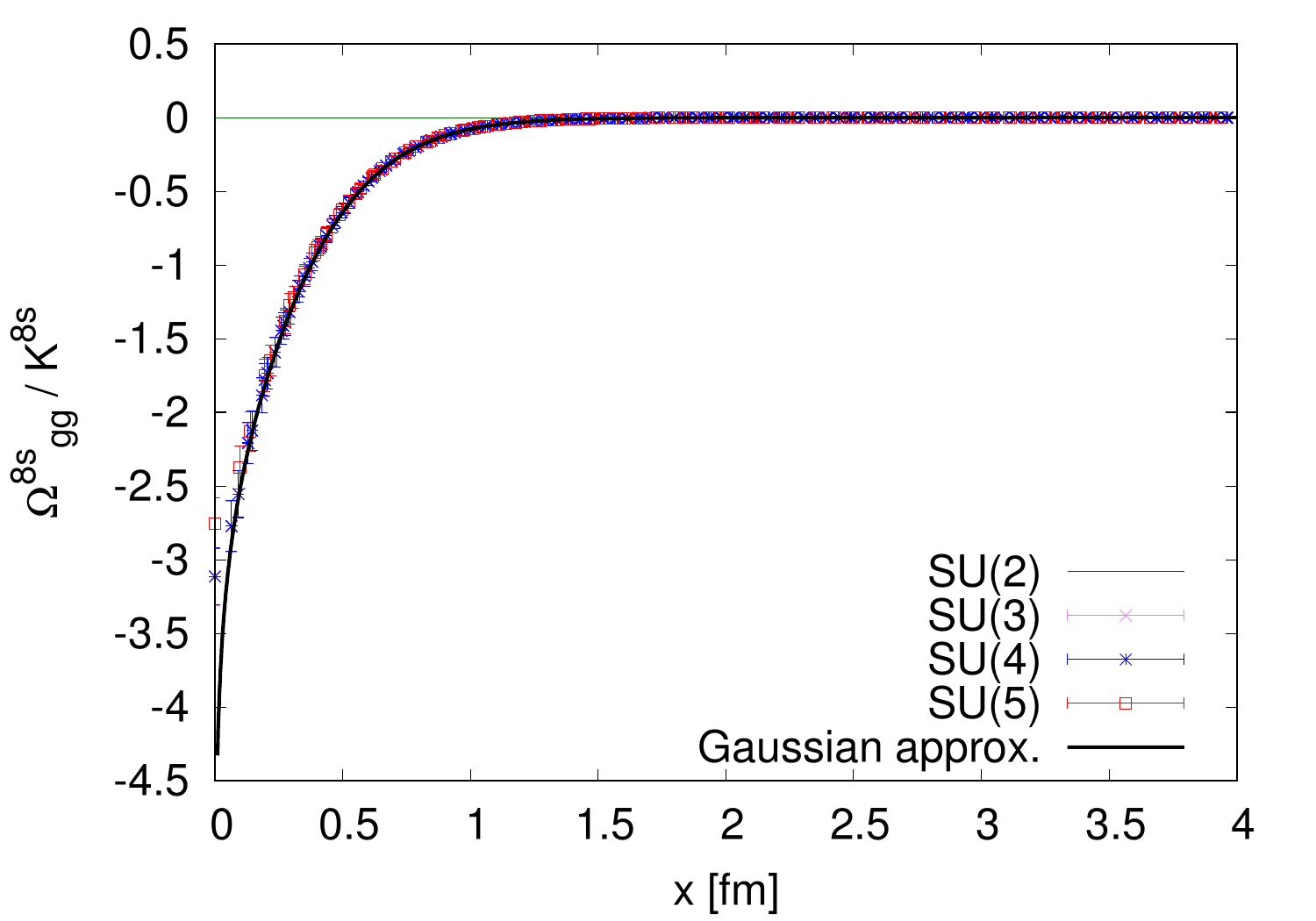}
    \includegraphics[width=0.35\textwidth, angle=270]{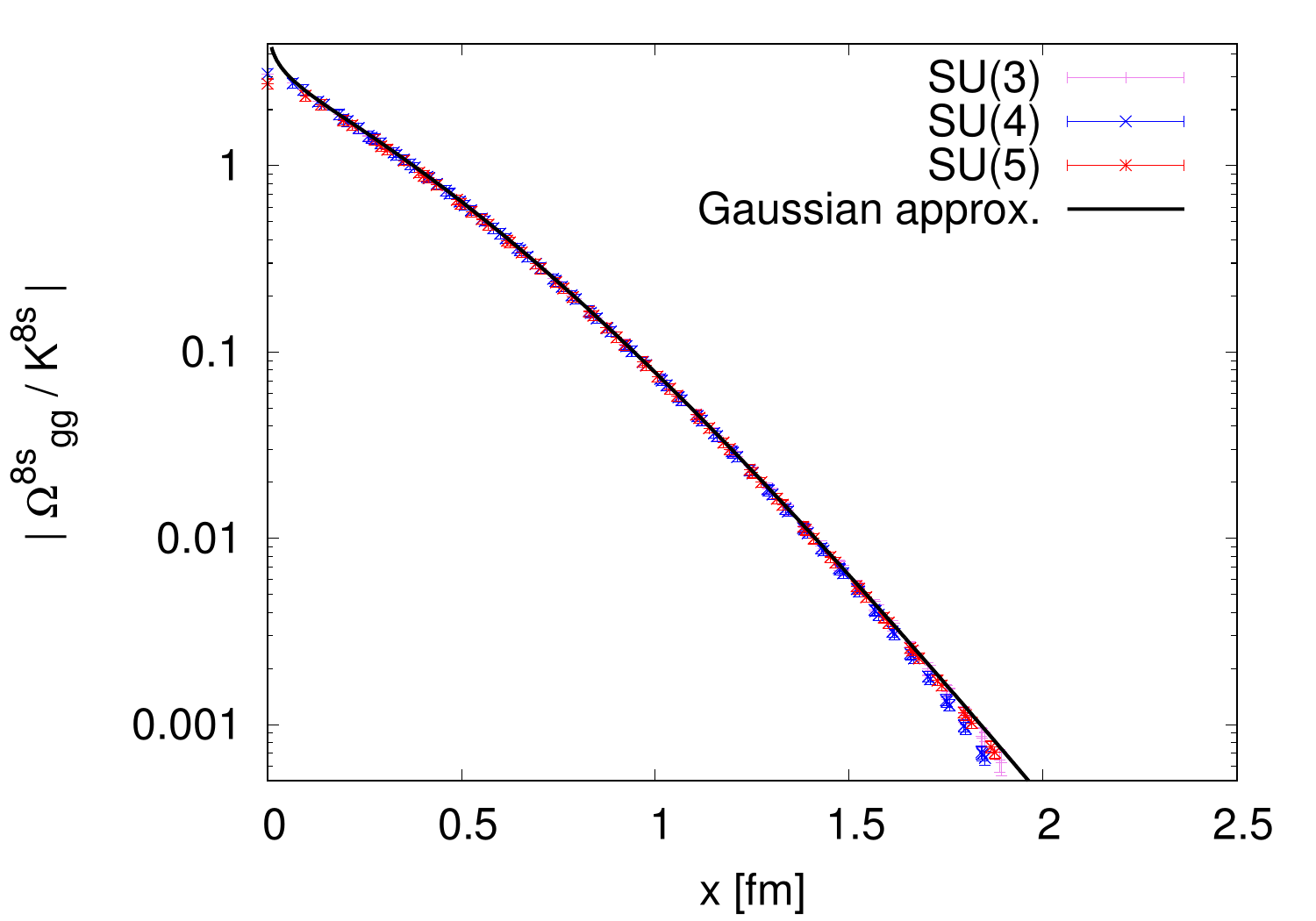}
    \caption{Dependence of $\Omega^{\omega}_{gg}/K_{\omega}$ on $N_c$ for $\omega \in \{1,10,8a,8s\}$. Additionally, the Gaussian approximation is shown with only one curve in black because of the scaling.
    \label{fig. omega gg scaling}
    }
\end{figure}

In {Fig.~\ref{fig. omega gg scaling}}, we see the agreement with $N_c$-scaling of $\Omega_{gg}$  for the irreps $\{1,10,8_a,8_s\}$.

\begin{figure}
    \includegraphics[width=0.35\textwidth, angle=270]{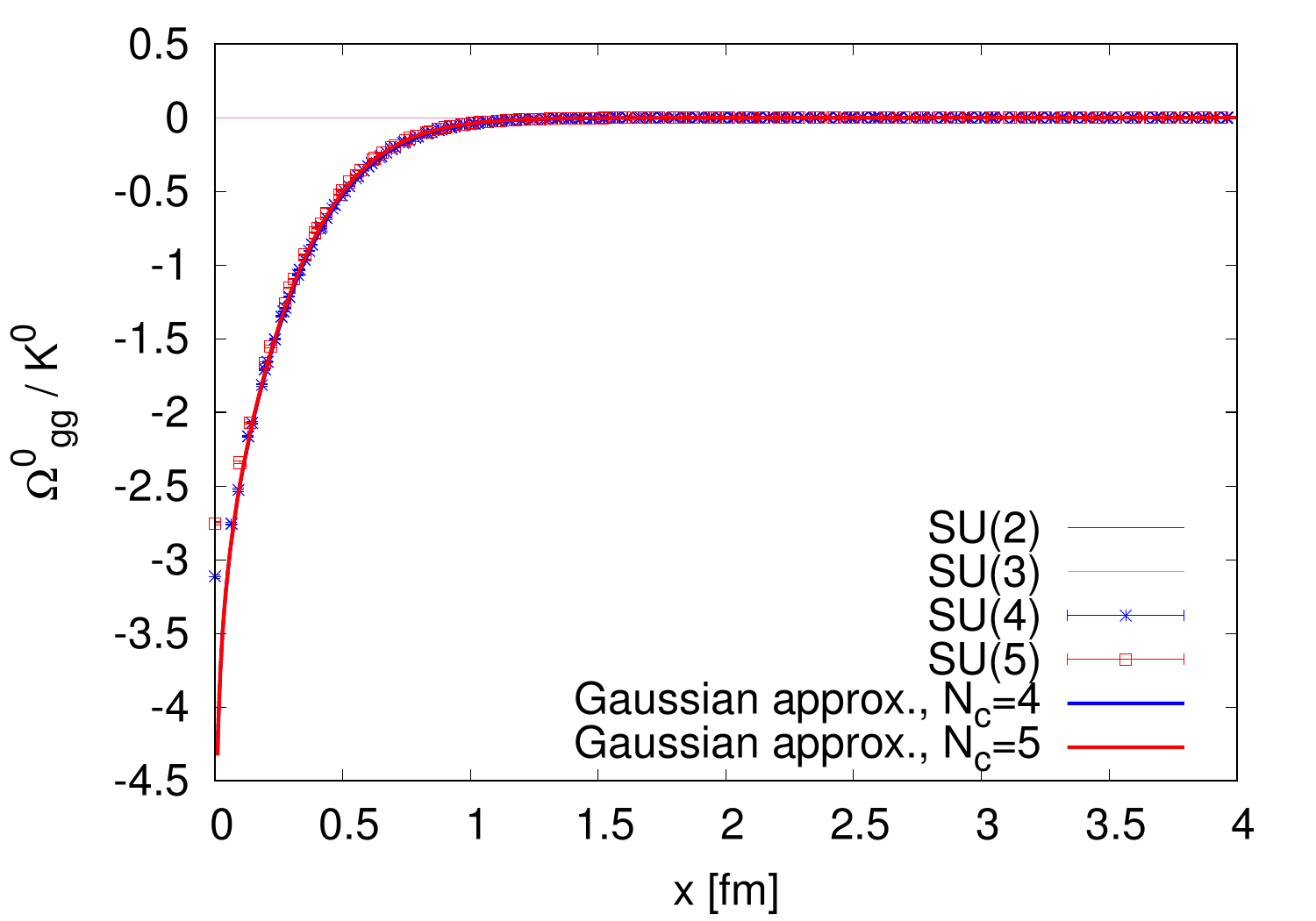}
    \includegraphics[width=0.35\textwidth, angle=270]{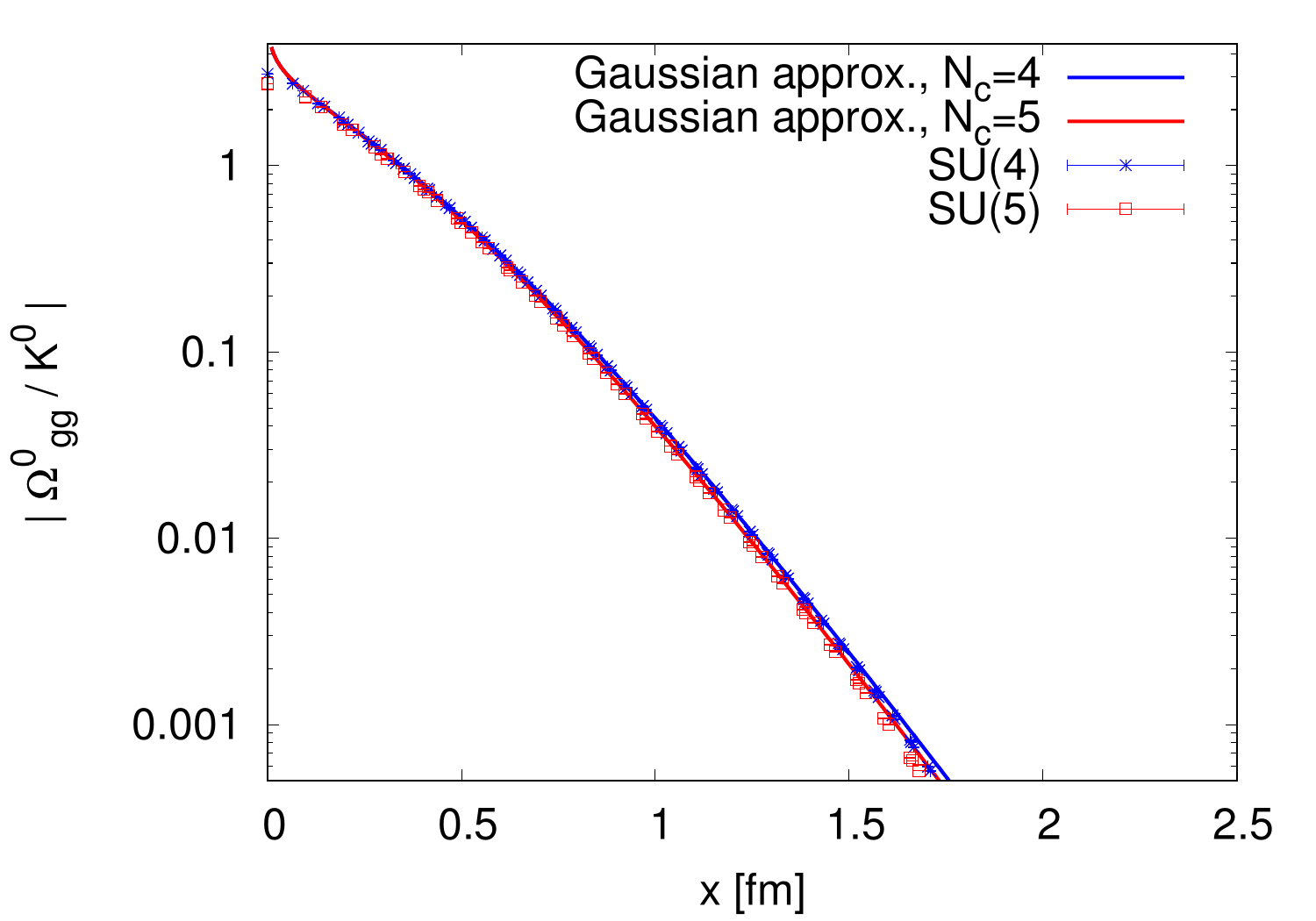}
    \includegraphics[width=0.35\textwidth, angle=270]{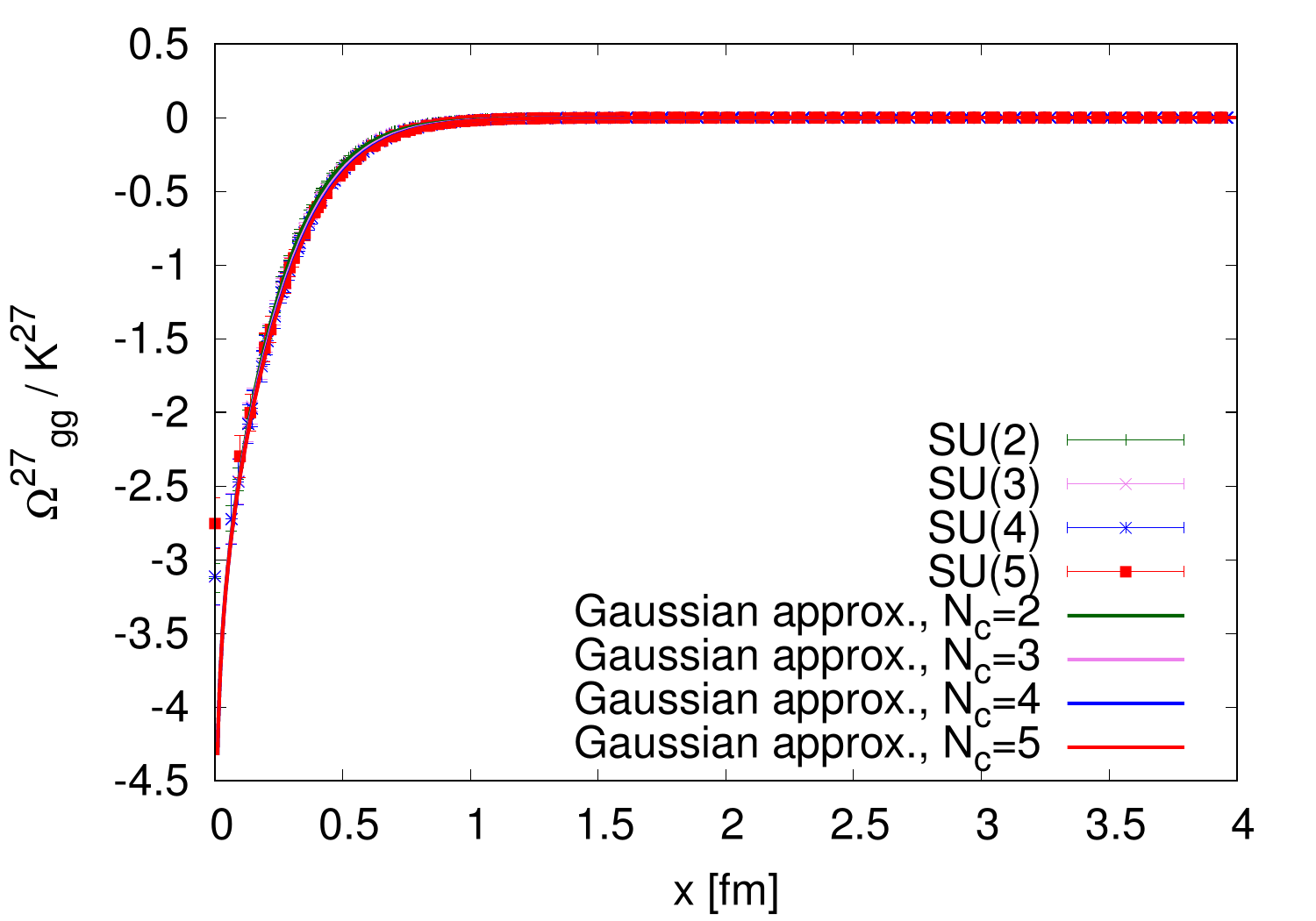}
    \includegraphics[width=0.35\textwidth, angle=270]{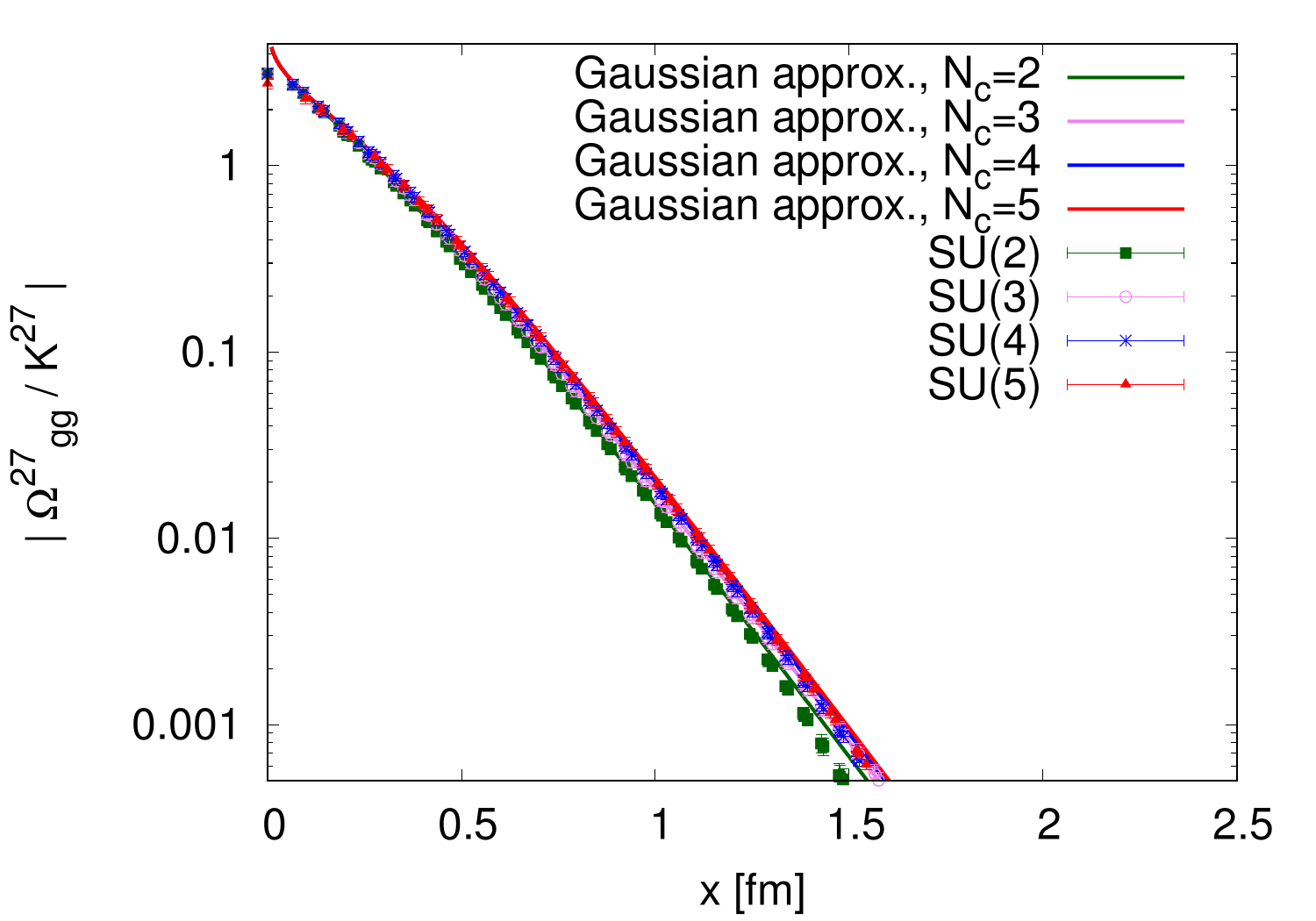}
\caption{Dependence of $\Omega^{\omega}_{gg}/K_{\omega}$ on $N_c$ for $\omega \in \{0, 27\}$. Additionally, the Gaussian approximation is shown for each $N_c$ with a continuous curve with color matching the color of the data. 
\label{fig. omega gg not scaling}
}
\end{figure}

\paragraph{Lack of scaling.}
In the case of the irreps $\{27,0\}$, the charges involved in \req{eq:Omega_gg_B} are
\begin{equation}
C_\omega \pm C_A =
\begin{cases}
    C_A(2 \pm 1) + 2, & \text{for } \omega = 27\\
    C_A(2 \pm 1) - 2, & \text{for } \omega = 0
\end{cases}\ .
\end{equation}
Factorizing $C_A$ implies that the remaining expression still carry an explicit $N_c$-dependence which formally breaks the $N_c$-scaling of the expectation values $\Omega_{gg}^{27}$ and $\Omega_{gg}^{0}$. In {Fig.~\ref{fig. omega gg not scaling}}, we observe a small deviation from $N_c$-scaling, in particular, in the tail of the $\Omega_{gg}^{27}/K_{27}$ within the range $r > 0.5\, \text{fm}$.

\begin{figure}[h]
    \centering
    \includegraphics[width=0.65\linewidth]{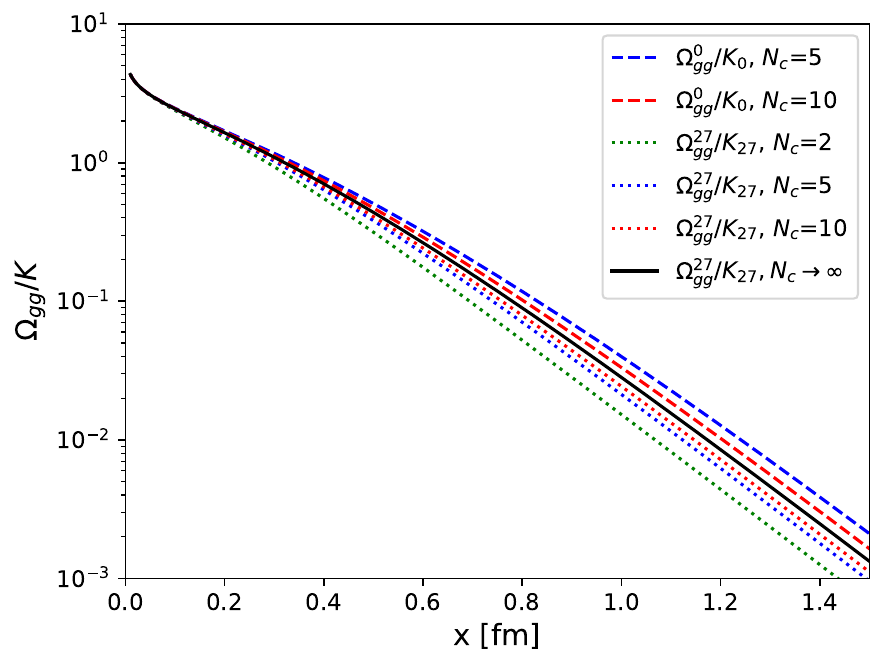}
    \caption{The comparison of $\Omega^\omega_{gg} / K_\omega$ for $\omega = \{27,0\}$ for various $N_c$ using the Gaussian approximation. 
    Both sets of data converge to a common limit: $\lim\limits_{N_c \rightarrow \infty} {\Omega_{gg}^{27}(\rm x)}/{K_{27}} = \beta_{27}(\rm x) = \lim\limits_{N_c \rightarrow \infty } {\Omega_{gg}^{0}(\rm x)}/{K_0} = \beta_0(\rm x)$. 
    \label{fig:Omega_large_Nc_gg}
    }
\end{figure}

For large values of $N_c$, one can approximate the charges according to
\begin{equation}
    \label{eq: scaling of 27 and 0}
    (C_\omega \pm C_A) \sim C_A(2 \pm 1)
\end{equation}
For which we can now properly factorize $C_A$ and observe scaling. This is shown in {Fig.~\ref{fig:Omega_large_Nc_gg}} where we can read the asymptotic behavior of $\Omega^\omega_{gg} / K_\omega$ for $\omega = \{27,0\}$ approaching a common curve along the trajectory for $N_c \in \{2,5,10\}$.

\paragraph{Sum rule.}
The irrep $0$ does not appear for $N_c = 3$, which is a case of particular interest for phenomenology. This implies that out of the seven $\calf_{gg}$, there are only six independent TMD distributions for $SU(N_c = 3)$. The linear combination relating those seven TMD operators is given in \req{eq:gg_27_and_0} for $\eta = -1$, and we write it here for $N_c = 3$:
\begin{align}\label{eq:sum_rule_gg_su3}
\left[
\frac{3}{2} \hcalf^{(1)} + \frac{3}{2} \hcalf^{(2)} - \frac{1}{4} \hcalf^{(3)} - \frac{1}{4} \hcalf^{(4)} - \frac{1}{4} \hcalf^{(5)} - \frac{9}{4}\hcalf^{(6)} +  \frac{3}{2} \hcalf^{(7)}
\right]_{N_c = 3} = 0
\end{align}
This holds true at the operator level, and this feature can serve as a constraint on the extraction of gluon-gluon TMD distributions.\footnote{In addition to this relation, our setup is invariant under time reversal. This introduces the constraint $\langle\hcalf^{(3)}_{gg} \rangle = \langle \hcalf^{(4)}_{gg}\rangle$ (note the expectation value used here, as opposed to operators used in \req{eq:sum_rule_gg_su3}). Thus, there are only five independent TMD distributions in the gluon-gluon sector in our current setup. In terms of $\Omega$s, this statement is a relation between expectation values of (a) the operator used for the transition $8 \rightarrow 1$ and (b) the operator used for the transition $1 \rightarrow 8$, between the initial state projection and the final state projection.}

\paragraph{Remark.}
We note some deviations between the Gaussian approximation and the MV-model data after the distributions span three orders of magnitude in both the quark-gluon sector and the gluon-gluon sector. This is elaborated in more detail within the Appendix \ref{App:Finite_volune} dedicated to the discussion of finite volume effects.

\section{Gaussian approximation for the TMD distributions}
\label{sec: tmds from omega}

In the following sections, we collect relations between the Gaussian approximation of the $\Omega$'s and the Gaussian approximation of the TMD distributions. Those are derived in {sec.~\ref{sec_birdtrack}} for both the quark-gluon sector and the gluon-gluon sector. 
This links the numerical results of {sec.~\ref{sec:num_result_omega}} to those of {sec.~\ref{sec:num_result_tmds}}.
We finally extract the large-$N_c$ limits of the Gaussian approximation of the TMD distributions. 
For this purpose, we make use of the invariants under scaling found in sec.~\ref{sec:num_result_omega}.
The obtained limits agree with the available results in \cite{Caucal:2025zkl}\footnote{The reader should note the different naming convention of TMD distributions in \cite{Caucal:2025zkl}. We are using convention from \cite{Marquet:2016cgx,Bury:2018kvg} which differs from the one used in \cite{Caucal:2025zkl}, compare our Eqs.~(\ref{def_TMDqg_WL}) and (\ref{def_TMDgg_WL}) with Eq.~(18) from \cite{Caucal:2025zkl}.}, where the TMD distributions are computed by first employing the mean-field approximation, then resulting expectation values are related to the dipole operator using the Gaussian approximation.

Finally, let us mention that the results obtained in this manuscript are fully in position space. Those distributions usually enter cross sections within a Fourier transform; thus, for a momentum-space representation of the Gaussian approximation of some of the TMD distributions, we refer to the study \cite{vanHameren:2016ftb}.

\subsection{Recovering TMD distributions from \texorpdfstring{$\Omega$s}{Omegas}}

In this section, we collect the expressions for TMD distributions from ${\Omega}_{qg}$ and ${\Omega}_{gg}$, in the form:
\begin{equation}
\hcalf_{ag} = M^{-1}_{ag} \cdot \hat{\Omega}_{ag} \quad \Longrightarrow \quad 
    \calf_{ag} = M^{-1}_{ag} \cdot \Omega_{ag} , \qquad a = \{q,g\}.
\end{equation}

\subsubsection{\texorpdfstring{$qg$}{qg} sector}
From the expressions given in \req{eq:Omega_to_Fqg}, we have
\begin{subequations}
\begin{align}
   - \Omega_{qg}^{\overline{3}} &= \frac{1}{2C_F} \calf^{(1)}_{qg},  \\
   - \Omega_{qg}^{\overline{15}/6} &= \frac{N_c}{2} \calf^{(2)}_{qg} \pm \thalf \calf^{(3)}_{qg} -\frac{1}{2(N_c\pm 1)}\calf^{(1)}_{qg}.
\end{align}
\end{subequations}
It gives us 
\begin{equation}
\calf_{qg} = M_{qg}^{-1} \cdot \Omega_{qg}
\qquad \Longleftrightarrow \qquad
\begin{pmatrix}
    \calf_{qg}^{(1)} \\[1ex]
    \calf_{qg}^{(2)} \\[1ex]
    \calf_{qg}^{(3)}
\end{pmatrix}
=
\begin{pmatrix}
    - \frac{N_c}{N_c^2-1} & 0 & 0 \\[1ex]
    -\frac{1}{N_c} & -\frac{1}{N_c} & -\frac{1}{N_c} \\[1ex]
    \frac{1}{2N_c} & \frac{1}{2} & - \frac{1}{2}
\end{pmatrix}
\cdot
\begin{pmatrix}
    \Omega^{\overline{3}}_{qg}  \\[1ex]
    \Omega^6_{qg} \\[1ex]
    \Omega^{\overline{15}}_{qg} 
\end{pmatrix}
\label{eq. qg tmd matrix}
\end{equation}
where we can use the expressions for $\Omega^\omega_{qg}$ given in \req{eq:recap_omega_qg},
{ which we recall here:
\begin{equation*}
    \Omega_{qg}^\omega =  \frac{N_c}{2}C_F(\nabla_r \Gamma_r)^2 e^{-C_F \Gamma_r}\ \delta_{\omega = \overline{3}}
    -\frac{K_\omega}{2}(\nabla^2_r {\Gamma}_r ) e^{-\frac{C_\omega + C_F}{2}\Gamma_r}\ \text{sinhc} \left[ \frac{1}{2}(C_\omega - C_F) \Gamma_r\right].
\end{equation*}
The charges $C_\omega$ and dimensions $K_\omega$ are given by
\begin{align*}
&C_{\overline{3}} = C_F = \frac{N_c^2-1}{2N_c}, \qquad &K_{\overline{3}} = K_F = N_c, \\
&C_{\overline{15}/6} = \Nc \pm 1 + C_F,\quad &K_{\overline{15}/6} = \thalf(\Nc \pm2)\Nc(\Nc\mp 1).
\end{align*}}

\subsubsection{\texorpdfstring{$gg$}{gg} sector}
Introduce the vector made of $\Omega_{gg}$'s where we order the irreps according to:
\begin{equation}
    \omega \in \{1,8_a,8_s,10+\overline{10},27,0,8_\ell\}.
\end{equation}
The first six irreps in the list are given by \req{omega_gg_recap}, while the last adjoint $8_\ell$ is given in \req{omega_gg_last_recap}. In the Gaussian approximation, the TMD distributions are given by
\begin{equation}
    \langle \hcalf_{gg} \rangle = M_{gg}^{-1} \cdot \Omega_{gg}
\end{equation}
with
\begin{equation}\label{eq:Fgg_from_Omega}
M_{gg}^{-1} = 
\begin{pmatrix}
 -\frac{N_c^2-1}{N_c^2} & -\frac{1}{2} & -\frac{N_c^2-4}{2 N_c^2} & 0 & 0 & 0 & -\frac{1}{N_c^2} \\
 -\frac{N_c^2-1}{N_c^2} & \frac{1}{2} & -\frac{N_c^2-4}{2 N_c^2} & 0 & 0 & 0 & -\frac{1}{N_c^2} \\
 0 & 0 & 0 & 0 & 0 & 0 & -1 \\
 -(N_c^2-1) & 0 & 0 & 0 & 0 & 0 & 0 \\
 -1 & 1 & -1 & 1 & -1 & -1 & 0 \\
 -\frac{1}{N_c^2} & -\frac{1}{N_c^2} & -\frac{1}{N_c^2} & -\frac{1}{N_c^2} & -\frac{1}{N_c^2} & -\frac{1}{N_c^2} & -\frac{1}{N_c^2} \\
 \frac{1}{N_c^2} & 0 & \frac{2}{N_c^2} & 0 & -\frac{1}{N_c} & \frac{1}{N_c} & -\frac{1}{N_c^2}
\end{pmatrix}
\end{equation}
{
Let us quote again the result of the Gaussian approximation obtained in section \ref{Sec:Gaussian_approx} for the gluon-gluon sector, Eqs.~(\ref{eq:Omega_gg}), (\ref{F4_F3_omega_rel}):
\begin{align*}
\Omega_{gg}^\omega &= \frac{K_A}{2} C_A (\nabla \Gamma_r)^2 e^{-C_A \Gamma_r}\  \delta_{\omega = 8_a}
-\frac{K_\omega}{2}(\nabla^2 {\Gamma}_r )\, e^{-\frac{C_A + C_\omega}{2}\Gamma_r}\ \text{sinhc} \left[ \frac{1}{2}(C_\omega - C_A) \Gamma_r\right],
\end{align*}
for $ \omega \in \{1,8_a,8_s,10+\overline{10},27,0\}$ and
\begin{equation*}
  \Omega_{gg}^{8\ell}= -(N_c^2-1) \Omega_{gg}^{1} .
  \label{omega_gg_last_recap}
  \end{equation*}
The charges $C_\omega$ and dimensions $K_\omega$ are given by:
\begin{align*}
    &C_1 =0,\quad 
    C_{10} = C_{\overline{10}} = 2N_c,\quad 
    C_{8} = C_A = N_c,\quad 
    C_{27/0} = 2(N_c \pm 1).\\
    &K_1 = 1, \quad K_{8a/8s} = K_A = \Nc^2-1, \\
    &K_{10/\overline{10}} = \frac{(\Nc^2-1)(\Nc^2-4)}{4}, \quad
    K_{27/0} = \Nc^2 \frac{(\Nc \mp 1)(\Nc\pm3)}{4}.
\end{align*}}

\subsection{Large-\texorpdfstring{$N_c$}{Nc} limit of the Gaussian approximation of TMD distributions}\label{sec:largeNc_of_Gaussian_approx}

Provided the transformation \req{eqs:Scaling_generic}, most $\Omega^{\omega}/K_{\omega}$ exhibit exact scaling, which means that they are $N_c$ independent when expressed in terms of $\textrm{x}$. 
As shown in {fig.~\ref{fig:Omega_large_Nc_qg} and {fig.~\ref{fig:Omega_large_Nc_gg} the non-scaling ones, $\Omega_{qg}^{6,\overline{15}}$ and $\Omega_{gg}^{27,0}$, have a mild $N_c$-dependence, and in particular, they share a common large-$N_c$ limit which we will make use of, i.e.
\begin{align}
    \lim_{N_c \rightarrow \infty} \Omega_{qg}^{\overline{3}}(\textrm{x})/K_{\overline{3}} &\equiv \beta_{\overline{3}}(\textrm{x}) \\
    \Omega_{qg}^{6}(\textrm{x})/K_6 &= \beta_6(\textrm{x}) + \frac{1}{C_F} \gamma_6(\textrm{x}) + \dots\\
    \Omega_{qg}^{\overline{15}}(\textrm{x})/K_{\overline{15}} &= \beta_6(\textrm{x}) - \frac{1}{C_F} \gamma_6(\textrm{x}) + \dots
\end{align}
with $\textrm{x}$ given by \req{Nc_scaled_distance}. We used the fact that $\beta_{\overline{15}} = \beta_6$ and $\gamma_{\overline{15}} = - \gamma_6$, which was anticipated in Eq.~\eqref{eq: scaling of 6 and 15} and is derived explicitly below. Note that each $\beta_\omega$ and $\gamma_\omega$ now scales according to \req{eqs:Scaling_generic}, the $N_c$ dependence is only carried by the pre-factors $(1/C_F)^n$.  Dropping the dependence on the rescaled separation $\textrm{x}$, Eq.~\eqref{eq. qg tmd matrix} can be written as
\begin{equation}
\begin{pmatrix}
    \calf_{qg}^{(1)} \\[1ex]
    \calf_{qg}^{(2)} \\[1ex]
    \calf_{qg}^{(3)}
\end{pmatrix}
=
\left( M^{-1}_{qg} \cdot K_\omega \right) 
\cdot
\begin{pmatrix}
    \beta_{\overline{3}}  \\[1ex]
    \beta_6 + \frac{1}{C_F} \gamma_6 + \dots \\[1ex]
    \beta_6 - \frac{1}{C_F} \gamma_6 + \dots
\end{pmatrix} 
=
\begin{pmatrix}
    - N_c^2 \ \beta_{\overline{3}} + \calo(1) \\[1ex]
    - N_c^2 \ \beta_6 + \calo(1) \\[1ex]
    - N_c^2 \ \big( \frac{1}{2} \beta_6 - \gamma_6 \big) + \calo(1)
\end{pmatrix}
\end{equation}
where we used $K_\omega$ as the diagonal matrix of dimensions. Indeed, the corrections in the above set of equations are all of the order $\calo(1)$ instead of the expected $\calo(N_c)$.  
In the large-$N_c$ limit, this allows us to define and calculate
\begin{equation}\label{eq:largeNc-limit-qg}
\begin{pmatrix}
    \calf_{qg}^{(1)} / N_c^2 \\[1ex]
    \calf_{qg}^{(2)} / N_c^2 \\[1ex]
    \calf_{qg}^{(3)} / N_c^2 
\end{pmatrix}
=
\begin{pmatrix}
    - \beta_{\overline{3}}  \\[1ex]
    - \beta_6 \\[1ex]
    - \frac{1}{2} \beta_6 + \gamma_6 
\end{pmatrix} + \calo(1/N_c^2)
\end{equation}
Proceeding similarly in the gluon-gluon case we have,
\begin{align}
    \lim_{N_c \rightarrow \infty} \Omega_{gg}^{\omega}(\textrm{x})/K_{\omega} &\equiv \beta_{\omega}(\textrm{x}), \quad \omega \in \{1,8a,8s,8l,10+\overline{10}\} \\
    \Omega_{gg}^{0}(\textrm{x})/K_{0} &= \beta_0(\textrm{x}) + \frac{1}{C_A} \gamma_0(\textrm{x}) + \frac{1}{C_A^2} \delta_0(\textrm{x}) + \dots, \\
    \Omega_{gg}^{27}(\textrm{x})/K_{27} &= \beta_0(\textrm{x}) - \frac{1}{C_A} \gamma_0(\textrm{x}) + \frac{1}{C_A^2}\delta_0(\textrm{x}) + \dots
\end{align}
where similarly to the quark-gluon case we find $\beta_{27} = \beta_0$, $\gamma_{27} = -\gamma_{0}$, and we also have $\delta_{27} = \delta_{0}$. In the gluon-gluon case, we additionally have\footnote{It is because of this constrain that we need to expand to order $\calo(1/C_A^2)$ in the expression for $\Omega^\omega/K_\omega$ with $\omega = \{27,0\}$, in order to find the proper large-$N_c$ limit of the Gaussian approximation for $\calf^{(5)}_{gg}$.} $\beta_{10+\overline{10}} =\beta_0$, and from the definition $\gamma_{10+\overline{10}} = 0$ since $\Omega_{gg}^{10+\overline{10}}(\textrm{x})/K_{10+\overline{10}}$ scales exactly.
We can work out the large-$N_c$ limit of the TMD distributions as
\begin{equation}\label{eq:largeNc-limit-gg}
\begin{pmatrix}
    \calf_{gg}^{(1)} / N_c^2 \\[1ex]
    \calf_{gg}^{(2)} / N_c^2 \\[1ex]
    \calf_{gg}^{(3)} / N_c^2 \\[1ex]
    \calf_{gg}^{(4)} / N_c^2 \\[1ex]
    \calf_{gg}^{(5)} / N_c^2 \\[1ex]
    \calf_{gg}^{(6)} / N_c^2 \\[1ex]
    \calf_{gg}^{(7)} / N_c^2 
\end{pmatrix}
=
\begin{pmatrix}
    - \frac{1}{2} \beta_{8s} - \frac{1}{2} \beta_{8a} \\[1ex]
    - \frac{1}{2} \beta_{8s} + \frac{1}{2} \beta_{8a} \\[1ex]
    - \beta_{8l} \\[1ex]
    - \beta_{1} \\[1ex]
    - \beta_{8s} - \beta_{0} + \beta_{8a} + \gamma_{0} - {\thalf \delta_0}\\[1ex]
    - \beta_{0} \\[1ex]
    - \beta_{0} + \frac{1}{2} \gamma_{0} 
\end{pmatrix} + \calo(1/N_c^2)
\end{equation}

The explicit form of $\beta_6$ and $\gamma_6$ can be worked out by expansion. 
In \req{eq:Omega_qg_B}, the two charges involved for each of the two irreps $\{\overline{15},6\}$ can be written as
\begin{align*}
    (C_\omega \pm C_F) &= C_F + N_c + \eta_\omega \pm C_F \\
    &= C_F\left[ (3\pm1) + \frac{1}{C_F}\eta_\omega + \calo(1/C_F^2) \right]
\end{align*}
with $\eta_\omega = \pm$ for $\omega = \{ \overline{15},6\}$.
For those two irreps, we can write in the Gaussian approximation
\begin{align}
    \frac{\Omega^{\omega}_{qg}}{K_\omega} &= -\frac{1}{2}(\nabla^2 {\Gamma}_r )\, e^{-\frac{C_F + C_\omega}{2}\Gamma_r}\ \text{sinhc} \left[ \frac{1}{2}(C_\omega - C_F) \Gamma_r\right] \\
    &= -\frac{1}{2}(\nabla^2 {\Gamma}_r )\, e^{-2C_F\Gamma_r}\left\{ \text{sinhc} \left( C_F \Gamma_r\right) \phantom{\frac{\Gamma}{\Gamma}} \right.\notag \\
    &\left.+\frac{\eta_\omega}{C_F} \frac{\left[C_F \Gamma_r\, \text{cosh} (C_F \Gamma_r) -(1+C_F\Gamma_r)\, \text{sinh}(C_F \Gamma_r) \right]}{2\,C_F \Gamma_r} \right\} + \calo(1/C_F^2) \\
    &\equiv \beta_{\omega} + {\frac{1}{C_F}} \gamma_{\omega} + \calo(1/C_F^2).
\end{align}
Notice that in the very last line, $\beta_{\omega}$ and $\gamma_{\omega}$ will scale exactly provided we perform the replacement described by \req{eqs:Scaling_generic}.

Similarly, in the gluon-gluon sector, we have the case of irreps $\{ 27,0\}$ with $\eta_\omega = \pm$ for $\omega = \{ 27,0\}$: 
\begin{align*}
(C_\omega \pm C_A) = 2N_c + 2\eta_\omega \pm N_c = N_c\left[ (2\pm 1) + 2\frac{\eta_\omega}{N_c} \right]
\end{align*}
thus
\begin{align}
    \frac{\Omega^{\omega}_{gg}}{K_\omega} &= -\frac{1}{2}(\nabla^2 {\Gamma}_r )\, e^{-\frac{C_A + C_\omega}{2}\Gamma_r}\ \text{sinhc} \left[ \frac{1}{2}(C_\omega - C_A) \Gamma_r\right] \\
    &= -\frac{1}{2}(\nabla^2 {\Gamma}_r )\, e^{-\frac{3}{2}N_c\Gamma_r}\left\{ \text{sinhc} \left( \tfrac{N_c}{2} \Gamma_r\right) \phantom{\frac{\Gamma}{\Gamma}} \right.\notag \\
    &+\frac{2\eta_\omega}{N_c} \frac{\left[\tfrac{N_c}{2} \Gamma_r\, \text{cosh} (\tfrac{N_c}{2} \Gamma_r) -(1+\tfrac{N_c}{2} \Gamma_r)\, \text{sinh}(\tfrac{N_c}{2} \Gamma_r) \right]}{\tfrac{N_c}{2} \Gamma_r} \notag \\
    &-\left. \frac{4\eta^2_\omega}{N_c^2} 
    \frac{ \tfrac{N_c}{2}\Gamma_r (1+\tfrac{N_c}{2}\Gamma_r)\, e^{-\tfrac{N_c}{2}\Gamma_r } - \text{sinh}\left( \tfrac{N_c}{2}\Gamma_r\right)}{\tfrac{N_c}{2} \Gamma_r}
    \right\} + \calo(1/N_c^3) \\
    &\equiv \beta_{\omega} + {\frac{1}{C_A}} \gamma_{\omega} + \frac{1}{C_A^2} \delta_\omega +\calo(1/C_A^3).
\end{align}

The limiting behavior obtained in this section agrees with the results obtained within the mean-field approximation, provided the fundamental dipole is written as $S_r = e^{-C_F \Gamma_r}$. Those expressions can be found in Refs.~\cite{Caucal:2025zkl}. The agreement between the to approximations can be interpreted as a consequence of the large $N_c$-limit for \textit{reasonable operators} \cite{Yaffe:1981vf}.
We did not find in the literature the mean-field approximation for $\calf^{(3)}_{qg}$ and $\calf^{(5)}_{gg}$, $\calf^{(7)}_{gg}$, however, we can anticipate the results (using the large-$N_c$ limit of the Gaussian approximation) to be given by \req{eq:largeNc-limit-qg} and \req{eq:largeNc-limit-gg}.

\section{Numerical results for TMD distributions}\label{sec:num_result_tmds}

In this section, we will compare the TMD distributions obtained from simulations for $N_c \in \{2,3,4,5\}$, to the result of the Gaussian approximation given in {Sec.~\ref{Sec:Gaussian_approx}}. We also plot the results of the mean field approximation of \cite{Caucal:2025zkl}, that agree with the large-$N_c$ limit of the Gaussian approximation. 

\subsection{TMD distributions for \texorpdfstring{$SU(3)$}{SU(3)}}

We start by presenting TMD distributions obtained for $N_c=3$. In Fig.~\ref{fig. tmd qg} we show quark-gluon TMD distributions and in Fig.~\ref{fig. tmd gg} gluon-gluon TMD distributions. The insets present a linear scale on the vertical axis.
Overall, we observe very good agreement for all TMD distributions between the massive MV model (magenta points) and the Gaussian approximation (black curve).\footnote{The evaluation of the TMD $\calf^{(3/4)}_{gg}$ agrees with the Gaussian approximation computed in \cite{Dominguez:2011wm}.}
When it comes to the mean-field approximation expressions (green dashed curves), the agreement with all-$N_c$ results is good at small distances, $\textrm{x} \lesssim 1$fm. 
At larger $\textrm{x}$, the mean-field approximation curve follows data points only when the TMD is proportional to one $\Omega$ -- this is the case for $\calf^{(1)}_{qg}$ and we anticipate it to also be the case for $\calf^{(3)}_{gg} = \calf^{(4)}_{gg}$. 
In other cases, the mean-field approximation fails to reproduce the correct tail of TMD distributions: for $\calf^{(2)}_{qg}$ it misses the node; for $\calf^{(1)}_{gg}$ it has a false node; for $\calf^{(2)}_{gg}$ and $\calf^{(6)}_{gg}$ the tail is too steep in $\textrm{x}$. 
Finally, note that $\calf^{(3)}_{qg}$, $\calf^{(5)}_{gg}$ and $\calf^{(7)}_{gg}$ were not required in \cite{Caucal:2025zkl} and thus we do not show mean-field result for those; however, we do not expect the large-distance tail to be reproduced in this approximation as it should map to the large-$N_c$ limit of the Gaussian approximation worked out in {sec.~\ref{sec:largeNc_of_Gaussian_approx}}, where said tail is $1/N_c$ suppressed.

Finally, let us note that some of the TMD distributions show numerical stability within a range of $\textrm{x}$ extending beyond $3\, \text{fm}$, which is outside the range of the dipole fit performed in {sec.~\ref{sec:dipole_fit}}.  
This is explained in more detail in appendix \ref{App:Finite_volune}.

\begin{figure}[h]
    \includegraphics[width=0.48\textwidth]{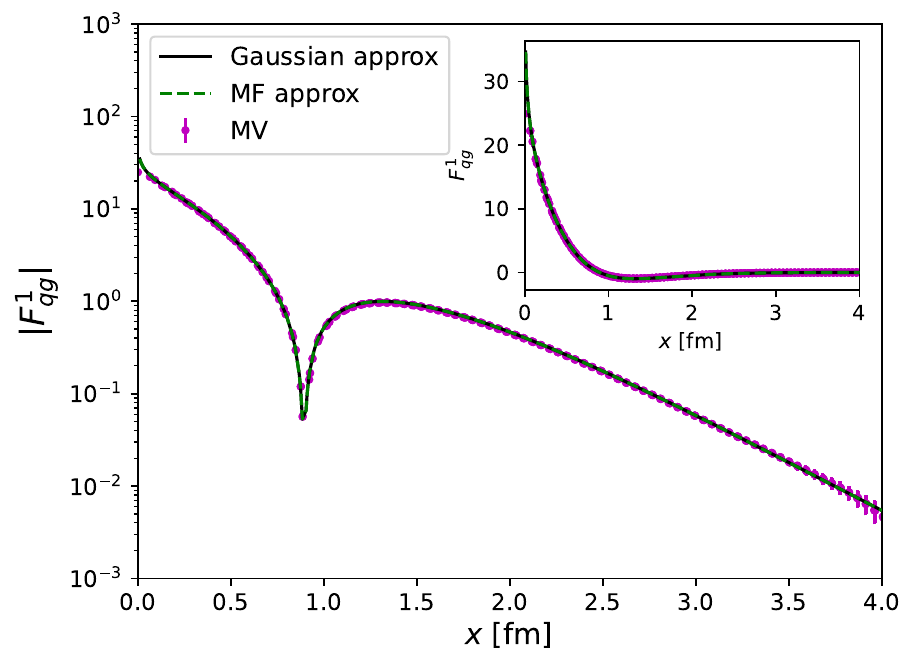}
    \hfill
    \includegraphics[width=0.48\textwidth]{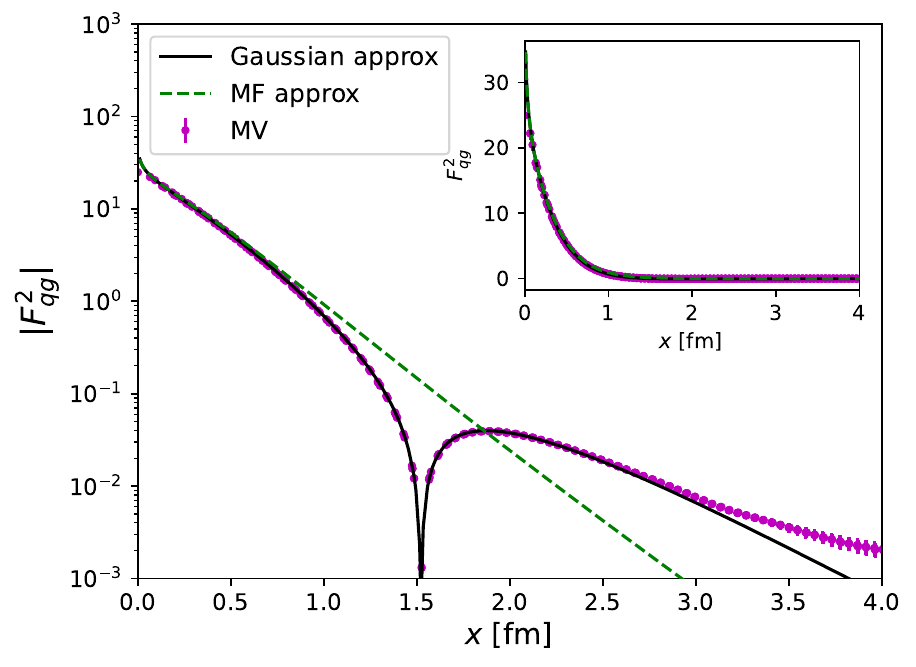}
    \includegraphics[width=0.48\textwidth]{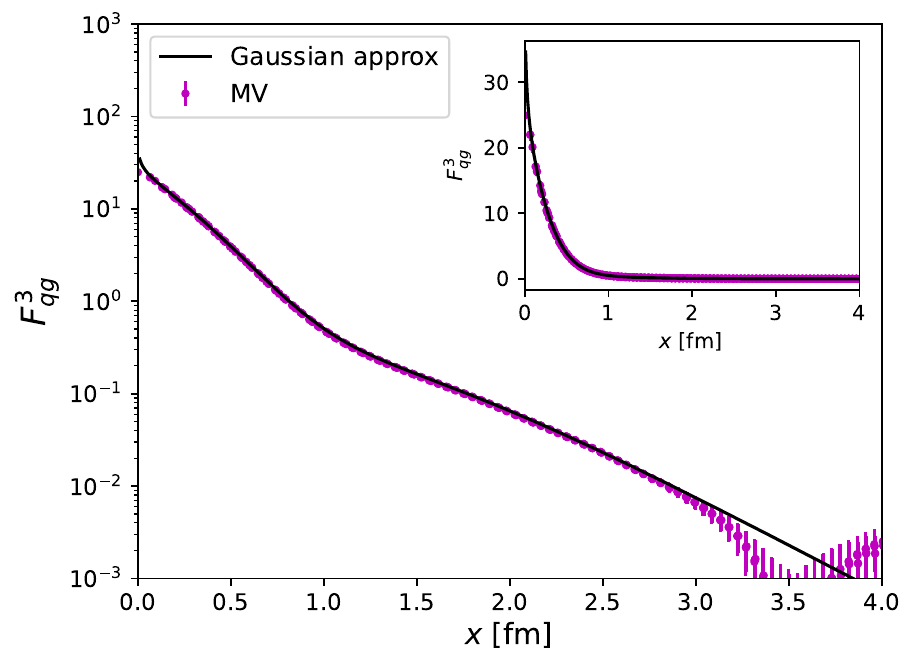}
    \caption{
    Evaluation of quark-gluon TMD distributions for $SU(3)$. The magenta data points are obtained from MV-model simulations. The black plain lines show the computed Gaussian approximations. The green dashed lines show results from \cite{Caucal:2025zkl} where the mean-field approximation has been used. The MF curve for $\calf^{(1)}_{qg}$ agrees with the Gaussian approximation and the data because of the exact scaling of this distribution.
    \label{fig. tmd qg}}
\end{figure}

\begin{figure}
    \includegraphics[width=0.48\textwidth]{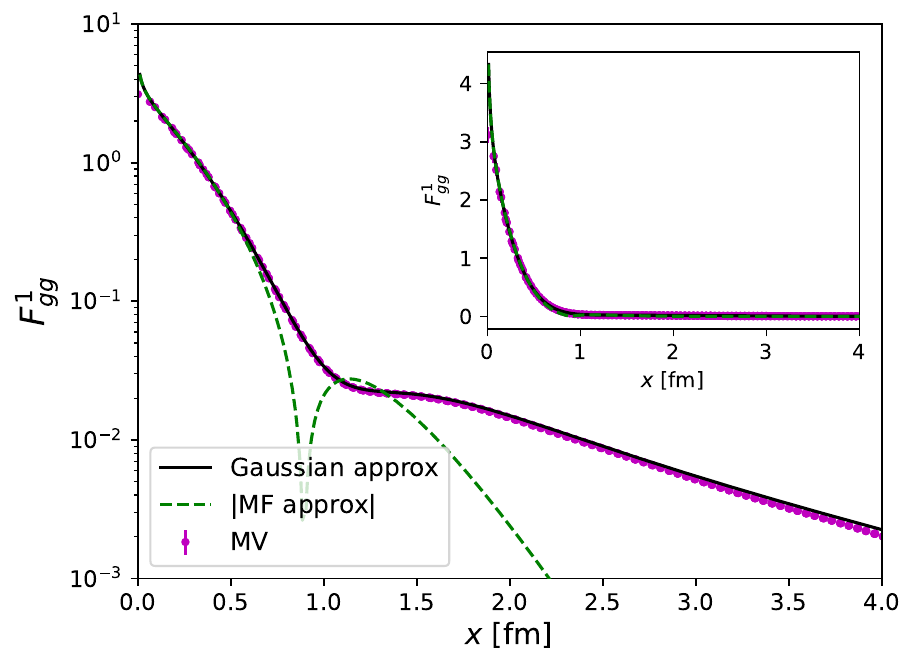}
    \hfill
    \includegraphics[width=0.48\textwidth]{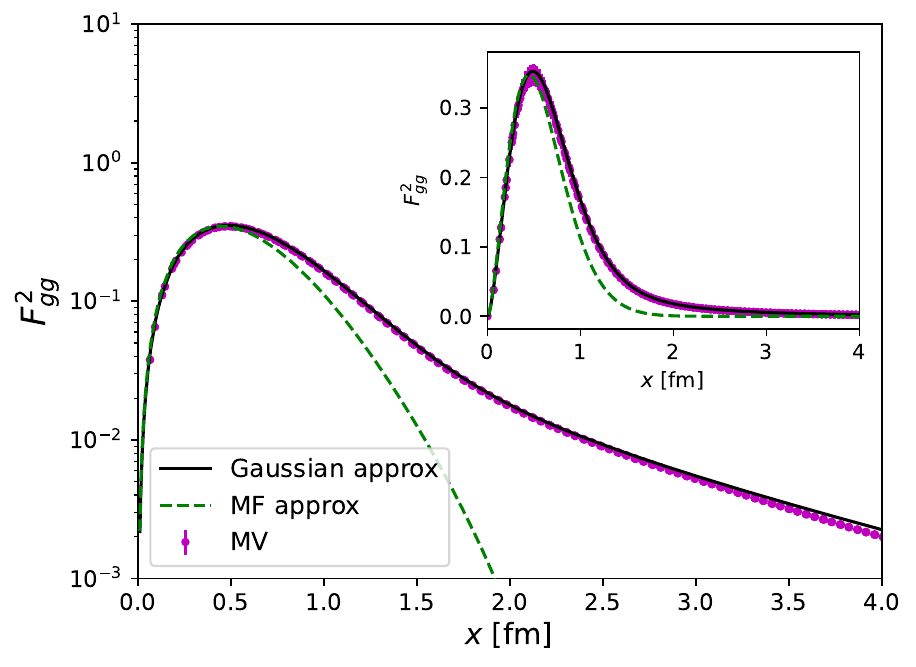}
    \includegraphics[width=0.48\textwidth]{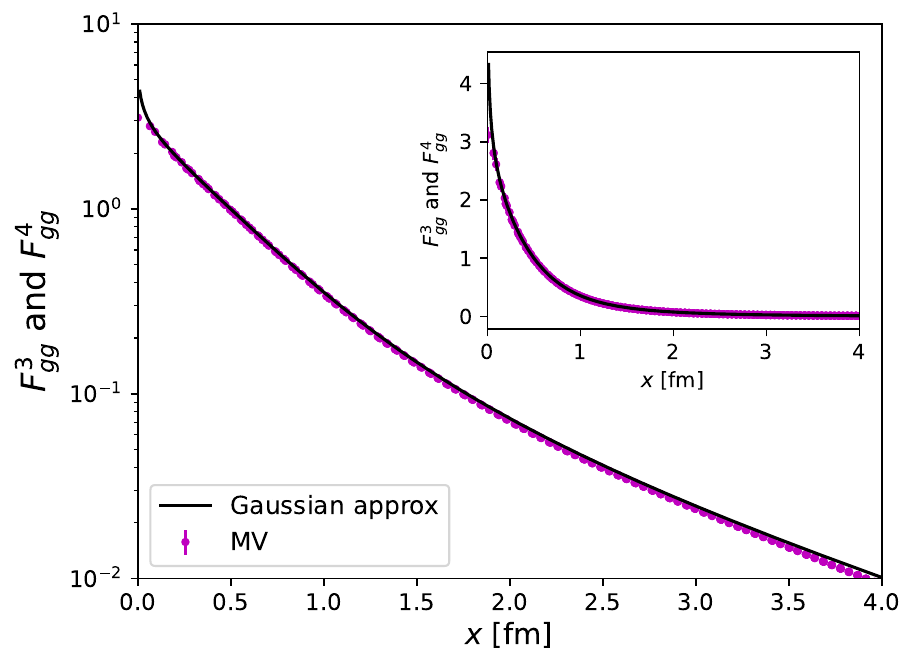}
    \hfill
    \includegraphics[width=0.48\textwidth]{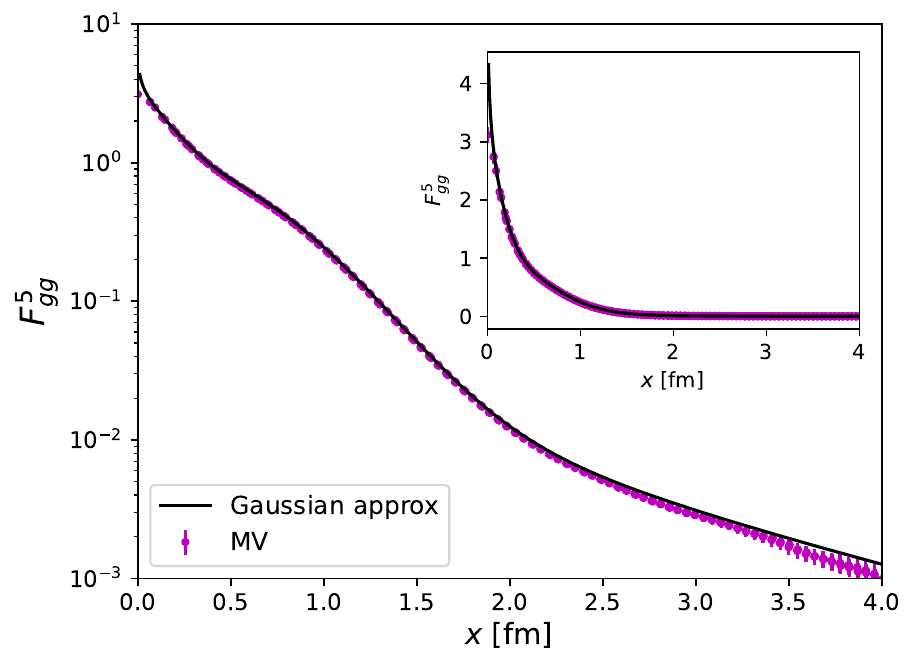}
    \includegraphics[width=0.48\textwidth]{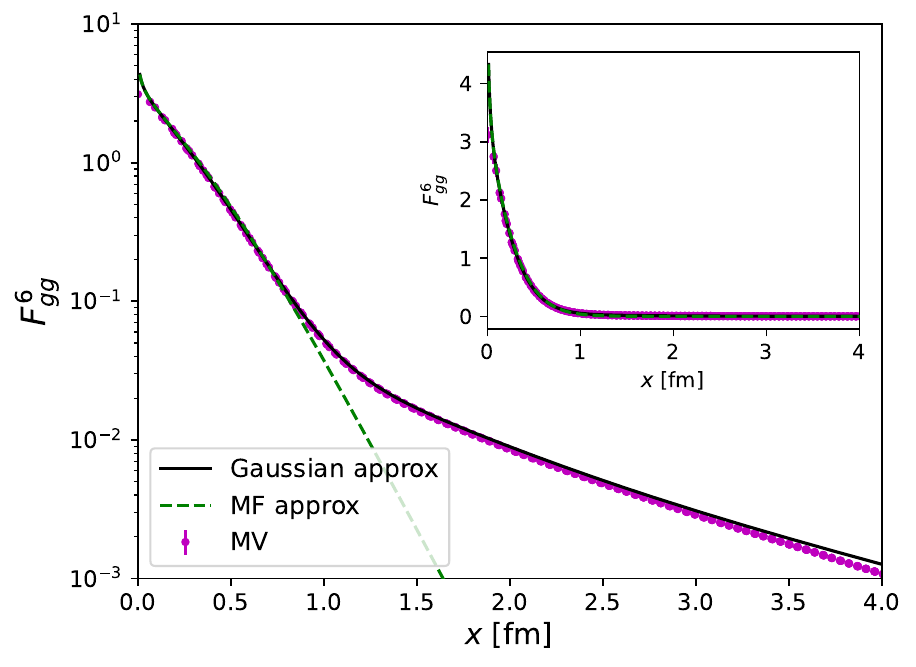}
    \hfill
    \includegraphics[width=0.48\textwidth]{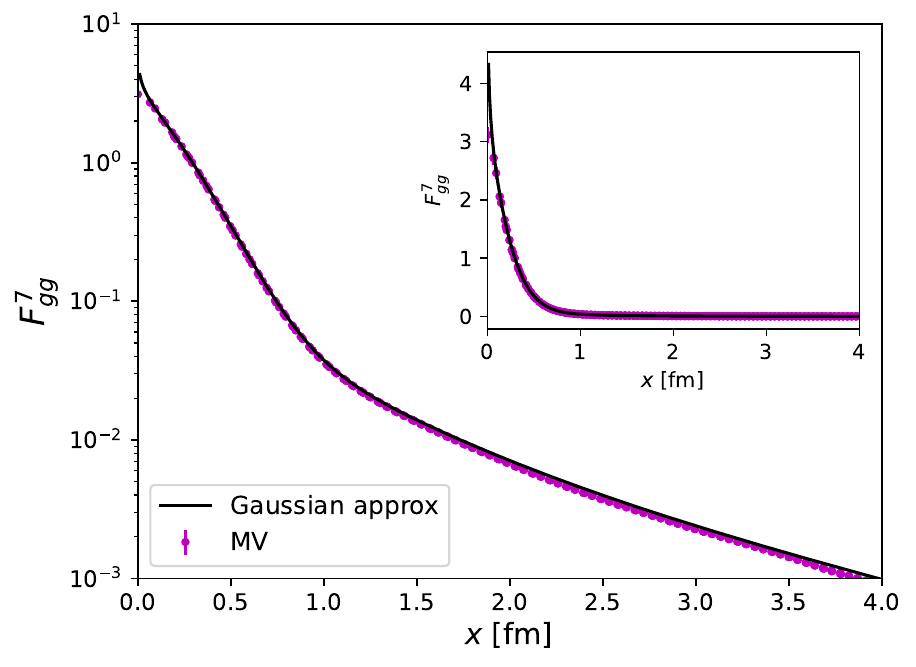}
    \caption{
    Evaluation of gluon-gluon TMD distributions for $SU(3)$. The magenta data points are obtained from MV-model simulations. The black plain lines show the computed Gaussian approximations. The green dashed lines show results from \cite{Caucal:2025zkl} where the mean-field approximation has been used. 
    \label{fig. tmd gg}}
\end{figure}

\subsection{TMD distributions at different \texorpdfstring{$N_c$}{Nc}}
Next, we will examine the dependence of the TMD distributions on $N_c$ for $N_c \in \{2,3,4,5\}$.
While previously shown $\Omega^\omega/K_\omega$ scale (exact or mild deviation to scaling) according to \req{eqs:Scaling_generic}, the TMD distributions in both the quark-gluon {Fig.~\ref{fig. tmd separate qg}} and in the gluon-gluon {Fig.~\ref{fig. tmd separate gg}} sectors show structures explicitly dependent on $N_c$. 
This is a consequence of the TMD distributions being linear combinations of $\Omega$s, with $N_c$-dependent coefficients, given by $M^{-1}_{ag}$.

In the quark-gluon sector, we observe exact $C_F$-scaling for $\calf_{qg}^{(1)} / (N_c^2-1)$. This is a consequence of this TMD being proportional to $\Omega_{qg}^{\overline{3}}/K_{\overline{3}}$.
The other two distributions, $\calf_{qg}^{(2)}$ and $\calf_{qg}^{(3)}$, can be normalized by the same factor $1/(N_c^2-1)$, such that in the small separation region $\textrm{x} < 0.5\, \text{fm}$ the resulting distributions are compatible for $N_c \in \{ 2,3,4,5 \}$.\footnote{This choice of normalization is also motivated by expressions in the Gaussian approximation using the GBW-model for the dipole, where $\Gamma_r^{GBW} \propto r^2$, instead of the functional form given in \req{eq:fit_func_form}. In such case, the $\textrm{x} \rightarrow 0$ limit is finite and the normalization makes the limit invariant of the $N_c$. }
Notably, in this region we see that the data points of $\calf_{qg}^{(2)}$, for all probed $N_c$,  agree with both the mean-field approximation (dashed pink curve) and the Gaussian approximation (plain black line). 
The tail behavior of the TMD distributions $\calf_{qg}^{(2/3)}$ show an $N_c$-ordering which we can describe well using the Gaussian approximation. 
It is clear that the data points favor the Gaussian approximation in the tail region $\textrm{x} > 1\, \text{fm}$ for values of $N_c$ close to the phenomenological point $N_c = 3$. 
The tail is dominated by a contribution which is explicitly $1/N_c$ suppressed and thus cannot be captured by the strict large-$N_c$ limit on which the mean-field approximation relies.
We conclude that, if the given cross-section is sensitive only to smaller separation scales, one can safely use mean-field approximation results. However, if the tail of the distribution is relevant, using the Gaussian approximation is necessary.
An interesting feature of $\calf_{qg}^{(2)}$ is the node present in the data points and the Gaussian approximation, which does not appear in the mean-field approximation.
As we increase $N_c$, we see that the node drifts toward larger values of $\textrm{x}$. In the limit of $N_c \rightarrow \infty$, the node disappears.

Similar features to those observed in the quark-gluon case can be seen in the gluon-gluon sector. 
The distribution $\calf_{gg}^{(3)}/N_c^2$ scales exactly with $N_c$. This follows from $\calf_{gg}^{(4)}$ being directly proportional to $\Omega^1_{gg}$, and the time-reversal symmetry of the setup which sets $\calf_{gg}^{(3)} = \calf_{gg}^{(4)}$.
In the small separation region $\textrm{x} < 0.5\, \text{fm}$, there is agreement between the data points from the simulations using the MV model, the Gaussian approximation (black curves), and the mean-field approximation (dashed pink curves, when available).
The tail of $\calf_{gg}^{(i)}$ for $i \in \{1,2,5,6,7\}$ shows an ordering in $N_c$ and we can see that the tails decay as the value of $N_c$ increases. 
Finally, the distribution $\calf_{gg}^{(1)}$ has an interesting structure as we increase $N_c$. 
For values of $N_c < 5$ the data from the MV simulation and the Gaussian approximation yield a positive distribution over the entire range of $\textrm{x}$, which contradicts the mean-field approximation, which has a single node. This node is in the range $\textrm{x} \sim 1\,\text{fm}$ (this is of the order of the saturation scale).
It is only for $N_c = 5$ that we see the appearance of two nodes. 
In the Gaussian approximation, we can check that the first node is close to the node obtained in the mean-field approximation. The second node drifts toward larger values of $\textrm{x}$ as we increase $N_c >5$.
In the limit $N_c \rightarrow \infty$, this node disappears, leaving only the single node obtained in the mean-field approximation. This node behaves similarly to the node of $\calf_{qg}^{(2)}$.

\begin{figure}[h]
\includegraphics[width=0.35\textwidth, angle=270]{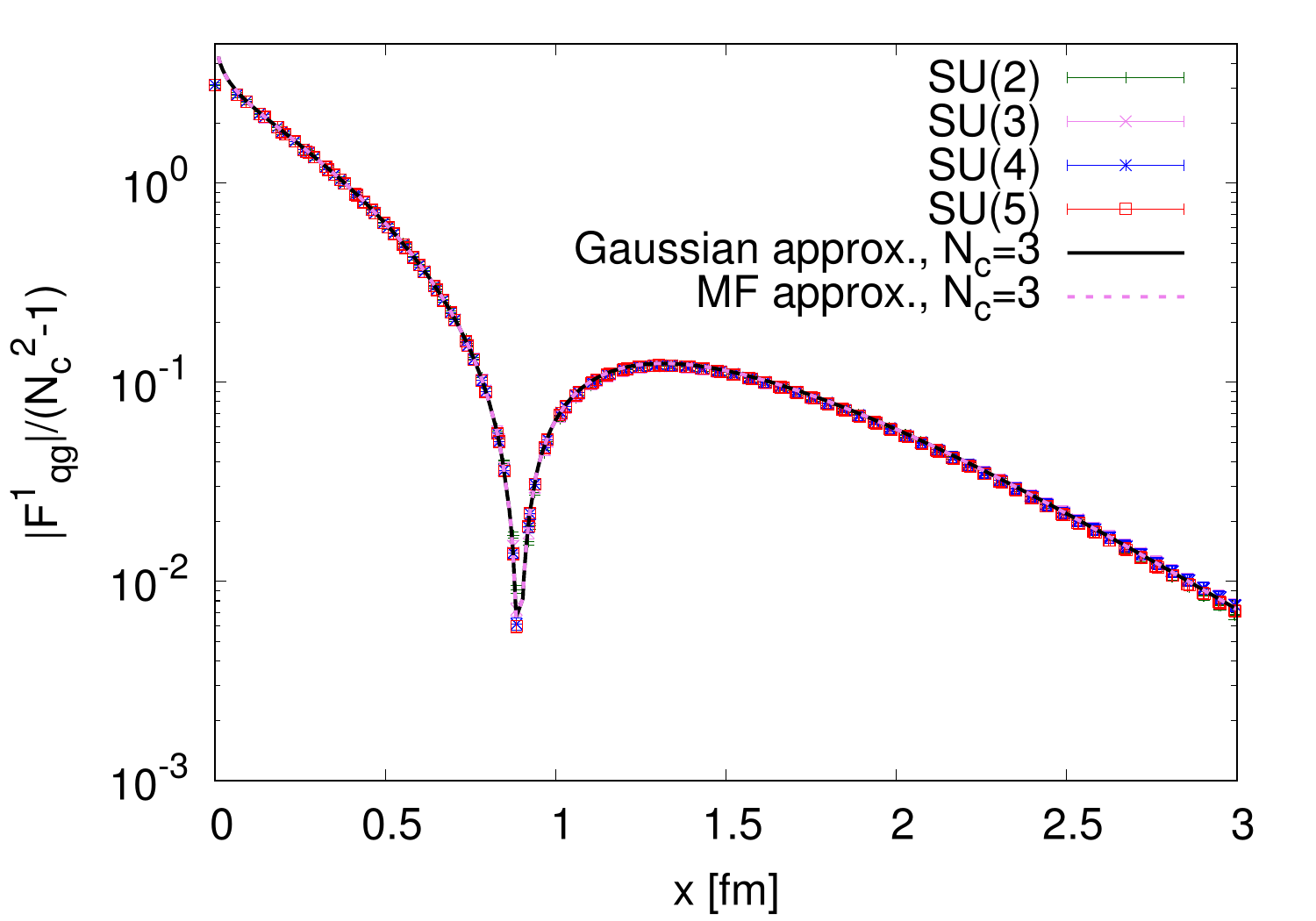}
\includegraphics[width=0.35\textwidth, angle=270]{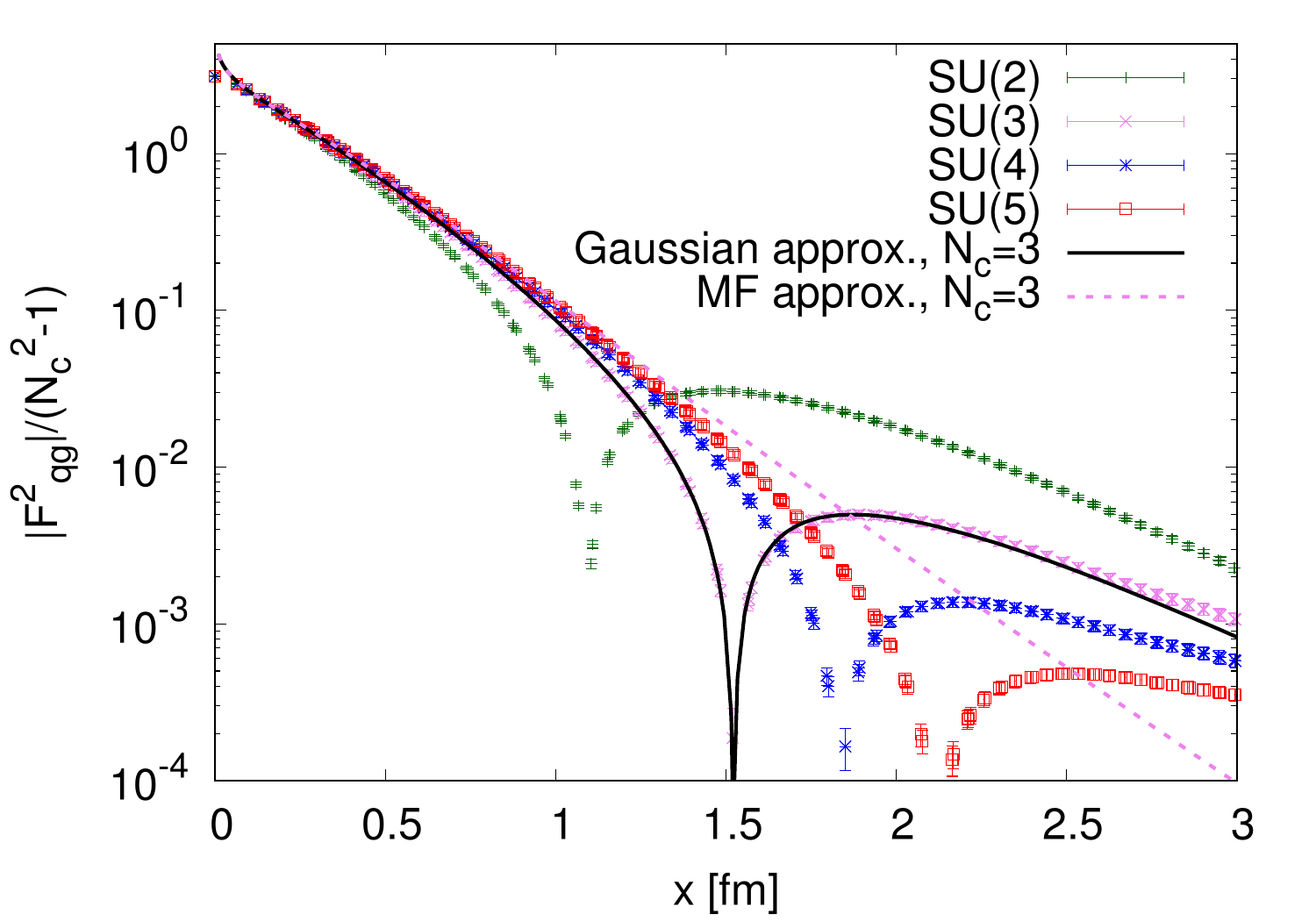}

\includegraphics[width=0.35\textwidth, angle=270]{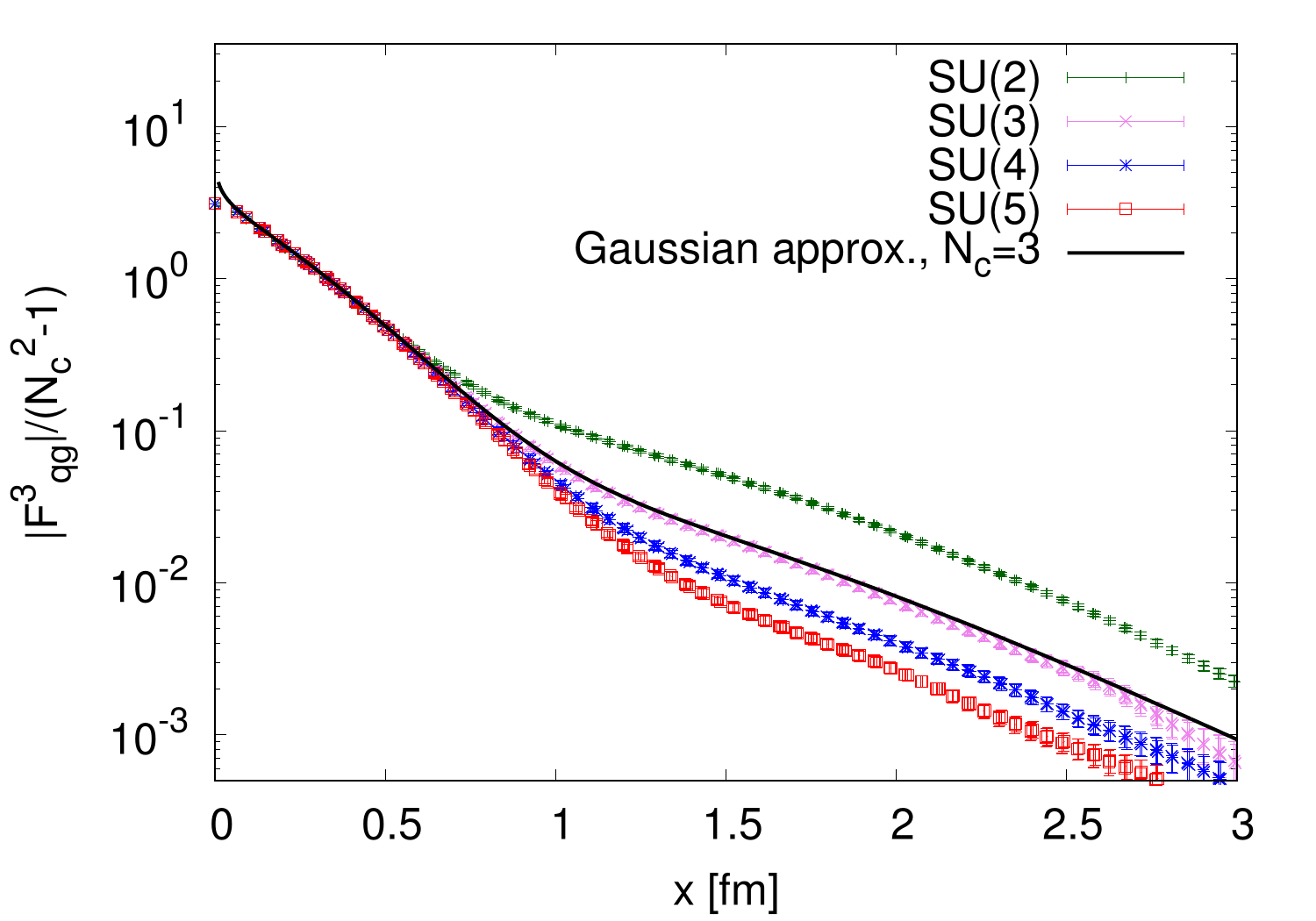}
\caption{$(N^2_c-1)$-normalized quark-gluon TMD distributions for different values of $N_c$ together with the Gaussian and mean-field approximations for $N_c=3$. Note that $\calf^{(1)}_{qg}/(N_c^2-1)$ is equivalent to $\Omega^3_{qg}/K_3$ which exhibits exact scaling with $N_c$ in the entire range of $\rm x$. As shown in Eq.~\eqref{eq:largeNc-limit-qg}, the convergence towards the mean-field approximation is of $\calo(1/N_c^2)$. 
\label{fig. tmd separate qg}}
\end{figure}

\begin{figure}[h]
    \includegraphics[width=0.35\textwidth, angle=270]{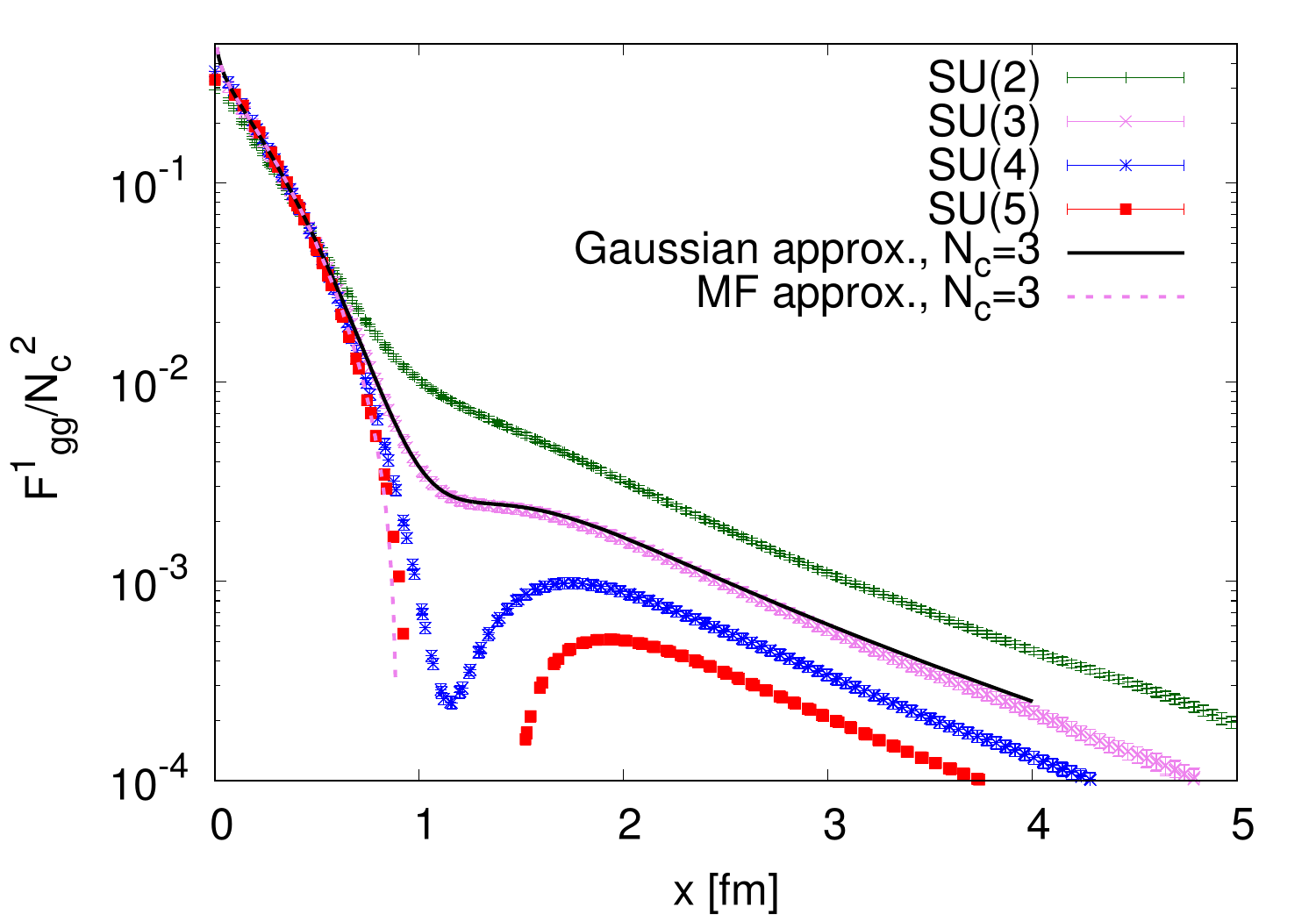}
    \includegraphics[width=0.35\textwidth, angle=270]{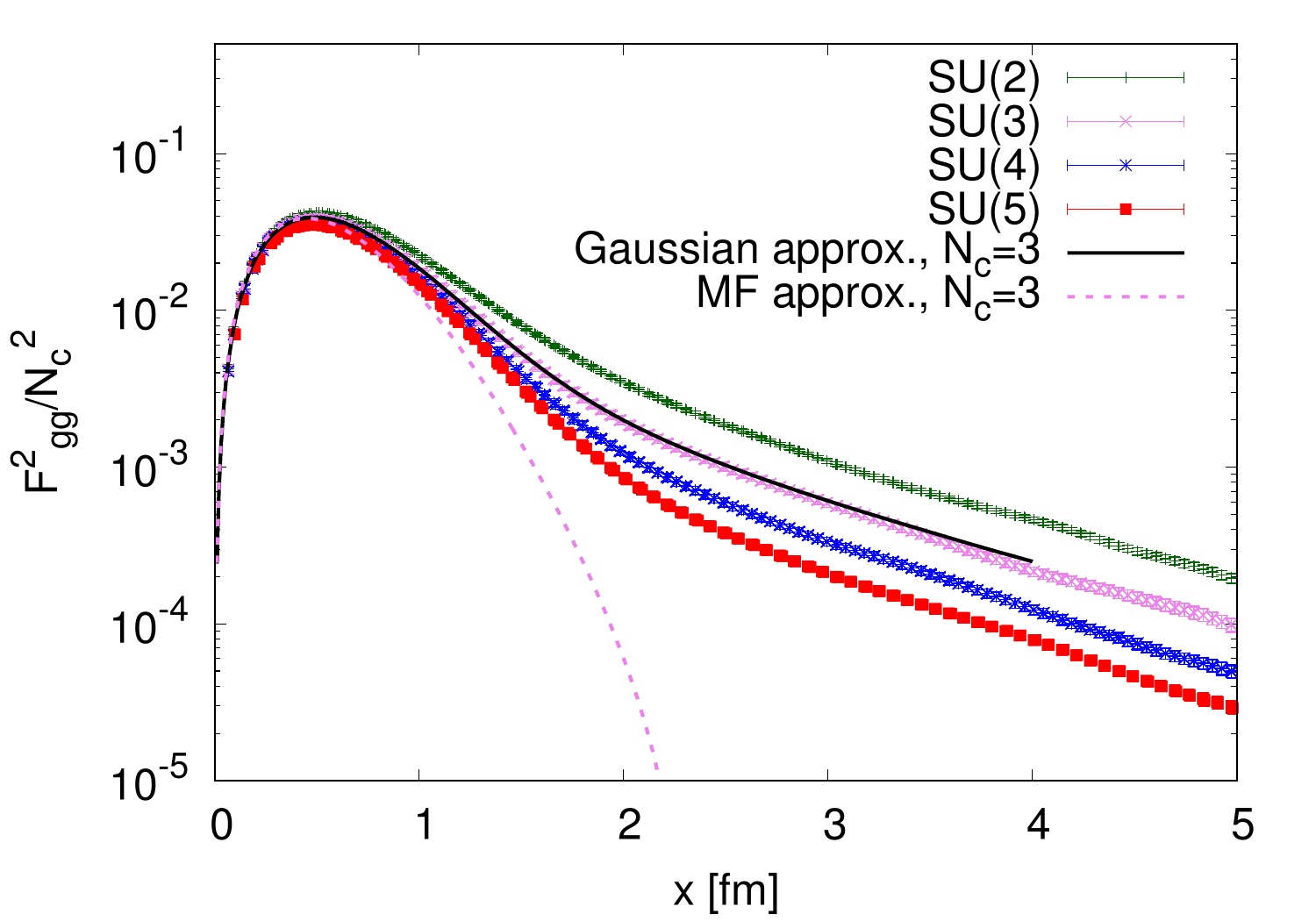}
    \includegraphics[width=0.35\textwidth, angle=270]{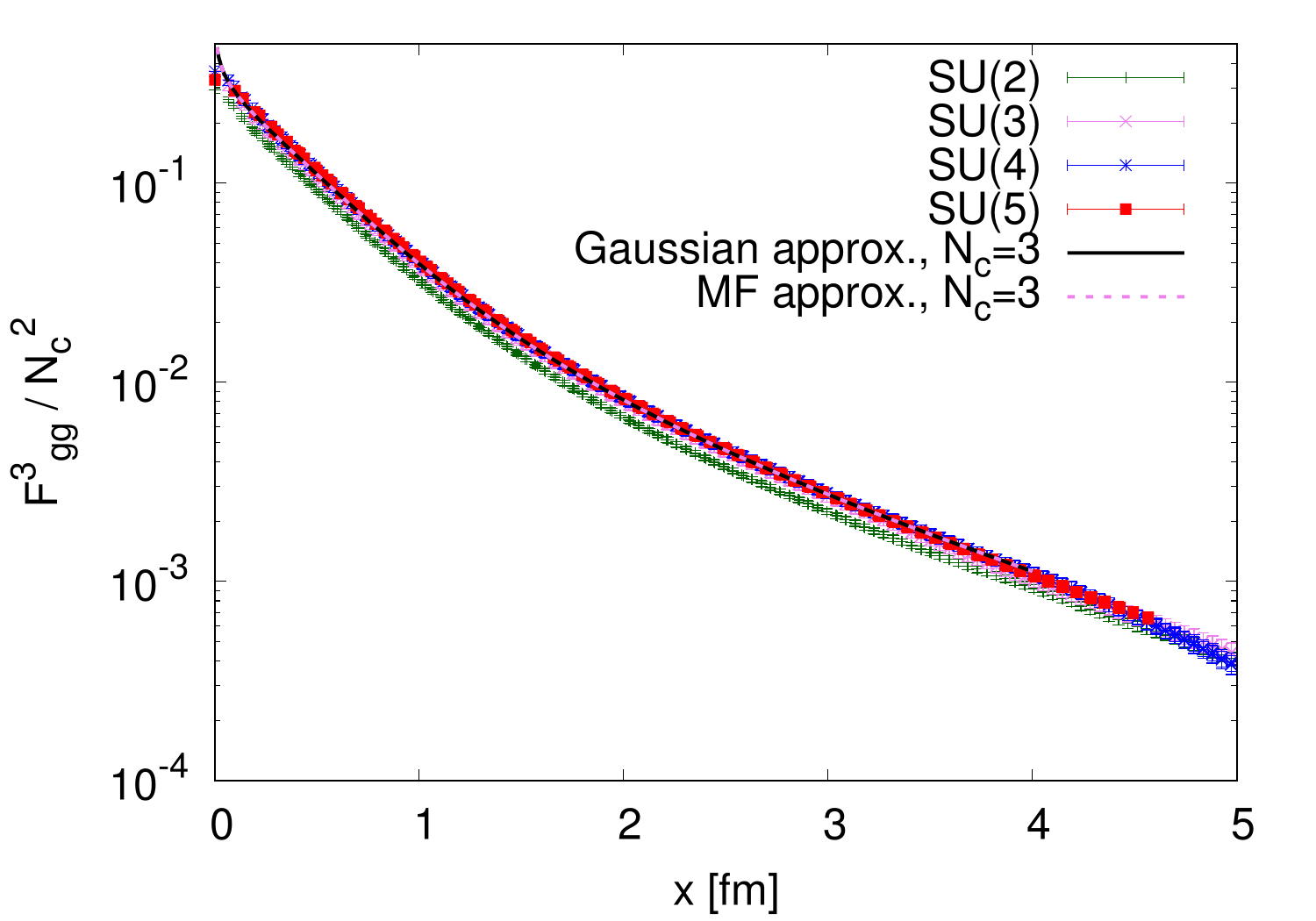}
    \includegraphics[width=0.35\textwidth, angle=270]{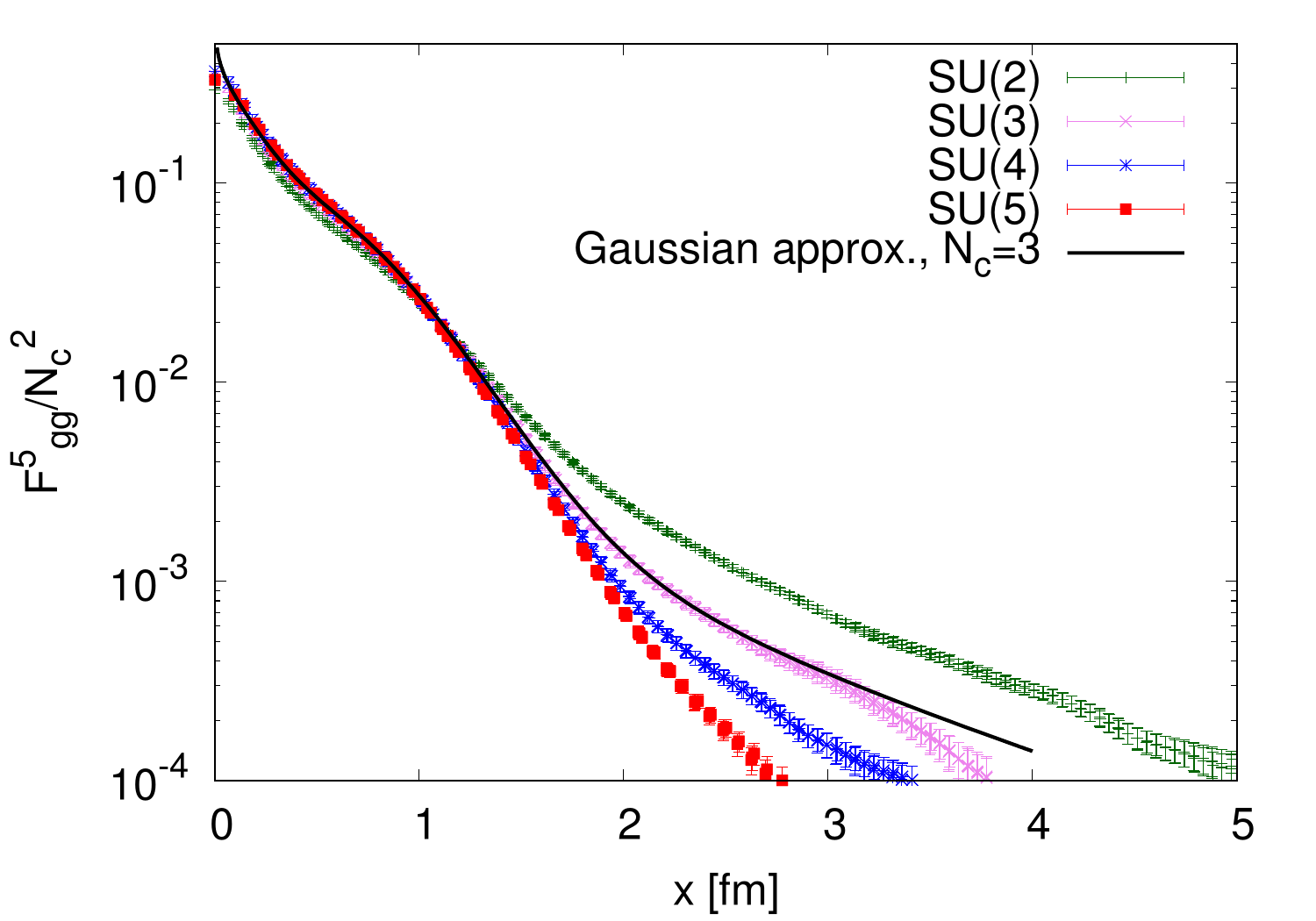}
    \includegraphics[width=0.35\textwidth, angle=270]{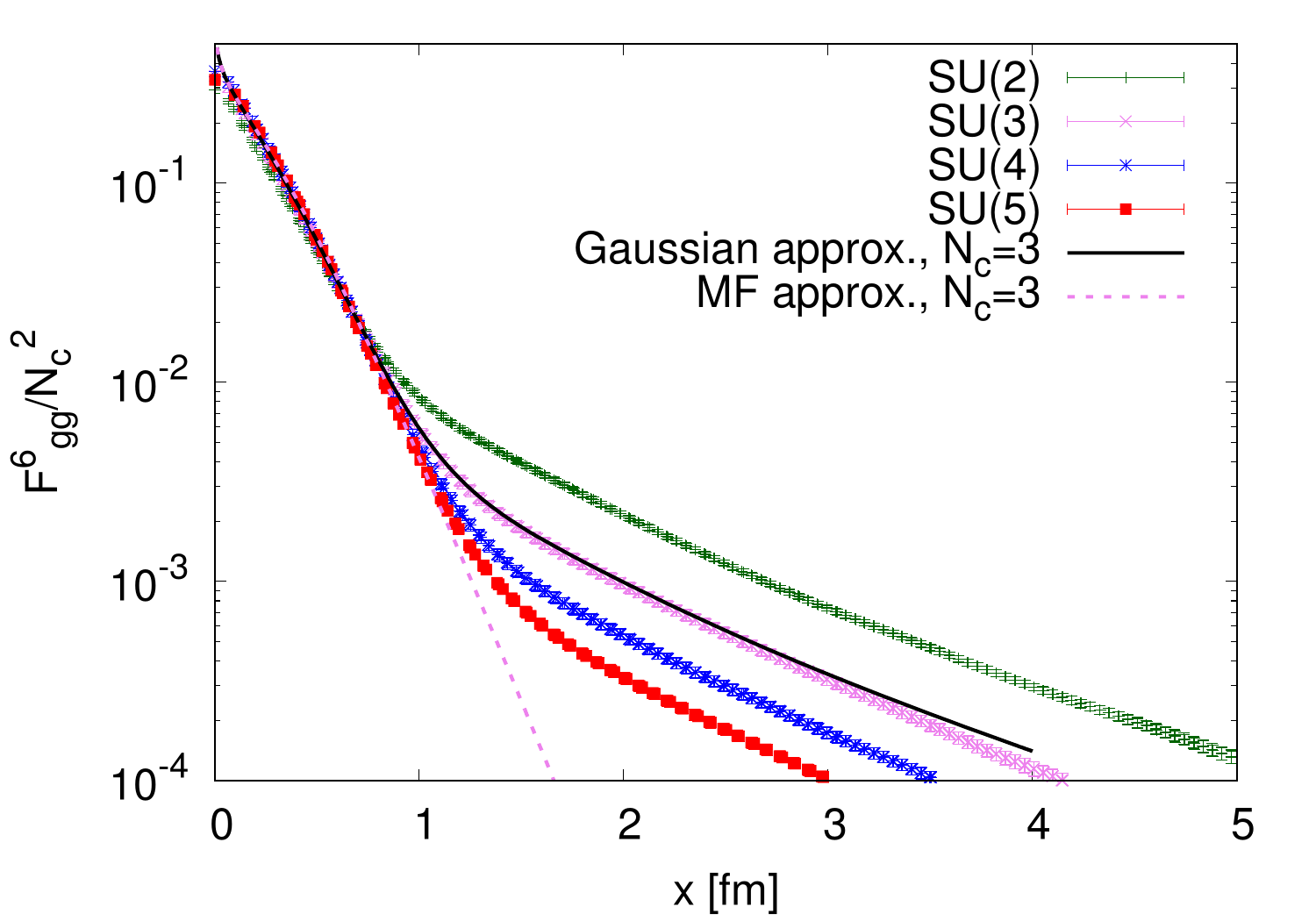}
    \includegraphics[width=0.35\textwidth, angle=270]{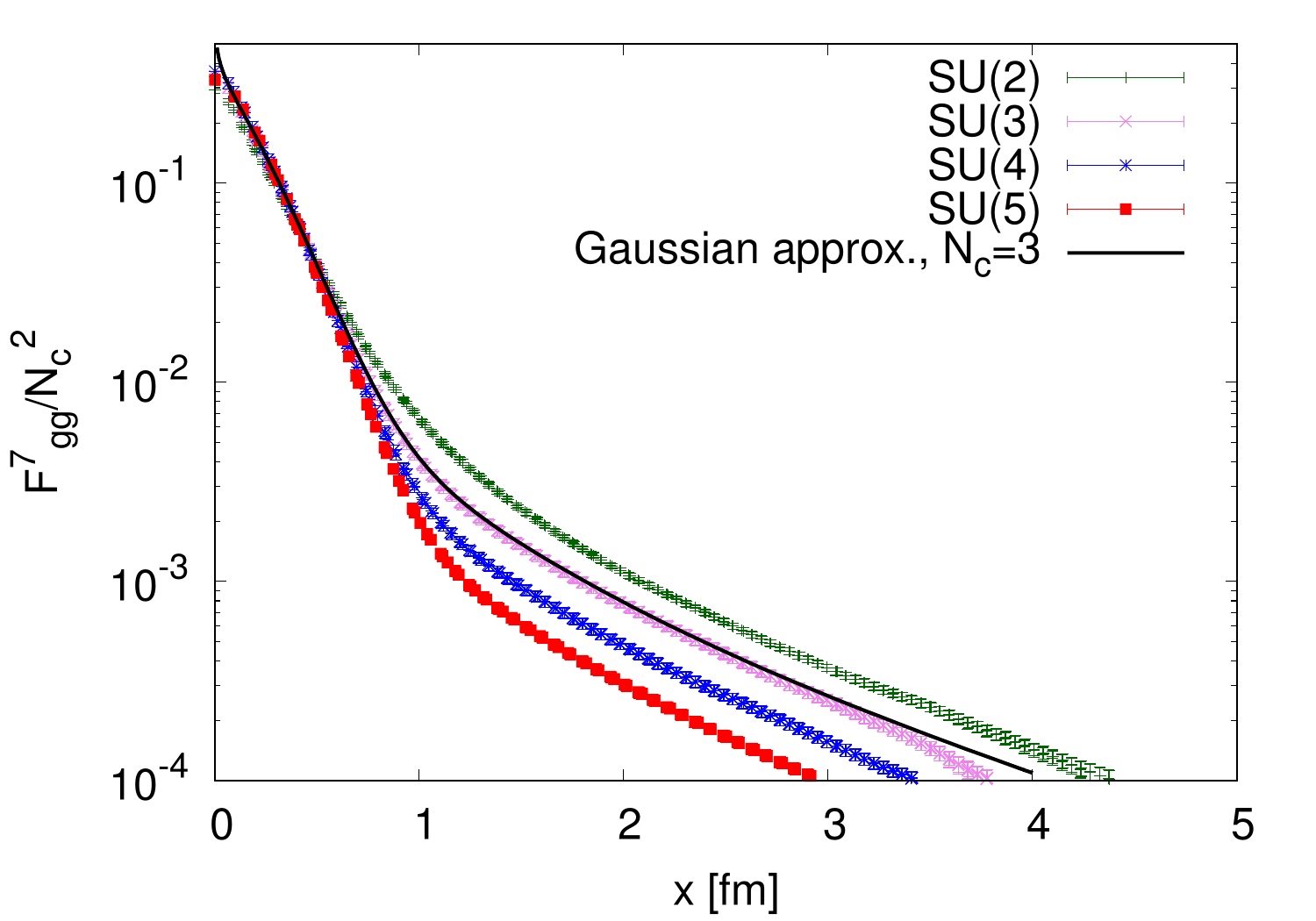}
    \caption{$N^2_c$-normalized gluon-gluon TMD distributions for the MV model for different values of $N_c$ together with the Gaussian and mean-field approximations for $N_c=3$ \cite{Caucal:2025zkl}. All TMD distributions show the scaling expected from Eq.~\eqref{eq:largeNc-limit-gg} and the convergence towards the mean-field approximation is of $\calo(1/N_c^2)$. For $\calf^{(3)}_{gg}/N_c^2$ all curves lie on top of each other due to exact scaling. 
    \label{fig. tmd separate gg}
    }
\end{figure}

\section{Conclusion and prospects}\label{sec: Conclusion and prospects}

In this work, we obtained the initial conditions for the leading twist TMD distributions, listed in \cite{Marquet:2016cgx,Bury:2018kvg}, using both a numerical implementation of the MV model for $N_c \in \{ 2,3,4,5 \}$ and analytical expressions in the Gaussian approximation for all $N_c$. 
These are illustrated in the quark-gluon sector by {Fig.~\ref{fig. tmd qg}} and {Fig.~\ref{fig. tmd separate qg}}; and in the gluon-gluon sector by {Fig.~\ref{fig. tmd gg}} and {Fig.~\ref{fig. tmd separate gg}}.
We found a good agreement between the two approaches.
We observed that the tails of the TMD distributions can be captured using the Gaussian approximation at fixed values of $N_c$, particularly at $N_c = 3$.

As far as the $N_c$ dependence is concerned, we confirmed that the Gaussian approximation at fixed $N_c$ and the mean-field + Gaussian approximation agree at small separation scales \cite{Dumitru:2011vk}, while the behavior of distribution's tails differs between the two approximations.
%
%
%
To explain these two features, we rely on distributions $\Omega_{qg}^\omega$ and $\Omega_{gg}^\omega$, for which content in terms of irrep $\omega$ is explicit.
We identify a set of irreps, see {Fig.~\ref{fig. omega qg}}, top row and {Fig.~\ref{fig. omega gg scaling}}, for which the associated distributions scale under separation redefinition provided the infrared behavior is scaled according to \req{eqs:Scaling_generic}.
We also identify a set of irreps, see  {Fig.~\ref{fig. omega qg}}, bottom rows, {Fig.~\ref{fig:Omega_large_Nc_qg}}, {Fig.~\ref{fig. omega gg not scaling}}, and {Fig.~\ref{fig:Omega_large_Nc_gg}}, for which there are slight deviation from $N_c$ scaling.
{We observe that the distribution width scales with the charge involved $C_{F/A} \pm C_\omega$. Therefore, the smaller the charge involved, the wider the associated distribution $\Omega_{ag}^{\omega}$.}
These distributions are in turn related to TMD distributions, at the operator level in {sec.~\ref{sec_birdtrack}} or in the Gaussian approximation in {sec.~\ref{sec: tmds from omega}.
{From the linear map, we can see that the distributions associated with irreps of small dimensions are effectively suppressed by powers of $1/N_c$.
These two features imply that, at small separations where the width has little impact, the TMD distributions are well approximated by the large $N_c$ result. 
Conversely, at larger distances, contributions from the large-$N_c$ limit have fallen off significantly, while distributions associated with smaller irreps are still within the range of their larger widths. In this case, the large-$N_c$ result becomes unreliable, so the finite $N_c$ result from the Gaussian approximation should be used instead.}
%
%

A side-product of our analysis is an interesting sum rule involving the seven gluon-gluon TMD operators for $SU(3)$, which is presented in \req{eq:sum_rule_gg_su3}. 
It follows from the fact that the irrep of dimension $K_0=\Nc^2 \tfrac{(\Nc + 1)(\Nc - 3)}{4}$ does not appear when $N_c = 3$. 

This study gives us a thorough understanding of the dependence of the initial conditions on the number of colors $N_c$ for the JIMWLK evolution over a large class of TMD distributions.
This ensures that the study of the JIMWLK evolution and its dependence on $N_c$ will not be misinterpreted due to an insufficient understanding of the initial conditions. Therefore, any potential deviation will be attributed solely to the evolution.

The next step in this project is to evaluate these TMD distributions after
JIMWLK evolution. For this evaluation, we plan to vary the value of $N_c$ in the same range $\{2,3,4,5\}$.
First, we should mention that the operators of interest in this manuscript are bilocal only.
The first step in evolution involves a three-point operator in the Balitsky hierarchy. Within the Gaussian approximation, this operator is a scalar in color space. Therefore, it can easily be obtained using the methods presented in this manuscript and could be useful to guide such analyses.
According to the MV model, an important assumption in the construction of the initial condition is time locality. This assumption follows from the fact that $t$-channel interactions are mediated by Glauber gluons.
This implies that there is no correlation between \textit{longitudinal slices of the target} along the projectile's light cone trajectory.
This assumption breaks down at the level of the JIMWLK Hamiltonian's action. This becomes clearer when the JIMWLK equation is written as a Langevin process, especially in symmetric form on either side of the shock wave \cite{Lappi:2012vw}.
A stochastic noise $\xi_\zvec^a$, is generated on one side of the shock wave and related to the other side of the shockwave $\tilde{\xi}_\zvec^a$, by a rotation given by an adjoint WL: $\tilde{\xi}_\zvec^a = U_\zvec^{ab, (\dagger)}\xi_\zvec^b$. 
This crossing of the shock wave explicitly correlates the different slices of the initial condition and previous steps of evolution with how the WLs are being rotated at a given step of evolution.
Because the evolution violates time locality, we can expect deviations between the JIMWLK-evolved operators and their Gaussian approximations expressed as functions of the evolved dipole.
This type of comparison is similar to the analysis performed for the quadrupole in \cite{Dumitru:2011vk}, which we plan to perform in the case of TMD distributions at several values of $N_c$.
It will be interesting to study whether the scaling behavior (and the mild deviation from it) observed in the expectation values of the operator, $\hat{\Omega}$, in the initial condition persists after several JIMWLK evolution steps.
From a phenomenological perspective, it is important to study whether the features of TMD distributions highlighted in this manuscript persist during JIMWLK evolution, especially the subleading-$N_c$ contributions to TMD distributions.

\acknowledgments

We thank Farid Salazar for his comments about mean-field results from \cite{Caucal:2025zkl}.

\noindent
We gratefully acknowledge Polish high-performance computing infrastructure PLGrid (HPC Center: ACK Cyfronet AGH) for providing computer facilities and support within computational grants no. PLG/2023/016656, PLG/2024/017690, PLG/2026/019139.

\noindent
We acknowledge Polish high-performance computing infrastructure PLGrid for awarding this project access to the LUMI supercomputer, owned by the EuroHPC Joint Undertaking, hosted by CSC (Finland) and the LUMI consortium through PLL/2025/08/018112.

\noindent
P.K. and F.C. acknowledge support from the Polish National Science Center (NCN) grant No. 2022/46/E/ST2/00346.
T.S. kindly acknowledges the support of the Polish National Science Center (NCN) grant No. 2021/43/D/ST2/03375. 

\appendix
\section{Numerical implementation details}

\subsection{Parameters used for the simulations}
\label{params_choice}

For the reference point for $SU(3)$ we used the following parameters:
\begin{equation}\label{eq:SU3_simu_ref_params}
    \mu = 0.7 \ \textrm{fm}^{-1}, \qquad m = 0.2 \ \textrm{fm}^{-1}, \qquad L=100.0 \ \textrm{fm}.
\end{equation}
Parameters for other $N_c$ can be found in Tab.\ref{tab:qg_gg_parameters}. The choice of those simulation parameters follows the discussion on scaling given in  {sec.~\ref{sec:dipole_fit}} and sec.~{\ref{N_c_scaling_omega_section}}.

\begin{table}[h]
\begin{center}
\begin{tabular}{|c|c|c|c|c|c|}
\hline
 &  & \multicolumn{2}{c|}{$qg$} & \multicolumn{2}{c|}{$gg$} \\
\hline
$N_c$ & $\mu$ [fm$^{-1}$] & $m$ [fm$^{-1}$] & $L$ [fm] & $m$ [fm$^{-1}$] & $L$ [fm] \\
\hline
2 & 0.7 & 0.15 & 133.33 & 0.1633 & 122.47 \\
\textbf{3} & \textbf{0.7} & \textbf{0.2} & \textbf{100} & \textbf{0.2} & \textbf{100} \\
4 & 0.7 & 0.2371 & 84.33 & 0.2309 & 86.60 \\
5 & 0.7 & 0.2683 & 74.54 & 0.2582 & 77.46 \\
\hline
\end{tabular}
\caption{Parameters for $qg$ and $gg$ cases.}
\label{tab:qg_gg_parameters}
\end{center}
\end{table}

\subsection{Wilson lines realization in the massive MV model}
\label{app:MV_impl}

A Wilson line at the transverse coordinate $\xvec$ is obtained on a per-realization basis according to:
\begin{equation}
    V_\xvec = \prod_{n=1}^{\text{erg}} e^{i\, A^a_n(\xvec) t^a}
\end{equation}
Each slice denoted by the subscript $n$ is independent, and the parameter ``erg'' corresponds to the number of small rotations that we perform in the gauge group. For our purpose, we found that a value of $n > 150$ provided stable results for the dipole and TMD distributions.  This parameter is fixed to $\text{erg} = 200$ for all the simulations used in this analysis.
The generators $t^a$ are represented as a set of $N_c^2-1$ matrices of dimension $N_c \times N_c$, which are normalized such that $\tr (t^a t^b) = \thalf \delta^{ab}$. 
In the case of $SU(2)$, the relevant matrices are the Pauli matrices, normalized by $1/2$. For $SU(3)$, the relevant matrices are the Gell-Mann matrices, normalized by $1/2$. For $SU(N_c > 3)$, the relevant matrices are the generalization of the Gell-Mann matrices.

We impose periodic boundary conditions on the grid in order to preserve translational invariance. 
Introducing dimensionless units (with a hat) on such a square grid as $\xvec = \hat{x} \ \frac{L}{N}$, where $L$ is the length of each of the circumferences of the torus and $N^2$ is the total number of nodes on the (square) torus, we have 
\begin{equation}
    V(\hat{x}) = \prod_{n=1}^{\text{erg}} e^{i\, A^a_n(\hat{x}) t^a}.
\end{equation}
The field at the node $\hat{x}$ is computed according to:
\begin{equation}
    A_n^a(\hat{x}) \equiv \text{Re}\ \frac{L}{2\pi} \frac{\mu}{\sqrt{\text{erg}}} \ 
    \text{FFTW}_{2d}\left\{
    \left[ \hat{k}^2 + \left( \frac{m\, L}{2\pi}\right)^2\right]^{-1}
    \nu^a_n(\hat{k}) \right\} (\hat{x})
\end{equation}
$\text{FFTW}_{2d}$ denotes the 2d Fast Fourier Transform implementation from the library \cite{FFTW05,FFTW98}.
For a function $f$ of two variables (or a vector in 2d space) stored as an array of size $N^2$, the normalization of the Fourier Transform is such that:
\begin{equation}
FT_{2d}^{-1}[FT_{2d}(f)] = f
\end{equation}
which we achieve by a normalization factor $1/N$ added to both $FT_{2d}^{-1}$ and $FT_{2d}$, since FFTW computes an \textit{unnormalized} Discrete Fourier transform.
We recognize within the square bracket the Green function of the Poisson equation expressed in lattice momenta $\kvec = \hat{k}\ \frac{N}{L}$. When expressed in the continuum (and infinite volume), this Green function takes the following form:
\begin{equation}
    \calg(\kvec) = \frac{1}{2\pi(\kvec^2 + m^2)},
\end{equation}
for which we recover the MV model in the limit $m \rightarrow 0$.
For each slice labeled by $n$, the quantity $\nu_n^a(\kvec)$ is a vector of dimension $K_A = N_c^2-1$. The distribution of each component is determined by the Gaussian distribution ${\cal N}(x ;\sigma)$, where the standard deviation $\sigma$ is set at a value of $1$:
\begin{equation}
    {\cal N}(x;\sigma) = \frac{1}{\sigma \sqrt{2\pi}}\ e^{-\thalf\left( \frac{x}{\sigma} \right)^2 }.
\end{equation}
This random variable is related to the color sources of the MV model by simply multiplying by $\mu$, which is a parameter of the simulation and characterizes the color charge density.
\begin{equation}
    \rho_n^a(\hat{k}) = \frac{\mu}{\sqrt{\text{erg}}} \nu_n^a(\hat{k})
\end{equation}
Thus, a realization is fully characterized by the set of parameters:
\begin{equation}
    \left\{ \text{erg}; N_c; N; L; m; \mu  \right\}
\end{equation}

\section{Finite volume effects}\label{App:Finite_volune}

Due to the long-range interaction, the control of finite volume effects becomes crucial in order to demonstrate the scaling with $N_c$ of the various distributions discussed in this work. In this Appendix, we gather some observations related to finite volume effects. We start with the discussion of the dipole amplitude, then we comment on the finite volume effects affecting the $\Omega_{ag}$ distributions, and eventually we conclude with the discussion of TMD distributions. Note, however, that a precise analysis of the relationship between the support of distributions and finite volume effects is beyond the scope of this manuscript. 

\subsection{Finite volume corrections to the dipole amplitude}

It was already alluded to in section \ref{app:MV_impl} that the numerical construction of the initial condition in the MV model can induce significant finite volume corrections. They originate from the long-range nature of gluon interactions present at tree-level. These may become important in the numerical setup where one imposes periodic boundary conditions in order to maintain translational symmetry, which turns out to be useful for the estimation of various correlation functions. In order to gain control over the long, power-law tails of the interaction potential, one introduces an effective mass parameter that suppresses them exponentially. In our setup, we have chosen a compromise between the infrared scale given by the volume $L = 100$ fm and the mass set to $0.2$ fm$^{-1}$. However, the dipole amplitude data were fitted with the infinite volume formula Eq.~\eqref{eq:fit_func_form}, and thus can exhibit some corrections due to the finite volume. They become visible at distances $2.5 - 3$ fm, as can be noticed in  Figs. \ref{fig:dip_fit_v2} and \ref{fig. dipole scaling}. Nevertheless, the hierarchy of scales chosen for the calculations makes these deviations small and irrelevant for the main points of our argumentation.

\subsection{Finite volume corrections to the \texorpdfstring{$\Omega_{ag}^{(i)}$}{Omega}}

Gaussian approximation formulae for $\Omega_{ag}^{(i)}$ being built out of the logarithm of the dipole amplitude may inherit finite volume distortions discussed in the previous section. They could be seen in our figures as deviations between the continuous curves corresponding to the Gaussian approximation and the numerical data at larger distances. However, a closer inspection of results for individual $\Omega_{ag}^{(i)}$ shows a more intricate situation. We observe stability in the data for the tail of $\Omega_{gg}^{(1)}$ over a wider range of $\textrm{x}$, and agreement with the curve obtained in the Gaussian approximation up to $\textrm{x} = 4 - 5\, \text{fm}$. 
This agreement spans three orders of magnitude on the vertical axis from $0\,\text{fm}$ to $5\,\text{fm}$. 
On the other hand, the data associated with $\Omega_{gg}^\omega$ where $\omega \neq 1$, all have a steeper slope. 
This is made explicit on the {r.h.s.} of {fig.~\ref{fig. omega gg scaling}} and {fig.~\ref{fig. omega gg not scaling}}, and it is understood to be a consequence of the higher charges involved in \req{omega_gg_recap}.
We observe a deviation between the data and the Gaussian approximation after the distributions also span three orders of magnitude.
Because of the different slopes of the distribution, this corresponds to different values of $\textrm{x}$.
For example, consider the bottom right panel of {fig.~\ref{fig. omega gg not scaling}} showing $\Omega_{gg}^{27}/K_{27}$.
In the region $\textrm{x} \sim 1.3 - 1.5\ \text{fm}$, deviations are noticeable, particularly for $SU(2)$, which is the case further away from the large-$N_c$ limit, where everything is expected to scale exactly, as illustrated in {fig.~\ref{fig:Omega_large_Nc_gg}}.
Conveniently for our study of TMD distributions, the tail behaviors are dominated by the contribution $\Omega_{gg}^{(1)}$, which appears to be the most insensitive to finite volume effects over a given range of $\textrm{x}$ of all $\Omega_{gg}$'s. A similar situation can be noticed in the quark-gluon sector, where  $\Omega_{qg}^{\overline{3}}$ has the longest tail and the longest agreement with the Gaussian approximation.

\subsection{Finite volume corrections to the TMD distributions}

Again, the deviations between the Gaussian approximation and the numerical data that are observed for some TMD distributions in the region $r>3$fm, e.g. $\calf_{qg}^{(2/3)}$ and $\calf_{gg}^{(5/6)}$, originate from the finite volume effects in the description of the dipole amplitude. This is the region where we expect our fit of the dipole to break down 
still, we can notice that some TMD distributions show stability in the behavior of the tail beyond $3 \, \text{fm}$. As explained in the preceding paragraph, the tails of the TMD distributions are dominated by the lowest irreps, either $\Omega_{qg}^{\overline{3}}$ or $\Omega_{gg}^1$.
These correspond to the lowest accessible charges and therefore have the smallest slope on a logarithmic scale.
Assuming finite volume effects appear in distributions spanning more than three orders of magnitude implies that these two distributions have the widest range of validity in $\textrm{x}$, a property that TMD distributions conveniently inherit.

\bibliographystyle{JHEP}
\bibliography{draft_bib}

\end{document}